\documentclass[12pt]{article}

\usepackage{arxiv}
\usepackage[utf8]{inputenc} %
\usepackage[T1]{fontenc}    %
\usepackage[hyperfootnotes=false]{hyperref} %
\usepackage{url}            %
\usepackage{booktabs}       %
\usepackage{amsfonts}       %
\usepackage{amsmath}        %
\usepackage{nicefrac}       %
\usepackage[nopatch=footnote]{microtype}      %
\usepackage{graphicx}
\graphicspath{{figures/}}
\usepackage[round]{natbib}
\usepackage{doi}
\usepackage{verbatim}        %
\usepackage{listings}        %
\usepackage{fancyvrb}       %
\usepackage{mdframed}       %
\usepackage{setspace}       %
\usepackage{makecell}       %
\usepackage{multirow}       %
\usepackage{svg}
\usepackage{transparent}
\usepackage{xcolor}
\usepackage{pdflscape}
\usepackage{rotating}        %
\usepackage{threeparttable}  %
\usepackage{tabularx}        %
\usepackage{siunitx}         %
\newcolumntype{Y}{>{\raggedright\arraybackslash}X}
\usepackage{longtable}
\usepackage{tikz}
\usetikzlibrary{positioning,arrows.meta,fit}
\usetikzlibrary{positioning,arrows.meta,fit,calc,backgrounds,shapes.geometric,shapes.symbols}

\definecolor{aelBlue1}{HTML}{1F4E79}   %
\definecolor{aelBlue2}{HTML}{2E75B6}
\definecolor{aelBlue3}{HTML}{7FAEDB}   %
\definecolor{aelBlue4}{HTML}{BDD7EE}   %
\definecolor{aelAccent}{HTML}{2F5C8F}  %
\definecolor{aelInk}{HTML}{333333}     %
\definecolor{aelGray}{HTML}{C9C9C9}    %
\definecolor{aelGrayFill}{HTML}{ECECEC}
\definecolor{aelGrayLine}{HTML}{9A9A9A}
\definecolor{aelStageA}{HTML}{F8FAFD}
\definecolor{aelStageB}{HTML}{F0F5FB}
\definecolor{aelStageC}{HTML}{E8EFF8}
\definecolor{aelStageD}{HTML}{E0E9F5}
\definecolor{aelStageLine}{HTML}{D3E0EF}
\definecolor{aelHitlFill}{HTML}{E6ECE6}
\definecolor{aelHitlLine}{HTML}{7C8F7E}

\newcommand{\aelHair}{0.5pt}   %
\newcommand{\aelThin}{0.8pt}   %
\newcommand{\aelBold}{1.0pt}   %

\newcommand{\aelNoHyphen}{\hyphenpenalty=10000\exhyphenpenalty=10000\relax}
\newcommand{\aelAgent}[2]{{\color{aelBlue1}\textbf{\textit{#1}}}\\[1.5pt]#2}
\newcommand{\aelTeam}[1]{{\color{aelBlue1}\textbf{#1}}}

\tikzset{
  ael/.style={
    >={Stealth[length=2.2mm,width=1.6mm]},
    line width=\aelBold,
    font=\sffamily\scriptsize,
    text=aelInk,
    node distance=5mm,
  },
  aelagent/.style={
    execute at begin node=\aelNoHyphen,
    draw=aelBlue3, fill=white, rounded corners=1.5pt, line width=\aelThin,
    text=aelInk, align=center, inner xsep=3pt, inner ysep=4pt,
    minimum width=21mm, minimum height=9mm, font=\sffamily\scriptsize,
  },
  aelagentcol/.style={aelagent, text width=29mm},
  aelagentshared/.style={aelagent, dash pattern=on 2.2pt off 1.8pt},
  aelagentopt/.style={aelagent, densely dotted},
  aelflowopt/.style={aelflow, densely dotted},
  aelart/.style={
    execute at begin node=\aelNoHyphen,
    draw=aelBlue3, fill=aelBlue4, rounded corners=1.5pt, line width=\aelThin,
    text=aelInk, align=center, inner xsep=3pt, inner ysep=4pt,
    minimum width=18mm, minimum height=9mm, font=\sffamily\scriptsize,
  },
  aelhitl/.style={
    execute at begin node=\aelNoHyphen,
    signal, signal to=east and west, signal pointer angle=115,
    draw=aelHitlLine, fill=aelHitlFill, line width=\aelThin,
    text=aelInk, align=center, inner xsep=2pt, inner ysep=3.5pt,
    minimum width=20mm, minimum height=8mm, font=\sffamily\scriptsize,
  },
  aelsource/.style={
    execute at begin node=\aelNoHyphen,
    cylinder, shape border rotate=90, aspect=0.28,
    draw=aelGrayLine, fill=aelGrayFill, line width=\aelThin,
    text=aelInk, align=center, inner xsep=3pt, inner ysep=5pt,
    text width=26mm, minimum width=29mm, minimum height=20mm,
    font=\sffamily\scriptsize,
  },
  aelout/.style={
    execute at begin node=\aelNoHyphen,
    draw=aelBlue1, fill=aelBlue1, rounded corners=2pt, line width=\aelThin,
    text=white, align=center, inner xsep=6pt, inner ysep=6pt,
    text width=36mm, minimum height=16mm, font=\sffamily\scriptsize\bfseries,
  },
  aelstage/.style={
    draw=aelStageLine, rounded corners=2.5pt, line width=\aelHair, inner sep=4.5mm,
  },
  aelstageA/.style={aelstage, fill=aelStageA},
  aelstageB/.style={aelstage, fill=aelStageB},
  aelstageC/.style={aelstage, fill=aelStageC},
  aelstageD/.style={aelstage, fill=aelStageD},
  aelstagenum/.style={
    circle, draw=aelBlue3, fill=white, text=aelBlue1,
    line width=\aelThin, inner sep=0pt, minimum size=3.8mm,
    font=\sffamily\scriptsize\bfseries,
  },
  aelstagetitle/.style={
    anchor=south, align=center, text=aelBlue1,
    font=\sffamily\footnotesize\bfseries,
    inner sep=0pt, xshift=2.7mm,
  },
  aelverdict/.style={
    font=\sffamily\scriptsize\itshape, text=aelInk, inner sep=1.5pt, align=center,
  },
  aelcheckframe/.style={
    draw=aelHitlLine, densely dashed, rounded corners=3pt,
    line width=\aelHair, inner sep=7mm,
  },
  aelflow/.style={draw=aelAccent, ->, line width=\aelBold},
  aelline/.style={draw=aelAccent, line width=\aelBold},
  aelfeed/.style={draw=aelAccent, ->, line width=\aelThin, densely dotted},
  aelcross/.style={draw=aelAccent, ->, line width=\aelThin, densely dashed},
  aellabel/.style={
    font=\sffamily\scriptsize, text=aelInk, inner sep=2pt, fill=white,
  },
}

\newcommand{\aelFigScale}{0.815}
\usepackage{pgfplots}
\pgfplotsset{compat=1.18}
\usepgfplotslibrary{polar}

\title{Agentic Workflows for Economic Research: Design, Implementation, and Evaluation}

\author{
  Herbert Dawid\textsuperscript{1},
  Philipp Harting\textsuperscript{2},
  Hankui Wang\textsuperscript{2,*},
  Zhongli Wang\textsuperscript{1,3},
  Jiachen Yi\textsuperscript{4} \\[3mm]
  {\normalfont\small\begin{tabular}{@{}c@{}}
  \textsuperscript{1}Faculty of Business Administration and Economics, Universität Bielefeld, Germany \\
  \textsuperscript{2}Université Côte d'Azur, CNRS, GREDEG, France \\
  \textsuperscript{3}Department of Economics \& Finance, Università Cattolica del Sacro Cuore, Italy \\
  \textsuperscript{4}School of Economics, University of Bristol, United Kingdom
  \end{tabular}}
}

\renewcommand{\shorttitle}{\small Agentic Workflows for Econ Research}

\hypersetup{
pdftitle={Agentic Workflows for Economic Research: Design, Implementation, and Evaluation},
pdfsubject={econ.GN, cs.AI},
pdfauthor={Herbert Dawid, Philipp Harting, Hankui Wang, Zhongli Wang, Jiachen Yi},
pdfkeywords={Economic research, AI Agents, Agentic Workflows, LLMs},
}

\begin{document}
\maketitle
{\renewcommand{\thefootnote}{*}\footnotetext{Corresponding author. Email: \href{mailto:hankui.wang@univ-cotedazur.fr}{\texttt{hankui.wang@univ-cotedazur.fr}}}}

\begin{abstract}
	Large Language Models (LLMs) can now assist across the economic research lifecycle, yet general-purpose LLMs and agent systems give no assurance that their outputs meet the discipline's bar. We present an agentic-workflow methodology that carries a research question through the full lifecycle, from ideation and literature review through modeling, data, implementation, estimation, and reporting, across seven specialized teams under a cross-team orchestrator, with human-in-the-loop review at each stage. The methodology incorporates two commitments. The first, \emph{trustworthy execution}, is a shared infrastructure that verifies outputs, records the provenance of every step, bounds each agent's authority, addresses agent-specific threats, and serves every LLM locally on open weights. The second, \emph{reasoning as an economist}, makes disciplinary checks explicit in the workflow: agents state identification strategies for causal claims, check structural models against stylized facts by simulation, and search for counterexamples to theoretical claims. Together they produce a first research prototype in hours, with every step's provenance and verification verdict recorded. A two-tier framework evaluates the workflows across fourteen dimensions, combining execution-log metrics with a dual-judge LLM-as-reviewer ensemble across all seven teams.
\end{abstract}


\section{Introduction}
\label{sec:Intro}

As computational capabilities expand and Large Language Models (LLMs) become more powerful, economists can use LLM agents in \textit{agentic workflows}, AI-assisted sequences of research steps with varying degrees of autonomy, to carry out parts of their research. Every stage of economic research, from ideation and literature review to data processing and empirical validation, involves time-consuming routine work, and researchers must propose novel questions while keeping up with a growing literature and increasingly complex datasets. Agentic workflows respond to these demands by automating routine tasks, integrating information from multiple online sources, and supporting complex analyses. The potential benefits for economists include proposing and refining research questions grounded in the existing literature, conducting literature reviews at a scale that would be infeasible through manual methods alone, and spending less time on data preprocessing. Furthermore, by documenting research steps in executable formats, these workflows enhance transparency and reproducibility.

A growing ecosystem of commercial and open-source AI research tools now exists, but few of these tools are built around the methodological needs of economics. General-purpose research agents generate hypotheses and run experiments end to end \citep{yamada2025aiscientistv2, agarwal2025autodiscovery, swanson2025virtuallab}. Within economics, LLM use has moved from a catalog of assistant-style use cases across the research lifecycle, covering ideation, background research, coding, data analysis, and writing \citep{korinek2023generative}, to guidance for economists building agents that plan, call tools, and carry out multi-step research tasks \citep{korinek2025aiagents}. The open question is no longer whether LLMs can help with individual research tasks but whether their output can be trusted, both within a single stage and when stages are chained into a pipeline. Within a stage, an LLM can produce an answer that reads plausibly but is wrong; across stages, an error in one becomes an input to the next, so that a fabricated citation can become a modeling assumption and a misread data series can become an estimate. A workflow therefore needs checks at every step and every hand-off, a record of which choices the agent made and which the economist made, and outputs held to the discipline's standards.

Economic research moves through a sequence of stages, from formulating a question and reviewing the literature to specifying a model, assembling data, implementing the model, estimating, and reporting, and each stage carries its own methodological standards, such as causal identification and the calibration of models against data. The general-purpose research agents above are organized around hypothesis search and experiment execution, not around these stages. The first version of this paper introduced agentic workflows to economic research and illustrated them with worked examples for ideation, literature review, modeling, and data \citep{dawid2025agenticv1}. We have since built the \textit{Agentic Econ Lab} (AEL), an open-source implementation of such a workflow\footnote{Code: \url{https://github.com/AgenticEconLab/agentic-econ-lab}; run outputs and expert ratings: \url{https://doi.org/10.5281/zenodo.23258640}.} that can run any single stage on its own or all of them in sequence, integrates the practice of human economists at each one, and is served on open-weight models. AEL covers seven applications, one for each stage: (i) proposing and refining research questions; (ii) reviewing and synthesizing the literature; (iii) specifying and calibrating economic models; (iv) collecting and preparing data; (v) translating models into verifiable code; (vi) estimating and validating empirical relationships; and (vii) interpreting and reporting results.

The contribution we claim for this framework is workflow-structural: it lies in how the workflow that produces research outputs is organized, checked, and recorded. Domain experts remain responsible for assessing the substantive validity of the models and interpretations it produces. Beyond the seven applications listed above, the paper makes four contributions to that structure. (viii) The architecture organizes the teams around the methodological dependencies of economic research, with a cross-team orchestrator, a bounded Model--Data feasibility cycle, and economist review in every stage. Unmet data requirements return to the modeler as revision requests or are recorded as limitations. (ix) The trustworthy-execution layer addresses reliability through output verification, accountability through provenance records and bounds on authority and budgets, safety through signed artifacts and sandboxed code execution, and privacy through local serving of every LLM on open weights. (x) Reasoning as an economist is built into the workflow through explicit steps: stating identification strategies for causal claims, checking structural models against stylized facts by simulation, reasoning within the theoretical frame the economist adopts, and searching for counterexamples with an adversarial Falsifier paired with the Theorist agent. (xi) A two-tier, fourteen-dimension evaluation framework adds Decision Quality, Economic Rigor, Simulation Fidelity, and Causal Validity to standard AI evaluation criteria. Six content dimensions are scored by two open-weight judges from model families different from the model that produced the outputs.

The framework serves two uses. The first is rapid research prototyping. Here a research prototype comprises a model specification, a dataset assembled for the research question, and a reduced-form estimate or a recorded explanation of why the proposed specification could not be estimated. The model carries a calibration verdict against cited targets, or a recorded reason why it could not be calibrated, and verification results and unresolved limitations accompany these outputs. The chained runs reported here completed in hours. The second use is onboarding: a researcher entering an unfamiliar subfield of economics can examine these outputs, their cited sources, and their verification records to study the subfield's methodological standards. Onboarding is the use we designed the framework for: a first prototype in an unfamiliar field can support research across the subfields of economics, between economics and the other social sciences, and in interdisciplinary work more broadly, and the workflow records the provenance, cited empirical targets, and calibration verdicts that others need before trusting such work.

The remainder of this paper is structured as follows. Section~\ref{sec:LR} reviews the literature on AI in economic research. Section~\ref{sec:method} presents the methodology and its five pillars, and Section~\ref{sec:workflow} describes the design and implementation of the seven teams and the pipeline that chains them. Section~\ref{sec:Examples} walks through one of 26 chained end-to-end runs on a live research question, Section~\ref{sec:evaluations} presents the evaluation framework and its results over these runs, and Section~\ref{sec:conclusion} concludes.

\section{Literature Review}
\label{sec:LR}

An agentic workflow for economic research is useful only if economists can trust what it produces. AI assistance can help on some tasks and hurt on others within a knowledge workflow. Access to an LLM raised the speed and quality of professional writing \citep{Noy2023}. On 18 consulting tasks within GPT-4's capability frontier, consultants given access to it completed 12.2\% more tasks and completed them 25.1\% more quickly on average, with higher-quality solutions, than consultants without AI. On a complex managerial task selected to lie outside that frontier, however, consultants with GPT-4 access were 19\% less likely to produce a correct solution \citep{DellAcqua2023}. AI can add more value when it executes contiguous chains of tasks than when it executes isolated ones \citep{demirer2026chaining}, and which tasks to delegate and which to reserve for human judgment can be set by a sufficient-statistic rule \citep{agarwal2025designing}. Applied to research, these results favor a workflow that chains the stages of a project and places human judgment at the points of delegation.

This section reads the literature for the conditions under which that trust is warranted. We first set out the four requirements that any research agent must meet, then follow the research lifecycle stage by stage, from ideation to reporting, turn to how such systems are evaluated, and close with the gaps the rest of the paper addresses.

\subsection{Reliability, Accountability, Safety, and Privacy}
\label{sec:LR-requirements}

An agent that handles confidential data, makes analytical decisions, and shapes a scholarly output has to meet four requirements before it is deployed: its outputs must be reliable, its use must be accountable, its actions must be safe, and the data it touches must stay private.

Reliability comes from the structure around the model more than from the model itself: hallucination falls when generation is grounded in observable outcomes and subjected to verification and external feedback \citep{Durante2024, schneider2025generative}. The components of such structure are well studied. Eliciting intermediate reasoning steps improves multi-step performance \citep{wei2022chain}; interleaving reasoning with actions lets a model consult its environment while keeping a reasoning trace \citep{yao2022react}; tool use ties outputs to external evidence and computation, through a browser \citep{nakano2021webgpt} or a modular set of tools \citep{karpas2022mrkl}; and multi-agent debate subjects claims to adversarial scrutiny before they are accepted \citep{du2023improving}. Surveys of agentic LLMs identify decomposition, role specialization, tool use, and critique as the means by which a task is broken into checkable steps \citep{plaat2025agentic, wang2024survey, guo2024large, xi2025rise}. General frameworks implement iterative refinement, self-correction, and code-grounded verification as workflow components \citep{wu2024autogen, tang2025autoagent, wang2024executable}, and domain deployments such as agentic auditing follow the same design \citep{Schreyer2025}. A research workflow built from these components makes each step checkable.

Accountability matters because socio-economic settings magnify the consequences of bias: biased systems lose user trust and with it the benefit of the technology \citep{ferrara2024fairness}, and generative AI may widen socioeconomic inequality and change how policy is made \citep{Capraro2024}. Such biases are documented in LLMs themselves: their outputs carry systematic decision and gender biases \citep{corazzini2025decision}, political bias and value misalignment appear in generative systems \citep{Motoki2025}, and LLM assistance can enable bidder-side manipulation in fair-division settings \citep{verma2025ai}. When an output passes through a pipeline of interconnected services, an ``AI supply chain'', responsibility for it becomes diffuse \citep{hopkins2025ai}, and effective use may require contracting and governance structures that reconcile productivity gains with control \citep{bastani2025human}. A workflow that delegates analytical choices to an agent must therefore record where each choice was made and by whom. The instruments for such a record exist: production agent platforms expose traces, token usage, tool invocations, and memory operations as inspectable records \citep{awsAgentCoreObservability2026, nvidiaAIQBlueprint2026}, and system-level observability links an agent's semantic trace to its process, filesystem, network, and runtime behavior \citep{agentsight2025}. For research, this telemetry belongs to the methodological record: it shows how evidence was retrieved, which code ran, and where a human intervened.

Safety evaluation has moved from static prompts to agents acting on real tools: on realistic multi-turn tasks with browsers, code execution, shells, messaging, and filesystems, frontier systems still show substantial safety failures \citep{vijayvargiya2025openagentsafety}. Evaluation must therefore cover the full action loop of memory, tool invocation, and collaboration, and its failure modes of prompt injection, action hallucination, and runaway execution \citep{arunkumar2026agenticarchitecture}. Standards bodies have begun to codify these threats. The Top~10 for Agentic Applications of the Open Worldwide Application Security Project lists ten threat categories specific to agents, among them goal hijacking, tool misuse, memory poisoning, and cascading failures \citep{owasptop10agentic2026}, and the US National Institute of Standards and Technology is developing standards for agent identity, authorization, and inter-agent communication \citep{nistAgentStandards2026}. The risk compounds for workflows that accumulate retrieved material across sessions: one poisoned document in a retrieval database can spread through an ecosystem connected through retrieval-augmented generation and extract data from it \citep{cohen2024unleashing}. The infrastructure through which agents reach tools and one another has developed alongside these evaluations. The Model Context Protocol (MCP) specifies how an LLM application connects to tools and data sources and is supported by the OpenAI Agents SDK \citep{openaiAgentsSDK2026} and the Microsoft Agent Framework \citep{microsoftAgentFramework2026}. The Agent2Agent (A2A) protocol specifies communication between agents across process boundaries \citep{a2aprotocol2025}. Each such channel, to a tool, a memory store, or another agent, is a route by which the threats above reach a workflow.

Privacy is the requirement that most directly shapes deployment, and three layers of protection are available. The first is contractual: providers publish data-use and retention policies \citep{OpenAI2025, Anthropic2025} within broader frameworks for accountability and ethical oversight \citep{floridi2019unified, leslie2021artificial, jobin2019global}, yet under any such policy, prompts and outputs sent to a hosted model are processed, and may be retained, on the provider's infrastructure on terms the provider sets. The second is API isolation, which separates inference from training; hosted agents nonetheless remain exposed to prompt injection \citep{kumar2024ethics}, which the agent-level evaluations above observe in practice \citep{vijayvargiya2025openagentsafety, agentsight2025}. The third is local deployment: on-device computation, as studied in edge intelligence and federated learning, keeps sensitive data from leaving the local environment \citep{bonawitz2022federated, wang2025empowering}. A workflow that serves its models locally keeps both the data and the model under the researcher's control.

\subsection{Across the Research Lifecycle}
\label{sec:LR-lifecycle}

\paragraph{Ideation.} Literatures are fragmented, time is scarce, and synthesis across fields is costly. LLMs can generate, refine, and screen candidate questions, and tools exist for parts of the stage: instruments for survey design and content analysis \citep{Bail2024}, JEL-code classification \citep{Heikkil2025}, and a human-in-the-loop system that carries an economic intuition to literature-grounded hypotheses and executable computational experiments \citep{chen2026agenteconomist}. For creative work the evidence cuts both ways: model-generated ideas raise the quality of individual output, most for the least creative writers, but reduce the diversity of what a group produces \citep{Doshi2024}. For research, a field converging on model-suggested questions would lose variety, so ideation inside a workflow needs a human editor.

\paragraph{Literature review.} A review is only as good as its retrieval and its citations. Without retrieval, GPT-4o hallucinated 78--90\% of its citations in a large literature-synthesis benchmark; a retrieval-augmented model over 45 million open-access papers reached citation accuracy on par with human experts, and experts preferred its answers to expert-written ones in half or more of comparisons \citep{asai2026openscholar}. A literature agent optimized for factuality matched or exceeded subject-matter experts on retrieval, summarization, and contradiction-detection tasks \citep{skarlinski2024paperqa2}, and multi-perspective question asking produces grounded, outline-first survey articles \citep{shao2024storm}. Two consequences follow: every claim in an automated review should be tied to a retrieved source, and the review can be no more complete than the corpus it retrieves from.

\paragraph{Modeling and simulation.} Human-subject experiments are costly and their samples narrow. LLMs reproduce economic behavior across risk, time, and social preferences \citep{Chen2023a}, mimic strategic reasoning in games \citep{Akata2023, Brookins2024}, replicate established behavioral results at low cost \citep{Horton2023}, and have been used to develop theory \citep{Tranchero2024, Zheng2024} and to synthesize policy evaluations \citep{Asatryan2024}. Simulation is moving from prompted mimicry to theory-grounded agents: competitive multi-agent environments let market dynamics among LLM agents be studied \citep{Zhao2024}; ``general social agents'' built from social-science theory, seed-game data, and pretrained knowledge forecast human play in new settings better than behavioral and game-theoretic baselines \citep{manning2026generalsocialagents}; and ``virtual agent economies'' turn agent interactions into measurable counterfactual outcomes \citep{tomasev2025virtual}. Related pipelines that combine structured data, news retrieval, and LLM reasoning forecast international events \citep{ye2024mirai}. An LLM can propose and populate a model, but its parameters and its fit to data must be established by the methods an economist would use without it.

\paragraph{Data.}\label{sec:LR-data} LLMs reduce the cost of turning unstructured material into measures that can enter an econometric model. From text they have produced indices of corporate sustainability commitments \citep{Bauer2024}, economic policy uncertainty narratives \citep{hartley2025narratives}, inflation sentiment that marginally improves a Federal Reserve nowcast \citep{allard2024inflationnowcasting}, central-bank mandates and stances \citep{bertsch2025central}, efficiency--equity framing in economics research \citep{galiani2026measuring}, the content of categorical labels \citep{Chen2024a}, AI-related products in trade data \citep{waugh2026tradeai}, and structured records from 3.3 million Chinese policy documents \citep{fang2025decoding}. Survey-style responses generated by LLMs can probe preference heterogeneity, although willingness-to-pay estimates are often inaccurate unless the model is fine-tuned on prior human data \citep{Brand2024}. Agents given a fixed research instrument, explicit evidence policies, and quality tracking construct datasets from public primary sources \citep{afonso2026deepresearch}. Other modalities are measured too, each with its own pipeline: night-time satellite imagery for growth \citep{henderson2012outerspace} and poverty \citep{jean2016satellite}, the Federal Reserve chair's voice \citep{gorodnichenko2023voice} and facial expressions \citep{curti2023faceit} at press conferences, image-based investor sentiment \citep{gu2025giffluence}, and multimodal inputs to investment research \citep{Li2023b}. Each constructed measure is only as credible as its validation against the source material, so a workflow must record how every variable it constructs was produced.

\paragraph{Coding.} Specialized econometric languages and hard-to-reproduce code make implementation costly. Programmers with GitHub Copilot completed a task 56\% faster than a control group \citep{Peng2023}, and a field experiment at a large technology firm found gains of similar size, concentrated among less experienced programmers \citep{Gambacorta2024}; a dataset of agent-authored pull requests lets adoption and review of coding agents be studied in real repositories \citep{aidev2026codingagents}. Agents carry specific tasks to completion: database interaction \citep{xue2024demonstration} and quantitative factor discovery through hypothesis generation, code, backtesting, and feedback \citep{li2025rdagentquant}. A second pattern treats expert-built software as a tool to be called rather than code to be written: an agent given eighteen expert-designed chemistry tools plans and executes multi-step syntheses without writing chemistry code \citep{bran2024chemcrow}, and a guide exists for economists building tool-using agents \citep{korinek2025aiagents}. An agent that calls an established economic toolbox runs code that economists have already tested and used; code that an agent writes anew has to be verified before its results are relied on.

\paragraph{Estimation.} LLM outputs used as data can contaminate inference: training-data leakage threatens identification, and an econometric framework incorporates LLM components into estimation while accounting for the leakage \citep{ludwig2025large}. LLMs can recover a decision maker's preferences from choice data \citep{Kim2024} and have been benchmarked against utility theory \citep{Ross2024}. An econometrics agent that plans an analysis, writes and executes code, and reflects on errors outperforms general-purpose LLMs on expert-level econometric tasks \citep{chen2025econometricsagent}. The estimator itself remains the economist's: the agent's role is to run established estimators and report their diagnostics.

\paragraph{Reporting.} LLMs draft policy advice \citep{Biancotti2023Loquacity}, generate finance studies that reviewers rate as acceptable \citep{Dowling2023}, and assist economic research from idea to write-up \citep{Leigh2024}. Agentic systems have produced whole manuscripts: one passed peer review at a workshop of the International Conference on Learning Representations \citep{yamada2025aiscientistv2}, and another spans literature review, experimentation, and report drafting \citep{schmidgall2025agentlaboratory}. When such manuscripts were examined by human reviewers, however, they contained hallucinated results and reproducibility failures, and most of those an automated reviewer accepted contained fabricated or unsupported claims \citep{gaddipati2026mlreplicate}. A report produced by a workflow should therefore trace every number to a computed result.

\subsection{Evaluation of LLMs and Research Workflows}

A trusted workflow needs an evaluation as structured as the workflow itself. LLM performance is multidimensional (capability, robustness, bias and safety, efficiency) and sensitive to task and measurement choices. Surveys organize evaluation around what, where, and how to evaluate and identify leakage and missing coverage as recurring pitfalls \citep{chang2024survey}, and a holistic evaluation framework reports trade-offs beyond accuracy, scenario by metric, which argues for multi-metric reporting over a headline score \citep{liang2022holistic}.

General capability is measured on standardized suites such as MMLU (Massive Multitask Language Understanding) \citep{hendrycks2020measuring}. Research assistants produce open-ended output, and for that the field uses preference- and rubric-based assessment: LLM judges agree substantially with human preference but show verbosity, position, and self-enhancement biases \citep{Zheng2023}, and learned metrics such as BERTScore and BLEURT match human judgments of generated text better than surface overlap \citep{zhang2019bertscore, sellam2020bleurt}. Factuality has its own instruments: TruthfulQA tests whether a model avoids common misconceptions \citep{lin2022truthfulqa}, FactScore verifies claim-level facts against trusted sources \citep{min2023factscore}, SelfCheckGPT detects hallucination by self-consistency sampling \citep{manakul2023selfcheckgpt}, and RAGAS scores retrieval and grounded generation without reference answers \citep{es2024ragas}. Given the self-enhancement bias of LLM judges, the judges of a workflow's output should come from model families other than the one that produced it.

A research workflow can be evaluated at three further levels. At the domain level, FinQA tests numerical reasoning over financial reports \citep{chen2021finqa}. At the agent level, AgentBench assesses tool use and sequential decision-making \citep{liu2023agentbench}, and practical guidance covers what to measure in agentic systems \citep{grace2026demystifying} and how to build agents that audit alignment \citep{bricken2025building}. At the workflow level, human--AI collaboration in the science of science has been measured as a workflow \citep{shao2025sciscigpt}, expert-level scientific tasks frame the evaluation of general scientific agents \citep{chai2025scimaster}, and a protocol for networks of autonomous scientific agents raises system-level evaluation questions \citep{jiang2025scp}. An economics workflow needs, in addition, criteria specific to the discipline, such as whether an effect is identified and whether a model is calibrated against data.

Single-run benchmarks miss reliability. Measured along consistency, robustness, predictability, and safety with twelve metrics across fifteen models, agent reliability has improved only slightly as capability has grown \citep{rabanser2026reliability}. Running twelve multi-agent systems under repeated runs, backend models, and tool configurations shows substantial run-to-run variance, with architecture the main driver of resource use and reproducibility \citep{ma2026maestro}. A research workflow must therefore be run more than once, with the variance reported alongside the mean.

\subsection{Research Gaps}

Agentic workflows that operate across several research stages with bounded autonomy are rare in economics. Elsewhere, end-to-end systems exist: agentic tree search carries machine-learning research from hypothesis through experiments \citep{yamada2025aiscientistv2}, and a six-agent platform covers chemical synthesis from literature search through screening, kinetics, optimization, scale-up, and purification \citep{ruan2024llmrdf}. Two gaps follow from the reading above.

The first gap is the absence of an implemented, open workflow spanning the lifecycle. The first version of this paper proposed such a workflow for economics \citep{dawid2025agenticv1}. The systems that have appeared since each cover part of it: data construction \citep{afonso2026deepresearch} and the path from an economic intuition to a simulated experiment \citep{chen2026agenteconomist}. As of this writing, none spans ideation, literature, modeling, data, code, estimation, and reporting with a public, runnable implementation. Whether such a workflow is reproducible, auditable, and methodologically valid under realistic constraints is the question the rest of this paper takes up.

The second gap concerns reliability, accountability, safety, and privacy (Section~\ref{sec:LR-requirements}) in systems that produce research. When six autonomous research systems, among them AI Scientist-v1 and v2 and Agent Laboratory, were run on papers from the 2025 International Conference on Machine Learning, human reviewers found methodological flaws, hallucinated experimental results, and reproducibility failures in all of them, and 59\% of the manuscripts an automated reviewer accepted contained fabricated or unsupported claims \citep{gaddipati2026mlreplicate}. Frontier coding agents reproduce a large share of published social-science findings from released materials, yet subtle prompt framing can nudge them toward confirmatory specification search \citep{alizadeh2026socsci}. A workflow that produces research must therefore make its outputs verifiable, record who made each choice and how each result can be reproduced, defend its infrastructure against the threats set out there, and keep its data under the researcher's control.

The lifecycle review above reads each stage on its own, but the stages also depend on one another: a literature review supplies the evidence for a modeling assumption, a model specification states what the data must measure, and an estimate carries identifying assumptions that limit its interpretation. A research workflow therefore needs checks on what passes between stages as well as on each stage's output, and economist review where a question, specification, or interpretation must be revised.

Sections~\ref{sec:method}--\ref{sec:evaluations} describe a workflow designed against both gaps and report its evaluation.

\section{Methodology}
\label{sec:method}

We introduce a methodology for economic research built on agentic workflows: autonomous, iterative processes in which LLMs carry out research tasks under human-in-the-loop oversight. Spanning the full lifecycle, from ideation and literature review through model specification, data collection, implementation, and empirical estimation, to interpretation and reporting, it lets agents act with substantial autonomy while economists retain control at the decisions that matter. The following subsections describe the five pillars of this methodology: agent design, human-in-the-loop oversight, workflow coordination, reasoning as an economist, and trustworthy execution.

\subsection{Agent Design}

Each agent in the system is engineered to manage a clearly defined responsibility within the economic research lifecycle.
The design is based on the following core principles:

 \textbf{Defined Responsibilities}: Each agent is assigned a specific scope of focus within the economic research pipeline. For example, \texttt{Ideator} surfaces candidate research concepts from economic phenomena documented in published research and emerging web discussions; \texttt{GreyScout} searches grey literature such as working papers, policy reports, and preprints; \texttt{DataCleaner} turns economic data from different sources into a validated, aligned, analysis-ready dataset; \texttt{ModelDesigner} formulates a chosen theoretical framework into precise mathematical form;
 \texttt{Estimator} proposes econometric specifications for a deterministic harness, a fixed rule-based procedure rather than an agent, to estimate and verify;
 and \texttt{Proofreader} suggests wording and clarity improvements against disciplinary conventions on an already-verified draft.

\textbf{Inter-Agent Communication}: Agents exchange structured, typed artifacts in a fixed handoff sequence that mirrors the economic research workflow. Each agent's output becomes a defined input for the next, rather than open-ended conversation between LLMs.
For instance, \texttt{Estimator} does not describe its regression to \texttt{Validator} in prose. It hands over an \texttt{EstimationSpec} that names the dependent variable, the regressors, their transformations, and the sign the theory predicts for each coefficient. The deterministic harness estimates exactly that specification, and \texttt{Validator} and \texttt{HypothesisTester} receive the coefficients, standard errors, and diagnostic statistics as fields of the same record, so every check is run on the specification that was estimated rather than on a restatement of it.

\textbf{Error and Escalation Pathways}: Each agent incorporates mechanisms to detect issues within its domain and either resolve them internally or escalate them. 
For example, when one of the literature databases the workflow searches stops responding, the search continues with the others and records which database failed; the problem is resolved within the stage. When a live data fetch fails, by contrast, \texttt{DataCollector} cannot recover the missing series itself, so it escalates: it flags the series as simulated, the estimation excludes it, and the report discloses it.

\textbf{Adaptive Mechanisms}: Agents are designed to revise a strategy or specification if the initial attempt proves inadequate, and this adaptation is not confined to a single agent; it can span stages.
For instance, when \texttt{ModelDesigner} specifies what data a model needs, the data agents report back which requirements the available data can actually support, which can only be approximated with a proxy, and which cannot be met at all.
Where the data fall short, \texttt{ModelDesigner} revises the specification accordingly.
This exchange is bounded, and each revision must make progress toward resolving the shortfall or the cycle stops, with the remaining gap documented as a limitation.

The design and implementation of these agents are fundamentally guided by human expertise. 
Economists define each agent's responsibilities, knowledge boundaries,
and interaction protocols based on established research methodologies and domain-specific requirements. 
Each agent operates only within the boundaries so defined.
Economists carefully structure the communication pathways between agents, whether sequential or parallel, to mirror the logical progression of economic research workflows.
What the system can do therefore depends on the economists' input at this design phase.
Furthermore, ethical considerations such as data privacy, citation integrity, and methodological transparency are embedded into agent protocols through human oversight.

\subsection{Human in the Loop}
Human economists are embedded at specific checkpoints throughout the workflow to guide and validate the autonomous processes.
As with the agents' responsibilities and communication pathways, the economists themselves decide where these checkpoints sit when they design the workflow.
Their roles include:

\textbf{Initial Review and Prioritization}: 
As raw research ideas first emerge, economists review and prioritize them,
and only the questions they select proceed. At this filter, economists weigh three things: whether a question is new relative to the literature, whether it can be answered with the data and identification strategies available, and what an answer would contribute, to economic understanding or to policy.

\textbf{Iterative Feedback}: 
At every stage, economists annotate and correct agent outputs before the next stage builds on them, steering the work toward their own standards of rigor and the economic debates they judge relevant.
During ideation, for example, they narrow a research question's scope, sharpen its theoretical grounding, and check that the available data can answer it; during modeling and estimation, they revise the assumptions and specifications the agents propose.

\textbf{Methodological Validation}:
Economists judge whether the chosen methods suit the research question and the evidence, by the standards of the subfield concerned: for example, whether a theoretical model's assumptions are consistent and its equilibrium well defined, whether an identification strategy supports a causal claim, whether a calibrated or simulated model reproduces the facts it targets, or whether an experiment has the power to detect the effect it seeks.

\textbf{Quality Control and Verification}:
Throughout analysis, interpretation, and documentation, economists check that each result is correct and that its interpretation stays within what was shown: for example, that a derivation holds, that a simulated result survives changes in its parameters, that an estimate is read within the limits of its identification, and that a welfare or policy statement follows from the model behind it. These checks target results that are correctly computed but wrongly interpreted or overstated.

\textbf{Ethical Oversight and Bias Mitigation}:
Throughout all stages, economists monitor for ethical concerns: data privacy in sensitive economic datasets, the distributional implications of policy recommendations, and algorithmic biases that might skew economic analyses along socioeconomic, racial, or gender lines.
This oversight matters most when research involves vulnerable populations or carries significant policy implications.

\textbf{Final Approval and Archiving}: 
In the final stages of formatting, replication, and version control, economists review and sign off each component against disciplinary standards, ethical guidelines, and replication requirements. 
The review includes examining data documentation, assessing code accessibility and clarity, and evaluating the comprehensiveness of sensitivity analyses.

The quality of these human checkpoints determines the value of the research output.
This human-in-the-loop (HITL) framework thus serves as both a quality assurance mechanism and an ethical safeguard against potential misuse or misapplication of AI in economic research.

\subsection{Workflow Coordination}

This methodology coordinates AI agents and human economists across an interconnected, multi-phase workflow that mirrors traditional economic research, assigning the work to agents and the judgment calls to economists.
Key phases include:

\textbf{Research Question Formulation and Initialization}: Various specialized agents work together to generate, enrich, and refine research questions within economic paradigms. The workflow supports both autonomous ideation, where agents discover research questions from scratch by analyzing economic trends and literature gaps, and human-initiated ideation, where economists provide initial ideas that are systematically enriched and developed by the agents. Human economists judge the proposed questions for novelty, answerability, and contribution before they proceed.

\textbf{Automated Literature Exploration}: Multiple agents systematically discover and synthesize relevant economic literature, while management agents keep the collection up to date as new work appears. However, these agents may occasionally hallucinate non-existent sources or misinterpret economic theories. Therefore, specialized supervision agents with conservative verification thresholds check citations and theoretical interpretations, with human economists providing the final validation of the literature foundation.

\textbf{Model Specification}: Agents translate the chosen theoretical framework into a formal model, specifying its equations, variables, functional forms, and the structure that identifies its parameters (e.g., DSGE systems, structural equations, or reduced-form instrumental-variable designs), and set out the data the model will require. A deterministic harness then estimates the model's parameters against cited external targets and returns a calibration verdict, including uncalibratable where the evidence does not identify them. Human economists check the specification for internal coherence and for whether its parameters are identifiable before any data are gathered.

\textbf{Data Collection and Preprocessing}: Specialized agents gather and prepare economic datasets (e.g., panel data, time series, cross-sectional surveys), followed by quality-assurance agents that check each dataset and record how it was produced. Human economists collaborate with these agents to address complex data issues, such as structural breaks in macroeconomic time series or selection bias in microeconomic surveys, that call for domain judgment.

\textbf{Model Implementation and Coding}: Agents derive runnable code from the model's equations and test it by running it; a component that cannot be translated is declined and reported, not approximated. Human economists review the code for fidelity to the specification, since an implementation that runs is not yet one that faithfully encodes the model; a discrepancy goes back to the agents for correction, or to the model specification when the equations themselves must change, and the stage ends when code and specification agree or the remaining gap is recorded as a limitation.

\textbf{Empirical Estimation}: Agents propose an estimation specification, reduced-form or structural, which a deterministic harness estimates; they then run the diagnostics the method calls for (e.g., residual tests for a regression, over-identification tests for a moment-based estimator) and carry out robustness and sensitivity checks. Human economists judge whether the identifying assumptions hold before the estimates are interpreted.

\textbf{Interpretation and Reporting}: Agents derive economic insights from the validated results (e.g., elasticity interpretations, standardized effect sizes, policy-relevant magnitudes), assemble supporting exhibits, proofread the draft, and format outputs to economic-journal conventions; they also package, track, and archive the project's assets so that every step can be traced to its inputs and its deterministic steps re-run. The economist owns the interpretation and the write-up, treats the agents' outputs as provisional (aware that both humans and AI can misread statistical results or overstate causal claims), and determines which version constitutes the authoritative research record.

Across these phases, the agents carry out the work and the economists make the judgment calls: a phase's output moves on to the next phase only after passing the checkpoint the economists have set for it, and what they return there, a correction, a revised question or specification, or an accepted limitation, directs the agents' next round of work.

\subsection{Reasoning as an Economist}
\label{sec:econ-intelligence}

Economic reasoning draws on a wide range of methods, from optimization and equilibrium analysis to causal inference and structural modeling, among many others. Whatever the technique, a claim must still meet standards general enough to build directly into a workflow: an empirical effect must be identified, the mechanism behind it must be structurally coherent, and the theory it rests on must survive deliberate refutation. Integrating economics into an agentic workflow means making these standards explicit and checkable at the point in the pipeline where they apply. The following practices implement these standards, drawn from the broader universe of economic reasoning.

\textbf{Causal identification}: Identification comes first, and the workflow makes it an explicit step in its own right: when an agent proposes an empirical relationship, it must state the strategy that identifies it (an instrumental-variable, natural-experiment, or selection-on-observables argument) and put it up for scrutiny; a result that fits the data well but rests on no such argument is not allowed to stand as causal. Under the stated strategy the workflow then obtains a second estimate by a different estimation method and compares it with the agent's estimate; any divergence between the two estimates is reported rather than assumed away.

\textbf{Structural coherence and simulation}: Some claims rest on a model of an economic mechanism, such as how the economy responds to a policy or a shock. For those, the workflow runs the model as a simulation and checks its behavior against known stylized facts before any counterfactual is drawn from it: for example, it compares a macroeconomic model's simulated moments and impulse responses, or a market model's simulated price and entry responses, with their empirical counterparts, so a model whose behavior misses the stylized facts is flagged before its counterfactuals are used, testing the mechanism itself rather than a reduced-form summary of the data.

\textbf{Theory-guided reasoning}: The agents should reason within the theoretical frame the economist chooses to adopt, whether general equilibrium, game theory, or a particular econometric tradition. The workflow lets the economists ground the reasoning layer in the theory their study assumes, so that the agents' judgments start from economic argument and stay consistent with the modeling choices already made.

\textbf{Adversarial falsification}: When an agent advances a theoretical claim, such as that a policy raises welfare or that an effect is monotone in a parameter, the workflow searches the agreed model space for a counterexample: a parameterization or agent composition, admissible under the model's assumptions, in which the claim fails. A claim that survives the search is retained, with the search recorded beside it; one for which a counterexample is found is returned for revision. Refutation is thus a routine, documented step in building theory.

\subsection{Trustworthy Execution}
\label{sec:foundations}

An autonomous workflow can be trusted in economic research only if its own conduct meets the same standard of accountability as the research it produces. Trustworthy execution addresses reliability, accountability, safety, and privacy (Section~\ref{sec:LR-requirements}) with safeguards that apply at every stage.

\textbf{Reliability through verification}: The workflow verifies each agent's output through a step independent of the agent that produced it. Every citation is checked against the source it points to before it reaches the report: a verification step flags references it cannot confirm, and the economist judges whether each source supports the claim it is attached to. The data passed between stages must conform to an agreed structure before the receiving agent acts on them. Where reasoning as an economist tests whether a claim meets the field's methodological standards, verification tests whether it is supported by its sources and inputs.

\textbf{Accountability through provenance and bounded authority}: Every model call, tool invocation, and stage transition is logged, and every artifact carries the provenance of the inputs and decisions that shaped it. From these records a result can be traced back to the process that produced it, and its deterministic steps re-run. Authority is bounded as well as recorded: each agent is given only what its role needs, stays within its declared responsibilities, and may use only the tools it has been granted; anything else is denied by default. Budgets on cost, tokens, and retries bound each stage and the workflow as a whole. These limits confine each agent to its own task and cap the resources a run can consume.

\textbf{Safety through an explicit threat model}: Agents that read untrusted material, write to shared memory, and call external tools face risks an ordinary prompt does not, so the workflow is designed against an explicit threat model for agentic systems. Agent memory is isolated and allowed to expire, which limits how long a poisoned input persists; messages between agents are signed, so an alteration in transit is detected; generated code runs in a sandbox; and human approval is required at the steps that would be hard to undo. The authority limits above bound what an agent can do by accident or over-reach; these safeguards bound what an adversary can make it do.

\textbf{Privacy through local serving}: Every LLM in the workflow can be served on open weights on hardware the researcher controls, so prompts, intermediate artifacts, and outputs need not leave the researcher's environment. Incoming data are screened for personal information, and credentials are kept out of the logs.

Together, these safeguards record each run, so a result can be traced to its recorded process and the human checkpoints it passed. As AI systems and the risks they pose continue to develop, economists may extend or adapt these safeguards to match.

\section{Workflow Design and Implementation}
\label{sec:workflow}
This section presents a sample design and implementation of an agentic workflow system for economic research that follows the methodology outlined in Section~\ref{sec:method}. 
We describe a typical architecture, components, and operational mechanisms that enable collaborative research between AI agents and human economists,  
noting that actual implementations can vary based on economists' goals and design preferences. 
The following subsections elaborate on each stage of this example workflow.  

\subsection{Ideation and Research Question Formulation}

 The ideation process in economic research presents challenges, including the need to navigate vast and rapidly expanding literature, 
 identify meaningful gaps in knowledge, connect different theoretical frameworks, and formulate questions with both academic precision and policy relevance. 
 To address these difficulties, we employ agentic workflows for research question formulation.
 Semi-autonomous agents identify emerging trends, scan the literature, and synthesize what they find,
 and HITL checkpoints let economists review and redirect the output at each stage.
 The following subsections describe the agents and outline their respective responsibilities.
 They also explain how these agents work together to generate and refine candidate research questions.
 
 \subsubsection{Agents}
 
 The ideation phase uses eight agents, each responsible for one task in formulating research questions. 
 These agents operate in a sequential-parallel architecture, sharing information through standardized data structures while maintaining independent processing capabilities. 
 Together, they form a system that spans from broad trend identification to precise research question formulation, with human oversight integrated at strategic checkpoints.

 \textbf{TrendSurfer} identifies emerging or trending topics in real time by searching recent arXiv economics and finance preprints and an open news search. 
 It runs in each search round of the sourcing stage.
 
 \textbf{CorpusScout} gathers preliminary academic literature from Semantic Scholar (with Crossref as the fallback) and, in the web-crawl modes, from an open web search restricted to academic domains, to build an initial knowledge base.
 It generates a structured literature database with basic metadata that serves as the foundation for deeper analysis.
 This agent runs in each search round of the sourcing stage.
 \texttt{CorpusScout} works in close coordination with \texttt{ScholarSearcher} and provides its initial literature findings for concept generation.

 \textbf{ScholarSearcher} surfaces foundational and highly cited papers by querying open bibliometric indices, ranking candidates by citation count and filtering for seminal, field-defining contributions. It produces standardized research data with removed duplicates, rankings, and organized tags. For economists or institutions with subscription access, the connector layer is designed to be extended with premier licensed databases such as Web of Science and Elsevier Scopus, broadening coverage beyond what open indices alone provide. The agent is triggered when detailed research is needed following \texttt{CorpusScout}'s initial scan. \texttt{ScholarSearcher} collaborates with \texttt{CorpusScout} and feeds organized scholarly literature, enabling better coverage of peer-reviewed research.
 
 \textbf{GreyScout} expands the search beyond peer-reviewed publications to grey literature: working papers, preprints, policy reports, and conference proceedings from institutions and think tanks. It creates a grey-literature database with relevance scoring that captures emerging ideas not yet formalized in academic journals. This agent runs in each search round of the sourcing stage. \texttt{GreyScout} complements \texttt{ScholarSearcher}'s academic focus and provides alternative perspectives, aiming for a balanced view of both established and emerging knowledge.
 
 \textbf{Ideator} aggregates insights from both human economists and the sources discovered by the above agents to produce initial ideas, hypotheses, and key concepts. It generates potential research directions based on the information gathered throughout the system. The agent runs on the sourcing stage's output and the economists' inputs. \texttt{Ideator} receives data from \texttt{CorpusScout}, \texttt{ScholarSearcher}, and \texttt{GreyScout}, and subsequently feeds its generated ideas to \texttt{Refiner} for evaluation and narrowing.
 
 \textbf{Refiner} evaluates the ideas generated by \texttt{Ideator}, filtering out redundancies, checking conceptual coherence, and narrowing broad topics into well-defined research questions. It produces focused research questions derived from initially broad concepts. The agent engages automatically whenever new or updated ideas arrive from \texttt{Ideator}, often incorporating human feedback or similarity checks during its refinement process. \texttt{Refiner} processes \texttt{Ideator}'s output and forwards its refined questions to \texttt{Contextualizer} for theoretical framing.
 
 \textbf{Contextualizer} adds depth to emerging questions by mapping them against established economic theories and highlighting relevant policy debates. It creates contextualized research questions with clear theoretical and policy relevance that situate the inquiry within the broader academic discourse. This agent is triggered after gap analyses, debate-detection outputs, or by economist request. \texttt{Contextualizer} processes \texttt{Refiner}'s output and passes its contextualized questions to \texttt{Finalizer} for prioritization; it frames the questions it is given and does not add, drop, or reword them.
 
 \textbf{Finalizer} ranks the contextualized questions and writes, for each, its rationale, methodology, expected impact, and feasibility, checking alignment with project objectives and related literature. Each of its outputs must cite the identifier of the question it ranks and reproduce that question verbatim; a deterministic check replaces any reworded or uncited output with the source question. It produces a prioritized shortlist of research questions ready for final selection that represents the main idea of the entire ideation process. The agent is generally initiated by economists seeking to finalize the research question set after iterative refinements. \texttt{Finalizer} receives input from \texttt{Contextualizer} and presents its final questions to human economists for ultimate decision-making, completing the ideation cycle.
 
 \subsubsection{Workflow and Human Checkpoints}
 
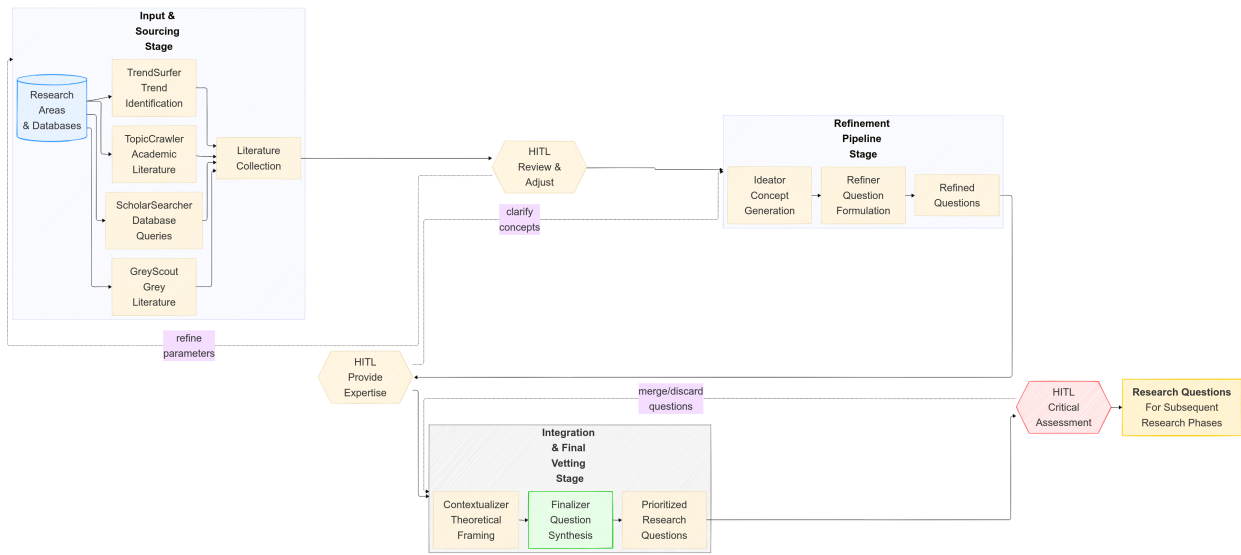
\begin{figure}[!htbp]
    \centering
    \scalebox{\aelFigScale}{\begin{tikzpicture}[ael]

\node[aelsource] (src) at (0,0)
  {Literature,\\Data \&\\Web Sources};

\coordinate (fanOut) at ($(src.east)+(6mm,0)$);
\coordinate (colL)   at ($(fanOut)+(4mm,0)$);
\node[aelagentcol, anchor=west] (trend)   at (colL |- 0, 2.4) {\aelAgent{TrendSurfer}{Trend Identification}};
\node[aelagentcol, anchor=west] (corpus)  at (colL |- 0, 0.8) {\aelAgent{CorpusScout}{Open-source Literature}};
\node[aelagentcol, anchor=west] (scholar) at (colL |- 0,-0.8) {\aelAgent{ScholarSearcher}{Foundational \&\\Highly Cited Papers}};
\node[aelagentcol, anchor=west] (grey)    at (colL |- 0,-2.4) {\aelAgent{GreyScout}{Grey Literature}};
\coordinate (fanIn) at ($(trend.east)+(4mm,0)$);
\coordinate (litL)  at ($(fanIn)+(4mm,0)$);
\node[aelart, anchor=west] (litcoll) at (litL |- src) {Literature\\Collection};

\node[fit=(src)(trend)(grey)(litcoll), inner sep=0pt] (stageonebody) {};
\node[aelstagetitle] (ttl1) at ($(stageonebody.north)+(0,5mm)$) {Input \& Sourcing};
\node[aelstagenum, left=1.3mm of ttl1] (num1) {1};
\begin{scope}[on background layer]
  \node[aelstageA, fit=(stageonebody)(ttl1)(num1)] (stageone) {};
\end{scope}

\draw[aelline] (src.east) -- (fanOut);
\draw[aelline] (fanOut |- trend) -- (fanOut |- grey);
\draw[aelline] (fanIn  |- trend) -- (fanIn  |- grey);
\foreach \t in {trend,corpus,scholar,grey}{
  \draw[aelflow] (fanOut |- \t) -- (\t.west);
  \draw[aelline] (\t.east) -- (fanIn |- \t);
}
\draw[aelflow] (fanIn |- litcoll) -- (litcoll.west);

\coordinate (colX) at ($(stageone.east)+(30.56mm,0)$);
\node[aelhitl, anchor=north] (hitlA)
  at (colX |- stageone.north) {\textbf{HITL}\\Review\\\& Adjust};
\coordinate (elbowA) at ($(stageone.east)+(4mm,0)$);
\draw[aelflow] (litcoll.east) -- (elbowA |- litcoll) |- (hitlA.west);

\node[aelagentcol, text width=24mm, anchor=north] (ideator) at ($(colX |- hitlA.south)+(0,-19.47mm)$)
  {\aelAgent{Ideator}{Concept generation}};
\node[aelagentcol, text width=24mm, below=4.5mm of ideator] (refiner) {\aelAgent{Refiner}{Question formulation}};
\node[aelart, text width=24mm, below=4.5mm of refiner] (refq) {Refined\\Questions};

\node[fit=(ideator)(refq), inner sep=0pt] (stagetwobody) {};
\node[aelstagetitle] (ttl2) at ($(stagetwobody.north)+(0,5mm)$) {Refinement Pipeline};
\node[aelstagenum, left=1.3mm of ttl2] (num2) {2};
\begin{scope}[on background layer]
  \node[aelstageB, fit=(stagetwobody)(ttl2)(num2)] (stagetwo) {};
\end{scope}

\draw[aelflow] (hitlA.south) -- (hitlA.south |- stagetwo.north);
\draw[aelflow] (ideator) -- (refiner);
\draw[aelflow] (refiner) -- (refq);

\coordinate (lane) at ($(stageone.west)+(-6mm,0)$);
\coordinate (busY) at ($(stageone.south)+(0,-4.5mm)$);
\coordinate (bus) at (lane |- busY);
\node[aelhitl] (hitlB) at (ideator |- busY) {\textbf{HITL}\\Provide Expertise};
\draw[aelflow] (refq.south) -- (hitlB.north);
\draw[aelline] (hitlB.west) -- (bus);

\coordinate (laneR) at ($(stagetwo.east)+(6mm,0)$);
\draw[aelfeed] (hitlB.east) -| (laneR |- ideator.east) -- (ideator.east);
\node[aellabel, align=center] (clarlbl) at (laneR |- refiner) {Clarify\\Concepts};

\coordinate (rowleft) at ($(stageone.west)+(4.76mm,0)$);   %
\coordinate (row2) at ($(busY)+(0,-25.62mm)$);
\node[aelagentcol, text width=24mm, anchor=west] (context) at (rowleft |- row2)
  {\aelAgent{Contextualizer}{Theoretical Framing}};
\node[aelagentcol, text width=24mm, right=4.5mm of context] (final) {\aelAgent{Finalizer}{Question Ranking}};
\node[aelart, text width=24mm, right=4.5mm of final] (prioq) {Prioritized\\Research Questions};

\node[fit=(context)(prioq), inner sep=0pt] (stagethreebody) {};
\node[aelstagetitle] (ttl3) at ($(stagethreebody.north)+(0,5mm)$) {Integration \& Final Vetting};
\node[aelstagenum, left=1.3mm of ttl3] (num3) {3};
\begin{scope}[on background layer]
  \node[aelstageC, fit=(stagethreebody)(ttl3)(num3)] (stagethree) {};
\end{scope}

\draw[aelflow] (bus) |- (stagethree.west |- context);
\draw[aelflow] (context) -- (final);
\draw[aelflow] (final) -- (prioq);

\node[aelout, anchor=east] (deliverable) at (clarlbl.east |- prioq)
  {Final Research Questions\\[2pt]\mdseries For Subsequent Phases};
\node[aelhitl] (hitlC) at ($(stagethree.east |- prioq)!0.5!(deliverable.west)$)
  {\textbf{HITL}\\Critical\\Assessment};
\draw[aelflow] (prioq.east) -- (hitlC.west);
\coordinate (retC) at ($(stagethree.north)+(0,4mm)$);
\draw[aelfeed] (hitlC.north) |- (context.north |- retC) -- (context.north);
\node[aellabel] at ($(context.north |- retC)!0.5!(hitlC.north |- retC)$)
  {Merge/Discard Questions};

\draw[aelflow] (hitlC.east) -- (deliverable.west);

\coordinate (refineY) at ($(hitlA.north)+(0,4mm)$);
\draw[aelfeed] (hitlA.north) -- (refineY) -| (src.north);
\node[aellabel] at ($(src.north |- refineY)!0.5!(refineY)$) {Refine Criteria};

\end{tikzpicture}}
    \caption{Agentic Workflow of Ideation in Economic Research}
    \label{fig:ideation}
\end{figure}

 The above agents function within a multi-stage process, each stage involving both automated actions and economist oversight. Figure~\ref{fig:ideation} shows the process.
 
 \paragraph{Input \& Sourcing Stage}
  Human economists can propose an initial research question or identify broad areas of interest. \texttt{TrendSurfer}, \texttt{CorpusScout}, \texttt{ScholarSearcher}, and \texttt{GreyScout} independently conduct literature searches, collecting pre-prints, peer-reviewed and grey literature. Economists review the aggregated output, adjusting search criteria or prioritizing particular topics. This HITL checkpoint checks alignment with specific research objectives.
 
 \paragraph{Refinement Pipeline Stage}
  The \texttt{Ideator} synthesizes all incoming data (economist prompts, existing literature) into a set of candidate research concepts, and the \texttt{Refiner} formulates each into a precise research question. The ranking is built from auditable component scores: each concept receives a novelty score combining its lexical distance from the retrieved literature, the breadth of themes it spans, and the breadth of literature it synthesizes, while each question receives a feasibility score reflecting its methodological tractability. The stage runs for up to three rounds. After the first round, economists review the ranked concepts and questions together with their component scores and supply feedback on promising directions, weak candidates, and missing angles at a HITL checkpoint; the second round reruns \texttt{Ideator} and \texttt{Refiner} with that feedback incorporated, sharpening the slate before it proceeds to contextualization. This keeps the refinement grounded in criteria an economist can inspect and contest, and anchored in human judgment between rounds.
 
 \paragraph{Integration \& Final Vetting Stage}
  The \texttt{Contextualizer} situates each refined question within relevant theoretical frameworks, policy debates, and the existing body of knowledge. The \texttt{Finalizer} compiles a shortlist of potential research questions, verifying their alignment with the project's strategic goals. Economists critically assess the \texttt{Finalizer}'s output, merging or discarding questions as necessary; when several reviewers answer, a question is merged or discarded only if a majority chose it, and at least three questions are kept, so that the modeling stage has alternatives to work from; if the economists judge none of the shortlisted questions adequate, they return the shortlist with their guidance for another round. The selected questions form the foundation for subsequent phases of the research.
 
 Throughout this process, economists also screen for algorithmic bias in what the sourcing and ranking agents surface, checking whether coverage is skewed toward more visible or more heavily cited work.

 \subsection{Automated Literature Review and Synthesis}
 \label{sec:LRS}
 
 This phase systematically finds, examines, and combines relevant literature using a network of specialized agents that retrieve targeted papers, summarize findings, detect research gaps, and verify citations.
 At HITL checkpoints, economists align the literature base with the research agenda.
 
 \subsubsection{Agents}
 
 A set of specialized agents (\texttt{TopicCrawler}, \texttt{TrendTracker}, \texttt{InsightSummarizer}, \texttt{PaperDecomposer}, \texttt{GapFinder}, \texttt{CiteKeeper}, \texttt{KnowledgeWeaver}, and \texttt{CitationVerifier}) retrieve, summarize, analyze, organize, and track academic research, extracting insights, identifying research gaps, and maintaining a structured knowledge base tailored for economic research. The paragraphs below describe each in turn.

 \textbf{TopicCrawler} retrieves literature directly targeted at the incoming research questions. It queries several open bibliographic sources in turn for each search query (Semantic Scholar, or Crossref when no Semantic Scholar key is set; arXiv; and OpenAlex) and merges the results into a single deduplicated candidate pool; where richer page content is useful, it further enriches results through an open web-crawl chain with commercial services available only as a disabled-by-default fallback. Running at the start of the gathering stage, it produces the initial literature batch that seeds every downstream agent in this team.

 \textbf{TrendTracker} serves as the field-pattern analyst for the batch \texttt{TopicCrawler} retrieves. Its role involves identifying emerging research directions, methodological innovations, and shifts in theoretical paradigms within that batch. Given the retrieved literature and the active research questions, it distills a handful of named trends, each with a description, supporting papers, an approximate emergence year, and a growth trajectory (growing, stable, or declining). \texttt{TrendTracker} runs once per gathering pass, immediately after \texttt{TopicCrawler} delivers a new batch. This trend information later shapes the team's literature graph and helps connect individual references to the emerging directions they support.

 \textbf{InsightSummarizer} distills key theoretical frameworks, econometric findings, and policy implications from the papers \texttt{TopicCrawler} retrieves, producing structured summaries that highlight methodological approaches, economic models employed, and empirical results with statistical significance. \texttt{InsightSummarizer} activates immediately after \texttt{TopicCrawler} retrieves new batches of economic papers or when economists explicitly request overviews of specific economic topics. These summaries later feed the team's gap-analysis and knowledge-mapping steps.

 \textbf{PaperDecomposer} segments economic studies into their core components (research questions, theoretical foundations, methodology, results, and limitations), extracting the underlying economic models, regression specifications, and identification strategies. By default it works from each paper's metadata and abstract, escalating to full text and supplementary materials only on request, since full-text decomposition is costly at scale. \texttt{PaperDecomposer} operates on demand when economists require deeper analysis of methodological approaches or when comparing similar economic studies. It collaborates with \texttt{InsightSummarizer} by providing the extracted components to enrich summary content, and its methodological details later help identify research opportunities in economics.

 \textbf{GapFinder} functions as a research opportunity detector within economic literature. Its role is to identify underexplored economic phenomena, methodological limitations, and contradictory findings across studies. It takes input from processed summaries and paper decompositions, as well as citation networks and publication metadata. The agent produces gap analyses highlighting promising research directions, methodological innovations needed, and potential contributions to economic theory. \texttt{GapFinder} activates after \texttt{InsightSummarizer} completes processing batches of literature or when economists highlight specific uncertainties in existing research. Its gap insights later shape the evolution of research questions in the team's synthesis step, and it works with \texttt{TrendTracker} to flag emerging areas that need monitoring.

 \textbf{CiteKeeper} acts as the bibliographic management system for economic research. Its role involves maintaining reference information and facilitating proper attribution practices. The agent ingests metadata from published economic papers, working papers, and conference proceedings, including author affiliations, journal impact factors, and citation counts. It produces standardized citation entries in a consistent bibliographic format, manages DOI information, and generates formatted bibliographies. \texttt{CiteKeeper} updates continuously as new economic literature enters the system, with particular attention to version tracking for working papers. It connects with all other agents by providing normalized reference data, particularly supporting \texttt{TrendTracker} with citation metrics and feeding bibliographic relationships into the team's synthesis step.

 \textbf{KnowledgeWeaver} functions as the integrative synthesis engine for economic literature. Its role is to construct maps of economic knowledge domains, showing theoretical lineages and empirical contradictions. The agent processes all outputs from other agents, including summaries, decompositions, gap analyses, citation data, and trend reports. It produces interactive literature graphs visualizing schools of economic thought, methodology clusters, and conceptual relationships, while maintaining research notes. \texttt{KnowledgeWeaver} activates on demand for literature reviews or automatically as new agent outputs accumulate. It integrates with all other agents by consolidating their outputs into one knowledge structure, particularly informing \texttt{GapFinder} about system-level patterns that might reveal new research opportunities.

 \textbf{CitationVerifier} acts as the factual-grounding gate for the \texttt{LiteratureTeam}, addressing the well-documented risk that LLM-generated text cites sources that do not exist~\citep{asai2026openscholar}. It extracts every identifiable citation from the synthesized review text and checks each one against pluggable bibliographic backends (e.g., Semantic Scholar, Crossref, arXiv, OpenAlex), marking each \emph{verified} (a resolvable identifier a backend confirms), \emph{suspect} (no identifier a backend can resolve), or \emph{hallucinated} (no backend can find it at all). A \emph{verified} citation exists; whether it supports the claim it is attached to is left to the economist at the Synthesis and Integration Stage checkpoint. Hallucinated citations are redacted with an inline marker and suspect ones are flagged inline, with the full verdict report and hallucination rate logged for audit. \texttt{CitationVerifier} is activated by default but can be disabled for fast-iteration development.

 \subsubsection{Workflow and Human Checkpoints}
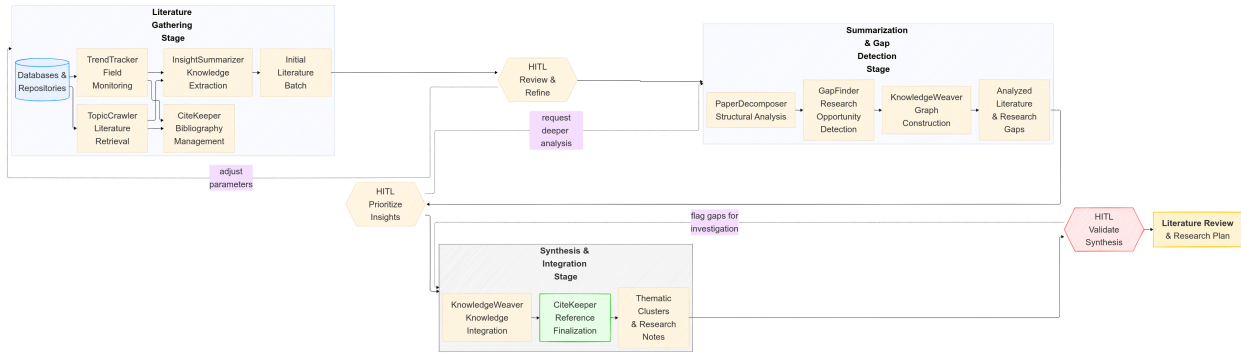
\begin{figure}[!htbp]
    \centering
    \scalebox{\aelFigScale}{\begin{tikzpicture}[ael]

\node[aelagentcol] (trend)   at (1.3,-2.1) {\aelAgent{TrendTracker}{Field Monitoring}};
\node[aelagentcol] (topic)   at (7.3,-2.1) {\aelAgent{TopicCrawler}{Literature Retrieval}};
\node[aelagentcol] (cite)    at (1.3,-4.1) {\aelAgent{CiteKeeper}{Bibliography Management}};
\node[aelagentcol] (insight) at (7.3,-4.1) {\aelAgent{InsightSummarizer}{Knowledge Extraction}};
\node[aelagentshared, text width=25mm] (kwA) at (4.3,-3.1)
  {\aelAgent{KnowledgeWeaver}{Knowledge Integration}};

\coordinate (fanL) at ($(4.3,0 |- trend.north)+(0,3.5mm)$);
\node[aelsource, anchor=south] (repos) at ($(4.3,0 |- fanL)+(0,3.5mm)$) {Literature\\Repositories};

\coordinate (joinL) at ($(4.3,0 |- cite.south)+(0,-3.5mm)$);
\node[aelagentcol, anchor=north] (verif) at ($(2.6,0 |- joinL)+(0,-4mm)$) {\aelAgent{CitationVerifier}{Citation Check}};
\node[aelart]      (batch) at (6.6,0 |- verif) {Initial Literature\\Batch};

\node[fit=(repos)(trend)(topic)(cite)(insight)(kwA)(verif)(batch), inner sep=0pt]
  (stageonebody) {};
\node[aelstagetitle] (ttlA) at ($(stageonebody.north)+(0,5mm)$) {Literature\\Gathering};
\node[aelstagenum, left=1.3mm of ttlA] (numA) {1};
\begin{scope}[on background layer]
  \node[aelstageA, fit=(stageonebody)(ttlA)(numA)] (stageone) {};
\end{scope}

\draw[aelline] (repos.south) -- (fanL);
\draw[aelline] (fanL -| trend) -- (fanL -| topic);
\draw[aelflow] (fanL -| trend) -- (trend.north);
\draw[aelflow] (fanL -| topic) -- (topic.north);

\draw[aelflow] (trend) -- (cite);
\draw[aelflow] (topic) -- (insight);

\draw[aelline] (cite.south)    -- (cite.south    |- joinL);
\draw[aelline] (insight.south) -- (insight.south |- joinL);
\draw[aelline] (cite.south |- joinL) -- (insight.south |- joinL);
\draw[aelflow] (verif.north |- joinL) -- (verif.north);
\draw[aelflow] (verif) -- (batch);

\draw[aelfeed] (kwA.north west) -| (trend.south);
\draw[aelfeed] (kwA.south west) -| (cite.north);
\draw[aelfeed] (kwA.north east) -| (topic.south);
\draw[aelfeed] (kwA.south east) -| (insight.north);

\node[aelhitl, anchor=west] (hitlRef)
  at ($(stageone.east |- batch)+(4mm,0)$) {\textbf{HITL}\\Review\\\& Refine};
\draw[aelflow] (batch.east) -- (hitlRef.west);

\coordinate (prioX) at ($(hitlRef.east)+(7mm,0)$);
\coordinate (colB)  at ($(hitlRef.east)+(33.05mm,0)$);
\node[aelagentshared, text width=33mm, anchor=north] (kwB) at (colB |- repos.north)
  {\aelAgent{KnowledgeWeaver}{Knowledge Integration}};
\node[aelart, text width=33mm] (gaps) at (colB |- batch) {Analyzed Literature\\\& Research Gaps};
\node[aelagentcol, text width=33mm] (decomp) at ($(kwB)!0.3333!(gaps)$) {\aelAgent{PaperDecomposer}{Paper Decomposition}};
\node[aelagentcol, text width=33mm] (gapf)   at ($(kwB)!0.6667!(gaps)$) {\aelAgent{GapFinder}{Research Opportunity Detection}};

\node[fit=(kwB)(decomp)(gapf)(gaps), inner sep=0pt] (stagetwobody) {};
\node[aelstagetitle] (ttlB) at ($(stagetwobody.north)+(0,5mm)$)
  {Summarization \&\\Gap Detection};
\node[aelstagenum, left=1.3mm of ttlB] (numB) {2};
\begin{scope}[on background layer]
  \node[aelstageB, fit=(stagetwobody)(ttlB)(numB)] (stagetwo) {};
\end{scope}

\draw[aelflow] (kwB)    -- (decomp);
\draw[aelflow] (decomp) -- (gapf);
\draw[aelflow] (gapf)   -- (gaps);
\draw[aelflow] (hitlRef.east) -- ++(3mm,0) |- (decomp.west);

\node[aelhitl, anchor=north] (hitlPrio) at ($(prioX |- stagetwo.south)+(0,-4.5mm)$) {\textbf{HITL}\\Prioritize};
\draw[aelflow] (gaps.south) |- (hitlPrio.east);

\draw[aelfeed] (hitlPrio.north) |- (decomp.south west);
\node[aellabel] at ($(hitlPrio.north)+(0,2.5mm)$) {Request Deeper Analysis};

\coordinate (rowleft) at ($(stageone.west)+(4.76mm,0)$);   %
\coordinate (row2) at ($(hitlPrio.south)+(0,-23.6mm)$);   %
\node[aelagentshared, text width=25mm, anchor=west] (kwC) at (rowleft |- row2)
  {\aelAgent{KnowledgeWeaver}{Knowledge Integration}};
\node[aelagent, text width=25mm, right=4.5mm of kwC] (citefin)
  {\aelAgent{CiteKeeper}{Reference Finalization}};
\node[aelart, text width=25mm, right=4.5mm of citefin] (clusters)
  {Thematic Clusters\\\& Research Notes};

\node[fit=(kwC)(citefin)(clusters), inner sep=0pt] (stagethreebody) {};
\node[aelstagetitle] (ttlC) at ($(stagethreebody.north)+(0,5mm)$)
  {Synthesis \& Integration};
\node[aelstagenum, left=1.3mm of ttlC] (numC) {3};
\begin{scope}[on background layer]
  \node[aelstageC, fit=(stagethreebody)(ttlC)(numC)] (stagethree) {};
\end{scope}

\draw[aelflow] (kwC)     -- (citefin);
\draw[aelflow] (citefin) -- (clusters);
\coordinate (laneL) at ($(stageone.west)+(-6mm,0)$);
\draw[aelflow] (hitlPrio.west) -- (laneL |- hitlPrio) |- (stagethree.west |- kwC);

\node[aelout, anchor=east] (out) at (stagetwo.east |- clusters)
  {Literature Review\\[2pt]\mdseries \& Research Plan};
\node[aelhitl] (hitlVal) at ($(stagethree.east |- clusters)!0.5!(out.west)$)
  {\textbf{HITL}\\Validate\\Synthesis};
\draw[aelflow] (clusters.east) -- (hitlVal.west);
\draw[aelflow] (hitlVal.east) -- (out.west);

\coordinate (retC) at ($(stagethree.south)+(0,-3.5mm)$);
\draw[aelfeed] (hitlVal.south) |- (citefin.south |- retC) -- (citefin.south);
\node[aellabel] at ($(citefin.south |- retC)!0.5!(hitlVal.south |- retC)$)
  {Flag Gaps for Investigation};

\draw[aelfeed] (hitlRef.north) |- (repos.east);
\node[aellabel] at ($(repos.east)!0.5!(stageone.east |- repos.east)$)
  {Adjust Criteria};

\end{tikzpicture}}
    \caption{Agentic Workflow for Literature Review and Synthesis in Economic Research}
    \label{fig:literature_review}
\end{figure}

 The literature discovery and synthesis phase runs across three stages, with economists guiding, validating, and refining the automated outputs at each one. Figure~\ref{fig:literature_review} shows the process.
 
 \textbf{Literature Gathering Stage} 
 In the initial stage, \texttt{TopicCrawler} retrieves the batch of studies most directly relevant to the incoming research questions. \texttt{TrendTracker} then distills emerging patterns from that batch, \texttt{CiteKeeper} compiles and formats its bibliography, and \texttt{InsightSummarizer} generates concise overviews of key findings.
 Economists then review initial batches of retrieved papers and summaries, refining search criteria and excluding irrelevant topics at a dedicated HITL checkpoint.
 
 \textbf{Summarization and Gap Detection Stage} 
 Once literature has been collected, \texttt{PaperDecomposer} segments each study into its components, enabling deeper analysis of its research question, methodology, and findings. \texttt{GapFinder} then examines study findings to identify methodological gaps and underexplored topics, while \texttt{KnowledgeWeaver} updates the literature graph to reflect emerging themes and overlapping areas. At this stage, economists engage with interactive summaries and gap analyses, merging or discarding duplicates and prioritizing directions for deeper inquiry.
 
 \textbf{Synthesis and Integration Stage} 
 With a refined literature base, \texttt{KnowledgeWeaver} compiles updated research notes and thematic clusters, showing how different studies connect or contradict each other. \texttt{CiteKeeper} finalizes reference lists in a standardized citation format, with export to external reference managers. Economists critically evaluate the synthesized literature map, validate key insights, and confirm the alignment of findings with the overarching research focus. Any newly identified gaps are flagged for further investigation.

 \subsection{Model Specification and Calibration}
 This phase turns a research question into a working model: a theoretical framework is developed, formalized mathematically, and calibrated against real-world data and established economic principles.
 \texttt{Theorist}, \texttt{ModelDesigner}, and \texttt{Calibrator} carry out these three steps under continuous human oversight, joined by an optional \texttt{Falsifier} and \texttt{Simulator}.
 
 \subsubsection{Agents}
 From theoretical formulation to mathematical implementation and calibration, each agent handles conceptual soundness, logical consistency, and empirical relevance.
 
 \textbf{Theorist} functions as the conceptual architect for economic models. Its primary role is to establish the theoretical foundation by selecting appropriate economic paradigms, defining assumptions, and outlining functional relationships between variables. The agent processes inputs from research questions, literature reviews, and human-specified constraints about economic phenomena being studied. It produces theoretical frameworks with clearly articulated economic assumptions, behavioral mechanisms, and expected relationships among variables. \texttt{Theorist} activates when preliminary research questions are sufficiently refined or when economists initiate the model development phase. It integrates closely with \texttt{KnowledgeWeaver} from the literature review phase by incorporating synthesized theoretical insights, and provides its theoretical framework directly to \texttt{ModelDesigner} for mathematical formalization, establishing a conceptual foundation for subsequent model development.
 
 \textbf{ModelDesigner} serves as the mathematical formalizer for economic theories. Its role involves translating theoretical constructs into precise mathematical structures such as utility functions and market equilibrium conditions. The agent processes the theoretical framework from \texttt{Theorist}, along with model specification requirements and methodological preferences from economists. It outputs formal mathematical models including equations (e.g., difference/differential, recursive, or simultaneous), parameter definitions, and computational implementation guidelines. \texttt{ModelDesigner} activates immediately after \texttt{Theorist} finalizes the theoretical framework or when economists request alternative mathematical formulations of existing theories. It collaborates both with \texttt{Theorist}, by translating its conceptual insights, and with \texttt{Calibrator}, by providing the formal structure that requires parameterization, connecting theory and implementation.
 
 \textbf{Calibrator} functions as the empirical alignment specialist for economic models. Its role involves determining parameter values intended to reproduce key empirical facts or stylized economic phenomena. The agent takes inputs from the formal model structure produced by \texttt{ModelDesigner}, empirical targets from the literature, and preliminary datasets provided by economists. It proposes candidate parameter sets, each with a qualitative sensitivity rating; the deterministic harness produces the goodness-of-fit metrics against the cited targets, and the \texttt{CodeTeam}'s comparative-statics sweep computes how parameter variations move the steady state. \texttt{Calibrator} activates once \texttt{ModelDesigner} completes the formal model structure and preliminary data or benchmarks become available. It integrates with \texttt{ModelDesigner} by operationalizing its mathematical framework, and its calibrated parameters later serve as a baseline the empirical estimation phase can compare its own estimated values against. Given observed data, the \texttt{Calibrator} can additionally produce a zero-shot causal estimate from a causal foundation model (under an instrumental-variable, natural-experiment, or selection-on-observables strategy, scored for identification soundness) as an independent cross-check on the classical parameter estimates; without observations it records a refusal instead of an estimate.

 \textbf{Falsifier} (optional) is paired with \texttt{Theorist}, within the theoretical foundation stage, to stress-test the theory adversarially before model design begins. For every claim, assumption, or proposition \texttt{Theorist} proposes, the \texttt{Falsifier} runs a bounded adversarial probe against the draft theoretical framework, searching for a counterexample the claim cannot survive. Each claim is reported as \emph{corroborated}, \emph{needs revision}, or \emph{falsified} (the last reserved for a validated, in-scope counterexample rather than any surface objection), and every report, whichever verdict it carries, is appended to the theoretical framework for the record. When the \texttt{Falsifier} is disabled, the theory stage reverts to single-pass behavior.

 \textbf{Simulator} (optional) executes the calibrated model through a pluggable backend to obtain simulation trajectories. It consumes the model specification from \texttt{ModelDesigner}, calibrated parameters from \texttt{Calibrator}, and scenario definitions from economists or from a library of built-in scenarios covering settings such as fiscal-policy transmission, monetary-policy counterfactuals, and strategic pricing under information asymmetry. It produces scenario-level trajectories, impulse responses, and macro-moment matching statistics. \texttt{Simulator} is invoked only when the optional simulation stage is enabled (below); it receives calibrated parameters from \texttt{Calibrator} and emits artifacts scored under the Simulation Fidelity dimension.

 \subsubsection{Workflow and Human Checkpoints}
 The model development process consists of three key stages: formulating the theoretical foundation, designing the model structure, and calibrating parameters. At each stage, specialized agents collaborate with human economists. Figure~\ref{fig:model_specification} shows the process.
 
 \textbf{Theoretical Foundation Stage} 
 In the initial stage, \texttt{Theorist} drafts the theoretical foundation. Economists then review the proposed framework for coherence with existing theories and project objectives, refining assumptions or selecting alternative theoretical constructs where the draft does not fit the intended research direction.
 
 \textbf{Model Design Stage} 
 Building on the theoretical framework, \texttt{ModelDesigner} produces the structured model. Human economists validate this theory-to-model translation for consistency and mathematical soundness, annotating components, refining variable definitions, and proposing alternative formulations where the representation of economic mechanisms falls short.
 
 \textbf{Calibration Stage}
 With the model structure in place, \texttt{Calibrator} proposes parameter values, which the deterministic harness estimates and scores against cited targets. Economists review the verdict and the targets behind it and may send the targets or the model back for revision, but cannot change the verdict.

 Calibration is deliberately split so that the language model only proposes parameters while a deterministic harness independently estimates and scores them. Against cited external targets, the harness returns a verdict: \emph{calibrated}, \emph{point-calibrated}, or \emph{partially calibrated} from the closed-form steady-state identities of the textbook structures it recognizes in the model's own parsed equations (an RBC or New Keynesian core, Cobb--Douglas or CES production, a hand-to-mouth household share); \emph{simulation-calibrated} by simulated method of moments on a simulator of the model's own equations, a path available when such a simulator is supplied, which the pipeline does not yet do; \emph{archetype-calibrated} by simulated method of moments on a registered canonical model that shares a mechanism with it, in which case only the parameters that model estimates are harness values; or \emph{uncalibratable}, marked non-applicable. An estimate counts only when it lies inside the parameter's declared range and off its bounds; a model is \emph{point-calibrated} only when every pinned moment lies within two standard errors of its target, and \emph{calibrated} only when its over-identifying restrictions are not rejected (Hansen's J test at 5\%). A registered canonical model is used only for a parameter whose name states the matching role and whose declared range overlaps the canonical model's admissible range. It is this verdict, not the model's narration, that the pipeline reports. The harness's tiered design, and a concrete run exhibiting these verdicts, are given in Appendix~\ref{appendix:example-model}.

 \paragraph{Simulation Stage (optional)}
 An optional simulation stage runs after calibration when enabled. The human role here is to choose the scenario and judge the result: an economist selects a scenario (built-in or their own), and once \texttt{Simulator} has executed the calibrated model, they review the results against known stylized facts and decide whether the simulated behavior is credible enough to draw counterfactuals from. Outputs that miss the stylized facts are sent back for re-specification rather than carried forward. The stage is disabled by default.

 \begin{figure}[!htbp]
    \centering
    \scalebox{\aelFigScale}{\begin{tikzpicture}[ael]

\node[aelagentcol, text width=28mm] (theorist)  at (6.4,-0.6) {\aelAgent{Theorist}{Conceptual Architect}};
\node[aelagentopt, text width=28mm, below=6mm of theorist] (falsifier)
  {\aelAgent{Falsifier}{Adversarial Probe (optional)}};
\node[aelart, text width=25mm, right=4.5mm of theorist] (framework) {Theoretical\\Framework};

\node[fit=(theorist)(falsifier)(framework), inner sep=0pt] (stageonebody) {};
\node[aelstagetitle] (ttlA) at ($(stageonebody.north)+(0,5mm)$) {Theoretical\\Foundation};
\node[aelstagenum, left=1.3mm of ttlA] (numA) {1};
\begin{scope}[on background layer]
  \node[aelstageA, fit=(stageonebody)(ttlA)(numA)] (stageone) {};
\end{scope}

\node[aelsource, anchor=east] (rq) at ($(stageone.west |- theorist)+(-6mm,0)$)
  {Research Questions\\\& Literature};

\draw[aelflow] (rq)       -- (theorist);
\draw[aelflow] (theorist) -- (framework);
\draw[aelfeed, <->] (theorist.south) -- (falsifier.north);

\node[aelhitl, anchor=west] (hitlThy)
  at ($(stageone.east |- framework)+(4mm,0)$) {\textbf{HITL}\\Validate\\Theory};
\draw[aelflow] (framework.east) -- (hitlThy.west);

\node[aelagentcol, text width=33mm, anchor=west] (md) at ($(hitlThy.east)+(10.5mm,0)$)
  {\aelAgent{ModelDesigner}{Mathematical Formalizer}};
\node[aelart, text width=33mm] (fmm) at (md |- falsifier) {Formal Mathematical\\Model};

\node[fit=(md)(fmm), inner sep=0pt] (stagetwobody) {};
\node[aelstagetitle] (ttlB) at ($(stagetwobody.north)+(0,5mm)$) {Model\\Design};
\node[aelstagenum, left=1.3mm of ttlB] (numB) {2};
\begin{scope}[on background layer]
  \node[aelstageB, fit=(stagetwobody)(ttlB)(numB)] (stagetwo) {};
\end{scope}

\draw[aelflow] (md) -- (fmm);
\draw[aelflow] (hitlThy.east) -- (md.west);

\node[aelhitl, anchor=north] (hitlChk) at ($(rq |- stageone.south)+(0,-7mm)$) {\textbf{HITL}\\Check};
\draw[aelflow] (fmm.south) |- (hitlChk.east);

\node[aelsource, anchor=north] (emp) at ($(hitlChk.south)+(0,-8.5mm)$)
  {Empirical Targets\\\& Benchmarks};
\node[aelagentcol, text width=28mm] (calib) at (theorist |- emp)
  {\aelAgent{Calibrator}{Empirical Alignment}};
\node[aelart, text width=27mm, right=4.5mm of calib] (calpar)
  {Parameters \&\\Calibration Verdict};

\node[fit=(calib)(calpar), inner sep=0pt] (stagethreebody) {};
\node[aelstagetitle] (ttlC) at ($(stagethreebody.north)+(0,5mm)$) {Calibration};
\node[aelstagenum, left=1.3mm of ttlC] (numC) {3};
\begin{scope}[on background layer]
  \node[aelstageC, fit=(stagethreebody)(ttlC)(numC)] (stagethree) {};
\end{scope}

\draw[aelflow] (emp)   -- (calib);
\draw[aelflow] (calib) -- (calpar);
\draw[aelflow] (hitlChk.south) -- (emp.north);

\node[aelhitl, anchor=west] (hitlAss)
  at ($(stagethree.east |- calpar)+(4mm,0)$) {\textbf{HITL}\\Assess};
\draw[aelflow] (calpar.east) -- (hitlAss.west);

\coordinate (row4left) at ($(rq.west)+(4.77mm,0)$);   %
\coordinate (row4) at ($(stagethree.south)+(0,-31.24mm)$);   %
\node[aelagentopt, text width=28mm, anchor=west] (sim) at (row4left |- row4)
  {\aelAgent{Simulator}{Scenario Execution (optional)}};
\node[aelart, text width=27mm, densely dotted, right=4.5mm of sim] (traj)
  {Scenario Trajectories\\\& Impulse Responses};

\node[fit=(sim)(traj), inner sep=0pt] (stagefourbody) {};
\node[aelstagetitle] (ttlD) at ($(stagefourbody.north)+(0,5mm)$) {Simulation (optional)};
\node[aelstagenum, left=1.3mm of ttlD] (numD) {4};
\begin{scope}[on background layer]
  \node[aelstageD, densely dotted, fit=(stagefourbody)(ttlD)(numD)] (stagefour) {};
\end{scope}

\draw[aelflowopt] (sim) -- (traj);
\coordinate (midCD) at ($(stagethree.south)!0.5!(stagefour.north)$);
\coordinate (lane) at ($(rq.west)+(-6mm,0)$);
\draw[aelflowopt] (hitlAss.south) |- (lane |- midCD) -- (lane |- sim)
  -- (stagefour.west |- sim);

\node[aelhitl, anchor=west] (hitlScn)
  at ($(stagefour.east |- traj)+(4mm,0)$) {\textbf{HITL}\\Review\\Scenario};
\draw[aelflowopt] (traj.east) -- (hitlScn.west);

\node[aelout, anchor=east] (out) at (stagetwo.east |- traj)
  {Specified Economic Model\\[2pt]\mdseries With Calibration Verdict};
\draw[aelflow]    (hitlAss.east) -| (out.north);
\draw[aelflowopt] (hitlScn.east) -- (out.west);

\coordinate (retA) at ($(stageone.north)+(0,4mm)$);
\draw[aelfeed] (hitlThy.north) |- (theorist.north |- retA) -- (theorist.north);
\node[aellabel] at ($(theorist.north |- retA)!0.5!(hitlThy.north |- retA)$)
  {Refine Assumptions};

\coordinate (laneR) at ($(stagetwo.east)+(6mm,0)$);
\coordinate (propose) at ($(hitlChk.north)+(0,3.5mm)$);
\draw[aelfeed] (hitlChk.north) -- (propose) -| (laneR |- md.east) -- (md.east);
\node[aellabel] at ($(propose)!0.5!(laneR |- propose)$) {Propose Alternatives};

\coordinate (retC) at ($(stagethree.north)+(0,4mm)$);
\draw[aelfeed] (hitlAss.north) |- (calib.north |- retC) -- (calib.north);
\node[aellabel] at ($(calib.north |- retC)!0.5!(hitlAss.north |- retC)$)
  {Adjust Methods};

\coordinate (retD) at ($(stagefour.south)+(0,-3.5mm)$);
\draw[aelfeed] (hitlScn.south) |- (sim.south |- retD) -- (sim.south);
\node[aellabel] at ($(sim.south |- retD)!0.5!(hitlScn.south |- retD)$)
  {Re-specify Scenario};

\end{tikzpicture}}
    \caption{Agentic Workflow for Model Specification and Calibration in Economic Research. An optional Simulation stage runs after Calibration and an optional \texttt{Falsifier} agent is paired with the \texttt{Theorist}; both are disabled by default.}
    \label{fig:model_specification}
\end{figure}

 \subsection{Data Collection and Preprocessing}
 \label{sec:data-collection}
 This phase systematically gathers, cleans, and prepares economic data for economists' empirical work. These data include time series, panel data, unstructured text, and data from alternative sources (e.g., satellite imagery, social media feeds). Because the agents are LLM-driven, the pipeline can ingest multilingual and multimodal data (non-English sources, tables, and figures) that would otherwise be costly to process at scale. Specialized agents with distinct roles retrieve, clean, validate, and document the data, and human economists remain in the loop through interactive feedback, guiding and validating each step of the process.

 The \texttt{DataTeam} draws on open-source APIs by default, escalating to premium subscribed sources (e.g., Bloomberg, Refinitiv) or user-uploaded datasets only when needed, and never silently substitutes a simulated series for a real one: a series from a source without a live connector enters the dataset only as a disclosed simulation. After its first retrieval, the \texttt{DataTeam} reconciles what the model needs against what the data can supply: the \texttt{ModelTeam} emits a \emph{Data Requirements Specification} (DRS), the \texttt{DataTeam} answers with a \emph{Data Availability Report} (DAR) and a feasibility verdict, and where a requirement cannot be met, a \emph{Model Revision Request} (MRR) carries the constraint back to the \texttt{ModelTeam} for re-specification. The full mechanics of this Model--Data feasibility exchange are described in the Feasibility Review Stage below.

 \subsubsection{Agents}
 
 The \texttt{DataTeam}'s agents form an automated data processing pipeline, facilitating dataset discovery, retrieval, cleaning, integration, validation, and documentation. Each agent has a specific function, from identifying relevant data sources to maintaining reproducibility.
 
 \textbf{AvailabilityScout} functions as the feasibility-search specialist. Its role is to determine, for each requirement in the DRS, whether a satisfying series exists and at what cost, searching in priority order across the three source tiers: open-source APIs (default), then premium subscribed sources, then user-uploaded data. For each requirement it searches the retrieved series and the series the open connectors can fetch, maps the requirement to the best-matching one, and grades it feasible, proxy-only, or unavailable; a feasible match outside the retrieved series is fetched in a supplemental retrieval. It consumes the DRS from the \texttt{ModelTeam} and passes its graded matches to \texttt{AvailabilityReporter}.

 \textbf{AvailabilityReporter} assembles the DAR from \texttt{AvailabilityScout}'s graded matches. For each requirement it records the best match and its source tier, the coverage gap relative to the DRS requirement, and a ranked list of candidate proxies, each annotated with an explicit bias caveat (what the proxy over- or under-measures), and it summarizes the share of essential requirements met alongside the overall feasibility verdict. It leaves the choice between a proxy and the ideal measure to the economist: it presents the options and their consequences for the economist to weigh at the Feasibility Review.

 \textbf{DataScout} functions as the economic data discovery specialist. Its primary role is to identify and evaluate relevant datasets that match the research requirements for economic analysis. The agent scans academic databases, APIs, and specialized repositories to locate both conventional economic indicators and alternative data sources. It processes inputs from research questions, model specifications, and economist-defined data requirements. \texttt{DataScout} produces dataset inventories with relevance scores, source reliability ratings, access requirements, and update frequencies for economic time series, cross-sectional, and panel data. It activates either through scheduled periodic scans of data repositories or when manually initiated by economists exploring new economic topics or requiring updated information. It integrates with \texttt{DataCollector} by providing prioritized dataset lists for acquisition, and its metadata on source characteristics later documents the data sources in economic research.
 
 \textbf{DataCollector} serves as the acquisition and standardization system for economic datasets. Its role involves retrieving identified datasets while maintaining access compliance and standardizing formats across sources. The agent processes targeted dataset URLs, API endpoints, and access credentials from \texttt{DataScout}, along with economist-specified collection settings. It outputs standardized raw datasets with harmonized formats addressing international differences in time zones, currencies, units of measurement, and periodicity commonly found in economic data. \texttt{DataCollector} activates immediately after \texttt{DataScout} compiles its inventory of relevant sources or upon direct economist request for specific economic series. It works in coordination with \texttt{DataScout} by executing its acquisition recommendations, and with \texttt{DataCleaner} by delivering consistently formatted raw datasets, establishing the foundation for subsequent cleaning operations in the economic data pipeline.
 
 \textbf{DataCleaner} functions as the quality control specialist for economic datasets. Its role involves identifying and resolving data quality issues while preserving the integrity of economic information. The agent processes raw datasets from \texttt{DataCollector}, along with domain-specific cleaning rules and economist-specified quality thresholds. 
 It produces processed datasets with documentation of all modifications, including missing value treatments, outlier handling, and duplicate removal. 
 \texttt{DataCleaner} activates following successful data collection or when economists identify quality issues in existing datasets. It integrates with \texttt{DataCollector} by processing its raw outputs, and with \texttt{DataIntegrator} by providing cleaned, individual datasets ready for merging.
 
 \textbf{DataIntegrator} serves as the dataset consolidation engine. Its role involves merging data from different sources into analysis-ready datasets while reconciling their formats, units, and frequencies. The agent processes multiple cleaned datasets from \texttt{DataCleaner}, along with economist specifications for desired integration outcomes. It produces unified datasets with harmonized variable names, time indices aligned deterministically (each series resampled to the coarsest common frequency by period mean, incomplete periods dropped), and consistent cross-sectional identifiers. \texttt{DataIntegrator} activates once \texttt{DataCleaner} completes processing individual datasets or when economists require a combination of disparate economic data sources. It works with \texttt{DataCleaner} by building upon its quality-assured outputs, and with \texttt{IndicatorBuilder} by providing integrated datasets ready for indicator construction, facilitating economic analyses that require multiple data sources.
 
 \textbf{IndicatorBuilder} functions as the variable creation specialist for economic analysis. 
 Its role involves developing theoretically grounded derived variables the analysis requires. 
 The agent processes integrated datasets from \texttt{DataIntegrator}, along with economic theory specifications and economist requirements for analytical indicators.
 It outputs enriched datasets containing derived economic indicators, transformed variables for statistical compliance, and dimensionality reductions for high-dimensional economic data. 
 \texttt{IndicatorBuilder} activates upon economist request for specific data analysis or when model specifications require particular variable forms not present in the raw data. 
 It integrates with \texttt{DataIntegrator} by enhancing its consolidated datasets, and with \texttt{DataAuditor} by providing indicator-rich datasets ready for quality verification, supporting econometric modeling with appropriately constructed variables.
 
 \textbf{DataAuditor} serves as the statistical verification system for economic datasets. 
 Its role is to check prepared data against quality and statistical requirements. 
 The agent processes indicator-enhanced datasets from \texttt{IndicatorBuilder}, along with predefined validation rules and economic benchmarks. The dataset's quality verdict comes from deterministic checks (constant series, no numeric values, duplicate dates, too few observations, simulated series), which the agent narrates; the estimation stage reads the same archived observations and excludes series that failed them.
 It produces validation reports detailing statistical properties (stationarity, normality, heteroskedasticity), variable correlations, benchmark comparisons against established economic facts, and recommendations for additional data processing if needed. 
 \texttt{DataAuditor} activates once data preprocessing is complete or periodically during updates. It works closely with \texttt{IndicatorBuilder} by validating its derived variables, and with \texttt{MetadataKeeper} by providing statistical assessment reports.
 
 \textbf{MetadataKeeper} functions as the documentation specialist. Its role involves creating metadata and maintaining processing records. 
 The agent collects information from all preceding data agents, including source characteristics, processing steps, and quality assessments. 
 It generates data dictionaries with economic variable definitions, units, and sources; transformation logs documenting all data analyses; and quality reports highlighting potential limitations or special considerations for economic interpretation. \texttt{MetadataKeeper} runs continuously throughout the data processing workflow, updating documentation with each modification or on scheduled intervals. It integrates with all other agents by centralizing their metadata and processing information, and its transformation logs and quality reports are the documentation \texttt{Archivist} (below) draws on for the replication manifest.
 
 \textbf{Archivist} is the replication-assurance agent for the \texttt{DataTeam}. It is deterministic (no language model call) and runs once the dataset is certified, consolidating what the pipeline already tracked into one artifact. The artifact holds a fingerprint of the Python environment (interpreter version, platform, and the installed versions of the packages the pipeline depends on), a content hash of the exact stage scripts that ran, a content hash of the certified dataset's record (its metadata, provenance, alignment summary, and last five complete-case rows), and a per-dataset audit trail assembled from the retrieval source, the merge and frequency-alignment methods applied, the certification verdict, and the sandbox-validation outcome. The result is a \emph{replication manifest}: given the same pipeline code and the same inputs, an economist can check whether a rerun reproduces the same stage code and dataset record, and which hashed component differs if it does not. \texttt{Archivist} activates automatically after \texttt{MetadataKeeper} completes documentation, drawing on its transformation logs and quality reports as the audit trail's source material.
 
 \subsubsection{Workflow and Human Checkpoints}
 
 The data preparation process consists of four key stages: discovering and retrieving data sources, preprocessing and cleaning, quality assurance and documentation, and reconciling model requirements with data availability (the Model--Data feasibility review), which the pipeline runs after the first retrieval. Figure~\ref{fig:data_collection} shows the process.
 At each step, economists review what the agents propose before it carries forward to the next stage.
 
 \textbf{Feasibility Review Stage}
 After the first retrieval, the economist reviews the DAR at a dedicated feasibility review checkpoint, the human trade-off gate. When the feasibility verdict is not feasible, the reviewer is shown the unmet essential requirements and their candidate proxies and chooses, per requirement, from a fixed menu: accept a named proxy (whose bias caveat is then recorded in provenance); escalate the source (authorize a premium subscription or request a user upload, re-running availability on that tier); revise the model (emit an MRR that loops back to the \texttt{ModelTeam} to drop, replace, or re-specify the variable); narrow the research question (the decision is recorded with the run's limitations; the run does not return to the \texttt{IdeationTeam}); or accept the specification as-is with a documented limitation. Because trading a proxy against a change to the model determines what the study can ultimately claim, this is a mandatory human checkpoint: the agents propose and the economist disposes. When the review revises the model or adds a fetchable series, the \texttt{DataTeam} retrieves, cleans, and validates the data once more; the downstream teams proceed even when the cycle ends without a feasible model--data pair, with the unmet requirements recorded.

 \textbf{Data Source Integration Stage}
 The open-source tier retrieves real observations through live connectors to seven services: FRED and Yahoo Finance via their native clients, plus five keyless open REST APIs (World Bank, DBnomics,\footnote{DBnomics federates more than 80 providers; the OECD, IMF, and BIS series are retrieved through it.} the ECB Data Portal, Eurostat, and BLS), consistent with the framework's open-stack default. Source discovery and series selection are \emph{truth-in-prompting}: agents are offered only sources that possess a live connector, together with each source's exact series-identifier grammar. Each retrieved series carries the source's own title; where that title differs materially from the variable the model requested (for instance, the World Bank's ``Gini index'', income-based for the United States, returned for a requested consumption Gini), the series is marked as a proxy, and the disclosure travels with it into the estimation and the report. Economists review identified and collected datasets, manually adjusting source priorities, verifying access protocols, and flagging data sources requiring special handling or additional review.
 
 \textbf{Data Preprocessing and Cleaning Stage} 
 During this stage, economists interact with tools that display data quality metrics and transformation logs generated by \texttt{DataCleaner} and \texttt{DataIntegrator}. They also access options for adjusting cleaning settings and constructing indicators via \texttt{IndicatorBuilder}, while anomalies or discrepancies detected by \texttt{DataAuditor} are highlighted for review. Human economists validate or override automated decisions, suggest alternative cleaning strategies, and propose additional derived variables.
 
 \textbf{Quality Assurance and Documentation Stage} 
 Economists review the reports \texttt{DataAuditor} and \texttt{MetadataKeeper} generate, checking whether all modifications are documented, reproducible, and meet the study's standards, and use the replication manifest \texttt{Archivist} then produces to request additional tests or annotate findings for follow-up.
 
 \begin{figure}[!htbp]
    \centering
    \scalebox{\aelFigScale}{\begin{tikzpicture}[ael]

\node[aelsource, anchor=north]              (src)     at (0,1.0) {Economic Data\\Sources};
\node[aelagentcol, below=5mm of src]        (avscout) {\aelAgent{AvailabilityScout}{Feasibility Search}};
\node[aelagentcol, below=4.5mm of avscout]  (avrep)   {\aelAgent{AvailabilityReporter}{Record Matches}};
\node[aelart, below=4.5mm of avrep]         (feasrep) {Availability\\Report};

\node[fit=(src)(avscout)(avrep)(feasrep), inner sep=0pt] (stageonebody) {};
\node[aelstagetitle] (ttlA) at ($(stageonebody.north)+(0,5mm)$) {Feasibility\\Review};
\node[aelstagenum, left=1.3mm of ttlA] (numA) {1};
\begin{scope}[on background layer]
  \node[aelstageA, fit=(stageonebody)(ttlA)(numA)] (stageone) {};
\end{scope}

\draw[aelflow] (src)     -- (avscout);
\draw[aelflow] (avscout) -- (avrep);
\draw[aelflow] (avrep)   -- (feasrep);

\node[aelhitl, anchor=west] (hitlSrc)
  at ($(stageone.east |- feasrep)+(4mm,0)$) {\textbf{HITL}\\Review\\Sources};
\draw[aelflow] (feasrep.east) -- (hitlSrc.west);

\coordinate (colB) at ($(hitlSrc.east)+(26.46mm,0)$);
\node[aelagentcol, anchor=north]            (dscout) at (colB |- 0,1.0) {\aelAgent{DataScout}{Data Discovery}};
\node[aelart] (rawds) at (colB |- feasrep) {Raw\\Datasets};
\node[aelagentcol] (dcoll) at ($(dscout)!0.5!(rawds)$) {\aelAgent{DataCollector}{Acquisition \& Standardization}};

\node[fit=(dscout)(dcoll)(rawds), inner sep=0pt] (stagetwobody) {};
\node[aelstagetitle] (ttlB) at ($(stagetwobody.north)+(0,5mm)$) {Data Source\\Integration};
\node[aelstagenum, left=1.3mm of ttlB] (numB) {2};
\begin{scope}[on background layer]
  \node[aelstageB, fit=(stagetwobody)(ttlB)(numB)] (stagetwo) {};
\end{scope}

\draw[aelflow] (dscout) -- (dcoll);
\draw[aelflow] (dcoll)  -- (rawds);

\draw[aelflow] (hitlSrc.east) -- ++(3mm,0) |- (dscout.west);
\draw[aelfeed] (hitlSrc.north) |- (avscout.east);
\node[aellabel] at ($(hitlSrc.north)+(0,1.1)$) {Adjust Conditions};

\node[aelhitl, anchor=west] (hitlPri)
  at ($(stagetwo.east |- rawds)+(4mm,0)$) {\textbf{HITL}\\Review\\Sources};
\draw[aelflow] (rawds.east) -- (hitlPri.west);

\coordinate (colC) at ($(hitlPri.east)+(28.27mm,0)$);
\node[aelagentcol, anchor=north]            (dclean) at (colC |- 0,1.0) {\aelAgent{DataCleaner}{Quality Control}};
\node[aelart] (featds) at (colC |- feasrep) {Indicator-Enhanced\\Datasets};
\node[aelagentcol] (dint) at ($(dclean)!0.3333!(featds)$) {\aelAgent{DataIntegrator}{Dataset Consolidation}};
\node[aelagentcol] (dind) at ($(dclean)!0.6667!(featds)$) {\aelAgent{IndicatorBuilder}{Variable Creation}};

\node[fit=(dclean)(dint)(dind)(featds), inner sep=0pt] (stagethreebody) {};
\node[aelstagetitle] (ttlC) at ($(stagethreebody.north)+(0,5mm)$) {Data Preprocessing\\\& Cleaning};
\node[aelstagenum, left=1.3mm of ttlC] (numC) {3};
\begin{scope}[on background layer]
  \node[aelstageC, fit=(stagethreebody)(ttlC)(numC)] (stagethree) {};
\end{scope}

\draw[aelflow] (dclean) -- (dint);
\draw[aelflow] (dint)   -- (dind);
\draw[aelflow] (dind)   -- (featds);

\draw[aelflow] (hitlPri.east) -- ++(3mm,0) |- (dclean.west);
\draw[aelfeed] (hitlPri.north) |- (dscout.east);
\node[aellabel] at ($(hitlPri.north)+(0,1.1)$) {Adjust Priorities};

\node[aelhitl, anchor=north] (hitlVal) at ($(colB |- stagetwo.south)+(0,-7mm)$) {\textbf{HITL}\\Validation};
\draw[aelflow] (featds.south) |- (hitlVal.east);
\coordinate (laneR) at ($(stagethree.east)+(6mm,0)$);
\draw[aelfeed] (hitlVal.north) -- ++(0,3.5mm) -| (laneR |- dclean.east) -- (dclean.east);
\node[aellabel] at ($(hitlVal.north)+(4.0,0.35)$) {Suggest Alternatives};

\coordinate (rowleft) at ($(stageone.west)+(4.76mm,0)$);
\coordinate (row2) at ($(hitlVal.south)+(0,-25.6mm)$);
\node[aelagent, text width=24mm, anchor=west] (daudit) at (rowleft |- row2)
  {\aelAgent{DataAuditor}{Statistical Verification}};
\node[aelagent, text width=24mm, right=4.5mm of daudit] (dmeta) {\aelAgent{MetadataKeeper}{Documentation}};
\node[aelagent, text width=24mm, right=4.5mm of dmeta] (darch) {\aelAgent{Archivist}{Replication Assurance}};
\node[aelart,   right=4.5mm of darch]  (docds) {Documented\\Datasets};

\node[fit=(daudit)(docds), inner sep=0pt] (stagefourbody) {};
\node[aelstagetitle] (ttlD) at ($(stagefourbody.north)+(0,5mm)$) {Quality Assurance \& Documentation};
\node[aelstagenum, left=1.3mm of ttlD] (numD) {4};
\begin{scope}[on background layer]
  \node[aelstageD, fit=(stagefourbody)(ttlD)(numD)] (stagefour) {};
\end{scope}

\draw[aelflow] (daudit) -- (dmeta);
\draw[aelflow] (dmeta)  -- (darch);
\draw[aelflow] (darch)  -- (docds);

\coordinate (lane) at ($(stageone.west)+(-6mm,0)$);
\draw[aelflow] (hitlVal.west) -- (lane |- hitlVal) |- (stagefour.west |- daudit);

\node[aelhitl, anchor=west] (hitlRev)
  at ($(stagefour.east |- docds)+(4mm,0)$) {\textbf{HITL}\\Review};
\draw[aelflow] (docds.east) -- (hitlRev.west);

\node[aelout, right=5.5mm of hitlRev] (out)
  {Analysis-Ready Data\\[2pt]\mdseries Validated \& Documented};
\draw[aelflow] (hitlRev.east) -- (out.west);

\coordinate (retD) at ($(stagefour.north)+(0,4mm)$);
\draw[aelfeed] (hitlRev.north) |- (daudit.north |- retD) -- (daudit.north);
\node[aellabel] at ($(daudit.north |- retD)!0.5!(hitlRev.north |- retD)$)
  {Request Additional Tests};

\end{tikzpicture}}
    \caption{Agentic Workflow for Data Collection and Preprocessing in Economic Research. The Feasibility Review (stage 1 in the figure) runs after the first retrieval of Data Source Integration (stage 2).}
    \label{fig:data_collection}
\end{figure}

 \subsection{Code Implementation and Testing}
 
 This phase derives an executable module from each model specification, validates it, runs a comparative-statics sweep on it, and versions and documents the result. Human oversight is embedded throughout to maintain quality, traceability, and alignment with the intended model.

 \subsubsection{Agents}

 The \texttt{CodeTeam}'s agents translate a model's equations into working code without ever letting a language model write that code itself: a fixed, rule-based procedure, not an agent, converts each equation directly into its corresponding piece of the program, much as a spreadsheet formula converts directly into a calculation. The agents' only job is to explain, check, time, version, or document what that procedure produced, never to write or edit the code. This guarantee is scoped to models whose equations reduce to a closed-form, steady-state-solvable system, such as DSGE or static equilibrium models; large agent-based simulations, which have no such system to parse, instead run through \texttt{ModelTeam}'s optional simulation stage on an established, pluggable backend. A likely medium-term path beyond a single custom parser is to wrap established, economist-trusted toolboxes such as Dynare, Stata, or MATLAB as callable tools once they expose programmatic APIs, rather than having an LLM re-author their functionality from scratch: economists already know these tools, and reimplementing what they already do well is slower and costlier than calling them directly.

 \textbf{Coder} reports on the executable module that the rule-based procedure derives for each formal model from the model's parsed equation system. The agent processes the parsed equations from \texttt{ModelDesigner} and the parameter values proposed by \texttt{Calibrator}, and narrates the result: which variables and parameters the module solves for, and, for any equation that could not be parsed into closed form, the parser's specific reason rather than a silent omission. \texttt{Coder} activates upon receiving finalized model specifications and proposed parameter values. It integrates with \texttt{ModelDesigner} and \texttt{Calibrator} by consuming their outputs, and with \texttt{Debugger} by passing the derived module forward for the executed check battery.

 \textbf{Debugger} narrates the executed, deterministic check battery run against each derived module: compilation, parameter completeness, residual finiteness, steady-state solvability, parameter sensitivity, and system completeness (whether the module is built from every equation of the model). Because the module is derived rather than hand-written, there is no free-hand code to debug: a failing check is a refusal to certify the module, not a bug the agent patches, and \texttt{Debugger}'s narration states what a failure implies about the model specification rather than proposing a code fix. \texttt{Debugger} activates whenever \texttt{Coder} delivers a new module. It works with \texttt{Coder} by reporting on its output, and certifies which modules are eligible for the comparative-statics sweep that follows.

 \textbf{BatchRunner} serves as the systematic experimentation engine for economic models. Its role involves sweeping every parameter of a certified module around its proposed value and reporting the resulting steady-state response, generating comparative statics for analysis. \texttt{BatchRunner} activates once \texttt{Debugger} certifies a module. It interfaces with \texttt{SweepProfiler}, which profiles its sweep's real execution time; with the versioning step, whose identifiers let its comparative-statics results be matched back to the exact module and parameters that produced them; and it provides structured simulation outputs to the empirical analysis phase downstream.

 \textbf{SweepProfiler} profiles the comparative-statics sweep \texttt{BatchRunner} runs: it times each model's sweep in wall-clock seconds and reports the total, mean, and slowest sweep time, alongside the share of individual parameter perturbations that failed to converge. This is diagnostic, not corrective: \texttt{SweepProfiler} identifies where computational cost concentrates; it does not rewrite the derived module, since LLM-authored edits to the derived module would break the rule that no LLM-authored code enters the pipeline. \texttt{SweepProfiler} activates automatically once \texttt{BatchRunner}'s sweep completes, and its findings are reported alongside the rest of that stage's output.

 \textbf{VersionManager} is deterministic and content-hash based: it fingerprints each derived module's exact source together with its exact parameter values into a single model version identifier, so identical code with different parameter values, or vice versa, is always a different, traceable version. It also fingerprints the pipeline's own stage code, the same technique the \texttt{DataTeam}'s \texttt{Archivist} uses, so the exact code version that produced a result is independently verifiable. \texttt{VersionManager} activates once \texttt{BatchRunner}'s sweep completes, versioning every module generated in that run. It connects with \texttt{BatchRunner} by identifying which version produced which comparative-statics results, and supports the \texttt{DataTeam}'s \texttt{Archivist} component by making exact model versions recreatable.

 \textbf{DocuAgent} narrates the combined record from generation, validation, experimentation, and versioning into a single technical-documentation summary (what was derived and from how many equations, what the validation checks found, what the comparative-statics sweep showed and how long it took, and each module's version identifier), under the same guardrail as \texttt{Coder}, \texttt{Debugger}, and \texttt{BatchRunner}: describe only what the pipeline actually produced, never invent capabilities the run did not exercise. \texttt{DocuAgent} activates once \texttt{VersionManager} completes versioning, drawing on all three stages' outputs plus the version manifest.

 \textbf{Logger} is the shared instrumentation layer used by every team in the pipeline, not an agent unique to \texttt{CodeTeam}. It records stage-level timing, item counts, errors, and status for each of the three \texttt{CodeTeam} stages into a structured execution log: a fixed, uniform audit trail rather than one an agent dynamically designs. \texttt{Logger} runs automatically at the start and end of every stage; economists consult its execution log to verify milestone completion and diagnose where a run failed.

 \subsubsection{Workflow and Human Checkpoints}

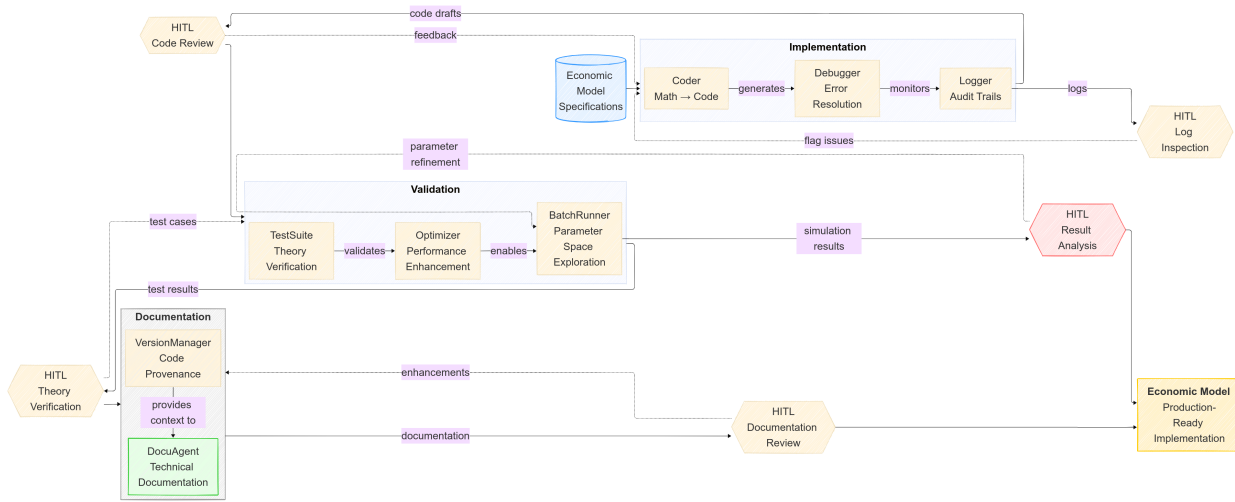
\begin{figure}[!htbp]
    \centering
    \scalebox{\aelFigScale}{\begin{tikzpicture}[ael]

\node[aelsource] (specs) at (1.5,0)
  {Economic Model\\Specifications};

\node[aelagentcol, text width=26mm, right=10.5mm of specs] (coder) {\aelAgent{Coder}{Derivation Report}};
\node[aelagentshared, text width=26mm, right=4.5mm of coder] (logA) {\aelAgent{Logger}{Audit Trails}};
\node[aelart, text width=14mm, right=4.5mm of logA] (logsA) {logs};

\node[fit=(coder)(logA)(logsA), inner sep=0pt] (stageonebody) {};
\node[aelstagetitle] (ttlA) at ($(stageonebody.north)+(0,5mm)$)
  {Initiation \&\\Code Generation};
\node[aelstagenum, left=1.3mm of ttlA] (numA) {1};
\begin{scope}[on background layer]
  \node[aelstageA, fit=(stageonebody)(ttlA)(numA)] (stageone) {};
\end{scope}

\draw[aelflow] (specs) -- (coder);
\draw[aelflow] (coder) -- (logA);
\draw[aelflow] (logA)  -- (logsA);

\node[aelhitl, anchor=west] (hitlCode) at (specs.west |- 0,-4.4)
  {\textbf{HITL}\\Code\\Review};

\node[aelagentcol, text width=26mm, anchor=west] (debug) at (coder.west |- hitlCode)
  {\aelAgent{Debugger}{Check-Battery Report}};
\node[aelagentshared, text width=26mm, right=4.5mm of debug] (logB) {\aelAgent{Logger}{Audit Trails}};
\node[aelart, text width=14mm, right=4.5mm of logB] (logsB) {logs};

\node[fit=(debug)(logB)(logsB), inner sep=0pt] (stagetwobody) {};
\node[aelstagetitle] (ttlB) at ($(stagetwobody.north)+(0,5mm)$)
  {Testing \&\\Validation};
\node[aelstagenum, left=1.3mm of ttlB] (numB) {2};
\begin{scope}[on background layer]
  \node[aelstageB, fit=(stagetwobody)(ttlB)(numB)] (stagetwo) {};
\end{scope}

\draw[aelflow] (hitlCode) -- (debug);
\draw[aelflow] (debug)    -- (logB);
\draw[aelflow] (logB)     -- (logsB);

\coordinate (rowleft) at ($(specs.west)+(4.76mm,0)$);
\node[aelagentcol, text width=24mm, anchor=west] (batch) at (rowleft |- 0,-8.66) {\aelAgent{BatchRunner}{Parameter Space Exploration}};
\node[aelagentcol, text width=24mm, right=4.5mm of batch] (sweep) {\aelAgent{SweepProfiler}{Sweep Timing}};
\node[aelagentcol, text width=24mm, right=4.5mm of sweep] (version) {\aelAgent{VersionManager}{Code Provenance}};
\node[aelagentcol, text width=24mm, right=4.5mm of version] (docu) {\aelAgent{DocuAgent}{Technical Documentation}};

\node[fit=(batch)(sweep)(version)(docu), inner sep=0pt] (stagethreebody) {};
\node[aelstagetitle] (ttlC) at ($(stagethreebody.north)+(0,5mm)$)
  {Experimentation, Versioning \& Documentation};
\node[aelstagenum, left=1.3mm of ttlC] (numC) {3};
\begin{scope}[on background layer]
  \node[aelstageC, fit=(stagethreebody)(ttlC)(numC)] (stagethree) {};
\end{scope}

\draw[aelflow] (batch)   -- (sweep);
\draw[aelflow] (sweep)   -- (version);
\draw[aelflow] (version) -- (docu);

\node[aelhitl, anchor=west] (hitlDoc)
  at ($(stagethree.east |- docu)+(4mm,0)$) {\textbf{HITL}\\Documentation\\Review};
\draw[aelflow] (docu.east) -- (hitlDoc.west);

\node[aelout, text width=26mm, right=5.5mm of hitlDoc] (out)
  {Economic Model\\[2pt]\mdseries Derived Module \& Documentation};
\draw[aelflow] (hitlDoc.east) -- (out.west);

\node[aelhitl, anchor=east] (hitlLogA) at (out.east |- logsA)
  {\textbf{HITL}\\Log\\Inspection};
\draw[aelflow] (logsA.east) -- (hitlLogA.west);
\node[aelhitl, anchor=east] (hitlLogB) at (out.east |- logsB)
  {\textbf{HITL}\\Log\\Inspection};
\draw[aelflow] (logsB.east) -- (hitlLogB.west);

\coordinate (laneA) at ($(specs.west)+(-6mm,0)$);
\coordinate (laneB) at (laneA);
\coordinate (midAB) at ($(stageone.south)!0.5!(stagetwo.north)$);
\draw[aelflow] (hitlLogA.south) |- (hitlCode.north |- midAB) -- (hitlCode.north);
\coordinate (midBC) at ($(stagetwo.south)!0.5!(stagethree.north)$);
\draw[aelflow] (hitlLogB.south) |- (laneB |- midBC) -- (laneB |- batch)
  -- (stagethree.west |- batch);

\coordinate (flagA) at ($(stageone.south)+(0,-3.5mm)$);
\draw[aelfeed] (hitlLogA.south) |- (coder.south |- flagA) -- (coder.south);
\node[aellabel] at ($(coder.south |- flagA)!0.5!(hitlLogA.south |- flagA)$) {flag issues};

\coordinate (flagB) at ($(stagetwo.south)+(0,-3.5mm)$);
\draw[aelfeed] (hitlLogB.south) |- (debug.south |- flagB) -- (debug.south);
\node[aellabel] at ($(debug.south |- flagB)!0.5!(hitlLogB.south |- flagB)$) {flag issues};

\coordinate (flagC) at ($(stagethree.south)+(0,-3.5mm)$);
\draw[aelfeed] (hitlDoc.south) |- (batch.south |- flagC) -- (batch.south);
\node[aellabel] at ($(batch.south |- flagC)!0.5!(hitlDoc.south |- flagC)$) {enhancements};

\coordinate (retA) at ($(stageone.north)+(0,4mm)$);
\draw[aelfeed] (hitlCode.west) -- (laneA |- hitlCode) |- (coder.north |- retA) -- (coder.north);
\node[aellabel] at ($(coder.north |- retA)+(-1.8,0)$) {feedback};

\end{tikzpicture}}
    \caption{Agentic Workflow for Code Implementation and Testing in Economic Research}
    \label{fig:code_implementation} 
\end{figure}

 The code development and validation process runs across three stages (deterministic code generation, testing and validation, and experimentation with versioning and documentation), with \texttt{Logger}'s instrumentation recording every stage automatically. At each step, deterministic computation does the work and an agent narrates it, with economists reviewing what was produced rather than co-authoring code. Figure~\ref{fig:code_implementation} shows the process.

 \textbf{Initiation and Code Generation Stage}
 Once model specifications and proposed parameter values are finalized, the rule-based procedure derives the module and \texttt{Coder} reports which models generated cleanly, which are partial, and which the parser refused, with reasons. Economists record a review of these verdicts at a HITL checkpoint before the module proceeds to validation.

 \textbf{Testing and Validation Stage}
 \texttt{Debugger} certifies each derived module or explains what a failing check implies about the specification. Human economists review these verdicts and the model specification they trace back to; the verdicts and the parser's reasons are carried into the report.

 \textbf{Experimentation, Versioning, and Documentation Stage}
 With a certified module in place, \texttt{BatchRunner} sweeps every parameter around its proposed value and reports the comparative-statics response of the steady state. \texttt{SweepProfiler} times this sweep and reports where computational cost concentrated; \texttt{VersionManager} fingerprints the exact module and parameters that produced the results into a traceable version identifier; and \texttt{DocuAgent} narrates the combined record into technical documentation. Economists review the comparative statics and the resulting documentation together, since one is a numerical result and the other is its record.
 
 \subsection{Empirical Estimation and Validation}
 This phase applies statistical and econometric techniques to estimate model parameters, then runs diagnostics and robustness checks on those estimates. Economic-magnitude interpretation is deliberately left to the next phase, Interpretation and Reporting, rather than duplicated here.
 
 \subsubsection{Agents}

 Four agents run this estimation and validation pipeline: an LLM proposes a specification or narrates already-computed statistics, while the reported verdict, diagnostics, and effects are always the deterministic numerical core's, never the LLM's.

 \textbf{Estimator} functions as the econometric specification specialist for economic models. Its role is to translate a model specification into one estimable reduced-form regression on the series actually available. The agent processes the model specification produced upstream (e.g., by \texttt{ModelDesigner}/\texttt{Calibrator}) and the validated dataset from \texttt{DataTeam}, proposing dependent/regressor variables, transforms, and theory-derived sign hypotheses as an \texttt{EstimationSpec}. The deterministic harness then estimates it and issues the verdict (\textit{estimated}, \textit{fragile}, or \textit{inestimable} with a machine-readable reason), never a substituted or fabricated fit. \texttt{Estimator} activates once a model specification and a prepared dataset are both available, and again for one bounded revision round after reviewer feedback.

 \textbf{Validator} serves as the empirical performance evaluator. It narrates the diagnostics battery the harness runs on the stored analysis panel: Breusch--Pagan (heteroskedasticity), Durbin--Watson and Ljung--Box (autocorrelation), Jarque--Bera (normality), a residual augmented Dickey--Fuller (ADF) test as a spurious-regression guard, Ramsey's RESET (functional form), variance inflation factors (VIF; multicollinearity), and a sample-size check. Severe failures (a residual AR(1) coefficient of at least 0.9, a Durbin--Watson statistic below 0.5, or a residual-ADF $p$-value of at least 0.10) downgrade the verdict from estimated to fragile, a harness decision the agent reports but cannot override. \texttt{Validator} activates immediately after an estimable specification is produced.

 \textbf{HypothesisTester} functions as the formal inference and stability specialist. It deterministically tests every theory-derived sign/zero restriction carried in the specification, classifying a sign restriction as \textit{supported} (one-sided $p<0.05$ in the predicted direction), \textit{contradicted} (two-sided $p<0.05$ in the opposite direction), or \textit{inconclusive} (neither), and a zero restriction as \textit{consistent} or \textit{rejected} at the 5\% level; it then runs robustness sweeps (e.g., drop-one-regressor, split-sample halves, and covariance-estimator swap), all re-estimated on the same stored analysis panel, reporting a sign-stability score. It narrates the computed hypothesis and robustness results; it does not run additional statistics beyond what the harness already computed.

 \textbf{LatencyProfiler} is a deterministic performance profiler: it aggregates the real wall-clock timing recorded across all three stages, split into deterministic-compute time (data loading, estimation, diagnostics, robustness sweeps) versus LLM proposal/narration time, so an economist can see where the workflow's actual bottleneck sits: at this pipeline's scale, LLM latency typically dominates the (comparatively cheap) econometric computation.

 \subsubsection{Workflow and Human Checkpoints}

\begin{figure}[!htbp]
    \centering
    \scalebox{\aelFigScale}{\begin{tikzpicture}[ael]

\node[aelagentcol, text width=28mm] (est) at (6.4,0) {\aelAgent{Estimator}{Translate Models}};
\node[fit=(est), inner sep=0pt] (stageonebody) {};
\node[aelstagetitle] (ttlA) at ($(stageonebody.north)+(0,5mm)$) {Parameter\\Estimation};
\node[aelstagenum, left=1.3mm of ttlA] (numA) {1};
\begin{scope}[on background layer]
  \node[aelstageA, fit=(stageonebody)(ttlA)(numA)] (stageone) {};
\end{scope}

\node[aelsource, anchor=east] (data) at ($(stageone.west |- est)+(-6mm,0)$)
  {Model \&\\Dataset};
\draw[aelflow] (data) -- (est);

\node[aelhitl, anchor=west] (hitlPar)
  at ($(stageone.east |- est)+(4mm,0)$) {\textbf{HITL}\\Review\\Specification};
\draw[aelflow] (est.east) -- (hitlPar.west);

\node[aelagentcol, text width=28mm, anchor=west] (val) at ($(hitlPar.east)+(10mm,0)$)
  {\aelAgent{Validator}{Evaluate Performance}};

\node[fit=(val), inner sep=0pt] (stagetwobody) {};
\node[aelstagetitle] (ttlB) at ($(stagetwobody.north)+(0,5mm)$) {Validation \&\\Diagnostics};
\node[aelstagenum, left=1.3mm of ttlB] (numB) {2};
\begin{scope}[on background layer]
  \node[aelstageB, fit=(stagetwobody)(ttlB)(numB)] (stagetwo) {};
\end{scope}

\draw[aelflow] (hitlPar.east) -- (stagetwo.west |- val);

\node[aelhitl, anchor=west] (hitlDia)
  at ($(stagetwo.east |- val)+(4mm,0)$) {\textbf{HITL}\\Assess\\Diagnostics};
\draw[aelflow] (val.east) -- (hitlDia.west);

\coordinate (rowleft) at ($(data.west)+(8.17mm,0)$);
\node[aelagentcol, text width=30mm, anchor=west] (hypo) at (rowleft |- 0,-3.9)
  {\aelAgent{HypothesisTester}{Statistical Tests}};
\node[aelagentcol, text width=30mm, right=4.5mm of hypo] (lat)
  {\aelAgent{LatencyProfiler}{Profile Performance}};

\node[fit=(hypo)(lat), inner sep=0pt] (stagethreebody) {};
\node[aelstagetitle] (ttlC) at ($(stagethreebody.north)+(0,5mm)$)
  {Inference, Robustness \& Performance Profiling};
\node[aelstagenum, left=1.3mm of ttlC] (numC) {3};
\begin{scope}[on background layer]
  \node[aelstageC, fit=(stagethreebody)(ttlC)(numC)] (stagethree) {};
\end{scope}

\draw[aelflow] (hypo) -- (lat);

\coordinate (midBC) at ($(stagetwo.south)!0.5!(stagethree.north)$);
\coordinate (laneL) at ($(data.west)+(-6mm,0)$);
\draw[aelflow] (hitlDia.south) |- (laneL |- midBC) -- (laneL |- hypo)
  -- (stagethree.west |- hypo);

\node[aelout, anchor=east] (out) at (hitlDia.east |- lat)
  {Estimation Results\\[2pt]\mdseries Diagnostics \& Hypothesis Tests};
\node[aelhitl] (hitlRef) at ($(stagethree.east |- lat)!0.5!(out.west)$)
  {\textbf{HITL}\\Final\\Review};
\draw[aelflow] (lat.east) -- (hitlRef.west);
\draw[aelflow] (hitlRef.east) -- (out.west);

\coordinate (retA) at ($(stageone.north)+(0,4mm)$);
\draw[aelfeed] (hitlPar.north) |- (est.north west |- retA) -- (est.north west);
\node[aellabel] at ($(est.north west |- retA)!0.5!(hitlPar.north |- retA)$)
  {adjust estimators};

\coordinate (retB) at ($(stagetwo.north)+(0,4mm)$);
\draw[aelfeed] (hitlDia.north) |- (val.north west |- retB) -- (val.north west);
\node[aellabel] at ($(val.north west |- retB)!0.5!(hitlDia.north |- retB)$)
  {adjust specifications};

\coordinate (retC) at ($(stagethree.south)+(0,-3.5mm)$);
\draw[aelfeed] (hitlRef.south) |- (lat.south |- retC) -- (lat.south);
\node[aellabel] at ($(lat.south |- retC)!0.5!(hitlRef.south |- retC)$)
  {request clarification};

\end{tikzpicture}}
    \caption{Agentic Workflow for Empirical Estimation and Validation in Economic Research}
    \label{fig:empirical_estimation}
\end{figure}
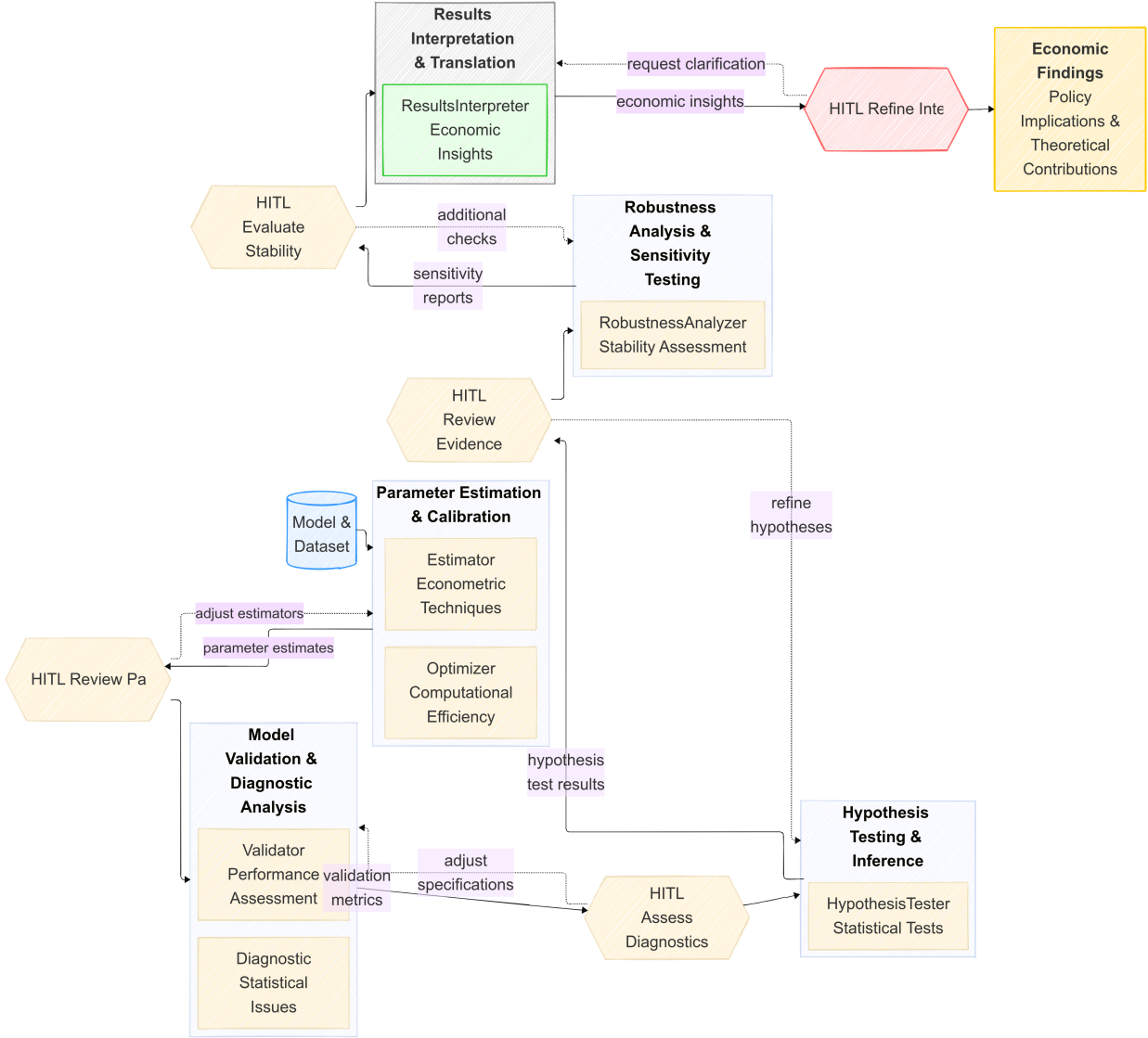

 The empirical estimation workflow runs in three stages, each pairing deterministic harness computation with LLM proposal or narration and, in the with-HITL configuration, a review checkpoint. Figure~\ref{fig:empirical_estimation} shows the process.

 \textbf{Parameter Estimation}
 \texttt{Estimator} proposes a specification. Before anything is estimated, economists review it against the available series and the observation budget, without seeing any estimate; a rejected proposal gets one revision, which is reviewed again, and if the revision is also rejected the original is estimated with the objections recorded. The harness then estimates the reviewed specification; it refuses any specification in which a series that builds the dependent variable enters a regressor in the same period. If the harness refuses the reviewed specification, one repair may be proposed: it must keep the reviewed dependent variable and hypotheses, it is reviewed before it is fitted, and it replaces the refused specification only if its verdict is better. Every fitted specification is recorded with its verdict and listed in the report. This HITL checkpoint reviews the specification against economic theory and institutional context before any estimate is seen.

 \textbf{Validation \& Diagnostics}
 The harness runs the full diagnostics battery on the stored analysis panel and applies any resulting verdict downgrade; \texttt{Validator} reports the results. Economists examine the battery results and the (possibly downgraded) verdict, determining whether the specification needs revision or the caveats can be carried forward into the final narrative.

 \textbf{Inference, Robustness \& Performance Profiling}
 \texttt{HypothesisTester} tests theory-derived restrictions and runs the robustness sweeps; \texttt{LatencyProfiler} aggregates the run's performance profile across all three stages. Economists conduct a final review of the hypothesis-test outcomes and stability results, recorded alongside the outputs, before the estimation artifact moves to \texttt{ReportingTeam} for economic interpretation and reporting.
 
 \subsection{Interpretation and Reporting}
 \label{sec:reporting}
 This stage converts technical model estimations and empirical data into interpretable economic knowledge, alongside the preparation of the final academic document.
 The stage computes economic magnitudes from the estimates, assembles the draft report, and checks the numbers in the draft against the upstream artifacts before the wording is polished.

 \subsubsection{Agents}

 Five agents run an automated report-writing pipeline built around a deterministic reporting harness: every number in the final report is either computed from the upstream artifacts or, unless exempt, looked up in them, and every exhibit is assembled directly from those artifacts. The LLM's narration is confined to captions, labels, and structural summaries; where the report calls for argumentative prose the pipeline cannot generate deterministically, such as the introduction and discussion, it hands the economist a bounded, numerically checked reference summary rather than inserting its own text.

 \textbf{ResultsInterpreter} functions as the economic significance translator for empirical findings. Its role is to convert statistical results into meaningful economic magnitudes. The agent processes the estimation artifact from \texttt{EstimationTeam} (coefficients, the stored analysis panel, and the fitted specification's transforms), computing one-standard-deviation effects, standardized betas, and elasticity-at-means for each regressor (deterministic; the interpretation kind follows the transform pair, as in \texttt{EstimationTeam}), and accompanying figures (e.g., a time-series plot and a coefficient plot) rendered directly from the same data. \texttt{ResultsInterpreter} then narrates the computed magnitudes into economic (not merely statistical) significance; it may not compute or invent a new number. An inestimable upstream verdict passes through with no invented magnitudes. \texttt{ResultsInterpreter} activates once \texttt{EstimationTeam}'s output is available.

 \textbf{Reporter} functions as the research narrative architect. It specializes in organizing empirical outcomes and their analyses into a research report. The agent deterministically assembles a structured report: a fixed section skeleton (research question, literature context, model, data, estimation results, discussion, limitations, references, provenance), with every table and figure formatted directly from the pipeline's own artifacts and a References list sourced from the literature corpus \texttt{LiteratureTeam} gathered. The model section tabulates each parameter's proposed value next to the harness's estimate, naming the stand-in simulator where one was used; the estimation section lists the module variables of the executable model that each series was linked to; and the limitations section is compiled from the artifacts: hypotheses that were inconclusive or contradicted, the scope of each calibration verdict, and every series used as a proxy. Where the skeleton calls for prose the pipeline cannot generate deterministically, such as the introduction and discussion, it leaves the slot marked for the economist and attaches a short, bounded reference summary of the relevant artifacts, grounded and numerically checked, for the economist to draw on while writing that section. \texttt{Reporter} activates once \texttt{ResultsInterpreter}'s magnitudes and figures are available, and again for one bounded revision round after reviewer feedback.

 \textbf{Proofreader} narrates wording and academic-convention suggestions on the checked draft; it has no authority over numbers, citations, or verdicts, and does not itself check them; that is the deterministic consistency check described next. \texttt{Proofreader} activates once the number-consistency check has run.

 \textbf{JournalAdvisor} serves as the publication strategy specialist; its shortlist is computed deterministically rather than freely recommended: it scores candidate journals by keyword overlap between the report's own content (research question, estimated dependent variable) and a small, static, illustrative reference table of journal scopes, explicitly not live database intelligence, so it never claims acceptance rates, impact factors, or current editorial preferences it cannot actually observe. For the top-ranked candidates it narrates an adaptation strategy (framing, emphasis) grounded only in the shortlist and the report's own content. \texttt{JournalAdvisor} activates after the number-consistency check has run.

 \textbf{Formatter} performs a deterministic, no-LLM re-render of the checked report into a specific journal reference style (AEA by default) and appends a Data \& Code Availability statement built from real pipeline facts (how many series were retrieved live versus disclosed as simulated; whether a code-implementation section is present). It writes a separate file and leaves the checked original unchanged. \texttt{Formatter} activates alongside \texttt{JournalAdvisor}, after quality verification.

 \subsubsection{Workflow and Human Checkpoints}

\begin{figure}[!htbp]
    \centering
    \scalebox{\aelFigScale}{\begin{tikzpicture}[ael]

\node[aelagentcol, text width=32mm] (results) at (6.8,0)
  {\aelAgent{ResultsInterpreter}{Economic Significance Translator}};
\node[fit=(results), inner sep=0pt] (stageonebody) {};
\node[aelstagetitle] (ttlA) at ($(stageonebody.north)+(0,5mm)$) {Interpretation \& Figures};
\node[aelstagenum, left=1.3mm of ttlA] (numA) {1};
\begin{scope}[on background layer]
  \node[aelstageA, fit=(stageonebody)(ttlA)(numA)] (stageone) {};
\end{scope}

\node[aelsource, anchor=east] (findings) at ($(stageone.west |- results)+(-6mm,0)$)
  {Empirical\\Findings};
\draw[aelflow] (findings) -- (results);

\node[aelart, anchor=west] (hitlInt)
  at ($(stageone.east |- results)+(4mm,0)$) {Economic\\Magnitudes};
\draw[aelflow] (results.east) -- (hitlInt.west);

\node[aelagentcol, text width=32mm, anchor=west] (reporter) at ($(hitlInt.east)+(10mm,0)$)
  {\aelAgent{Reporter}{Research Narrative Architect}};

\node[fit=(reporter), inner sep=0pt] (stagetwobody) {};
\node[aelstagetitle] (ttlB) at ($(stagetwobody.north)+(0,5mm)$) {Drafting \& Tables};
\node[aelstagenum, left=1.3mm of ttlB] (numB) {2};
\begin{scope}[on background layer]
  \node[aelstageB, fit=(stagetwobody)(ttlB)(numB)] (stagetwo) {};
\end{scope}

\draw[aelflow] (hitlInt.east) -- (stagetwo.west |- reporter);

\node[aelhitl, anchor=west] (hitlStr)
  at ($(stagetwo.east |- reporter)+(4mm,0)$) {\textbf{HITL}\\Review\\Structure};
\draw[aelflow] (reporter.east) -- (hitlStr.west);

\coordinate (rowleft) at ($(findings.west)+(4.76mm,0)$);
\node[aelagentcol, text width=28mm, anchor=west] (proof) at (rowleft |- 0,-4.1)
  {\aelAgent{Proofreader}{Quality Control Specialist}};
\node[aelagentcol, text width=28mm, right=4.5mm of proof] (advisor)
  {\aelAgent{JournalAdvisor}{Publication Strategy Specialist}};
\node[aelagentcol, text width=28mm, right=4.5mm of advisor] (fmt)
  {\aelAgent{Formatter}{Technical Document Specialist}};

\node[fit=(proof)(advisor)(fmt), inner sep=0pt] (stagethreebody) {};
\node[aelstagetitle] (ttlC) at ($(stagethreebody.north)+(0,5mm)$)
  {Quality, Journal Targeting \& Formatting};
\node[aelstagenum, left=1.3mm of ttlC] (numC) {3};
\begin{scope}[on background layer]
  \node[aelstageC, fit=(stagethreebody)(ttlC)(numC)] (stagethree) {};
\end{scope}

\draw[aelflow] (proof)   -- (advisor);
\draw[aelflow] (advisor) -- (fmt);

\coordinate (midBC) at ($(stagetwo.south)!0.5!(stagethree.north)$);
\coordinate (laneL) at ($(findings.west)+(-6mm,0)$);
\draw[aelflow] (hitlStr.south) |- (laneL |- midBC) -- (laneL |- proof)
  -- (stagethree.west |- proof);

\node[aelout, anchor=east] (out) at (hitlStr.east |- fmt)
  {Annotated Report};
\draw[aelflow] (fmt.east) -- (out.west);

\coordinate (retB) at ($(stagetwo.north)+(0,4mm)$);
\draw[aelfeed] (hitlStr.north) |- (reporter.north west |- retB) -- (reporter.north west);
\node[aellabel] at ($(reporter.north west |- retB)!0.5!(hitlStr.north |- retB)$)
  {adjust organization};

\end{tikzpicture}}
    \caption{Agentic Workflow for Interpretation and Reporting in Economic Research}
    \label{fig:interpretation}
\end{figure}
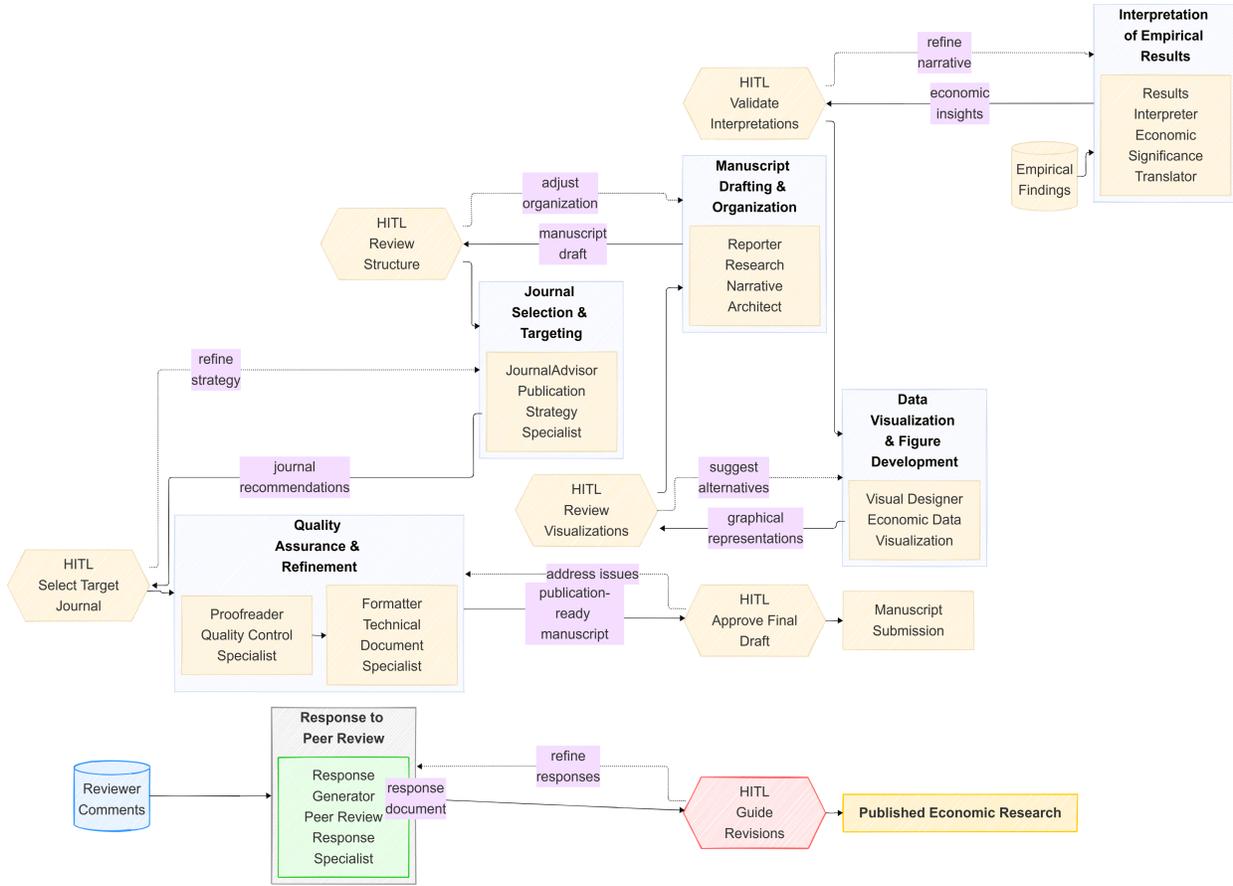

 The research dissemination process runs in three stages: each pairs deterministic harness computation with either LLM narration or, where the report calls for argumentative prose, the economist's own writing. Figure~\ref{fig:interpretation} shows the process.

 \textbf{Interpretation \& Figures}
 \texttt{ResultsInterpreter} converts \texttt{EstimationTeam}'s statistical results into economic magnitudes and the accompanying publication figures, narrating what each effect means in economic terms. An inestimable upstream verdict carries through unchanged. This stage runs without a dedicated human checkpoint; its output feeds directly into drafting, where \texttt{Reporter} tabulates it alongside the raw estimates and economists review the assembled draft.

 \textbf{Drafting \& Tables}
 \texttt{Reporter} assembles the report skeleton, tables, and figure placements, and attaches reference summaries for the sections the economist still needs to write. Economists draft the introduction and discussion themselves, informed by those summaries, and review the skeleton's framing, emphasis, and any missing caveats; substantive feedback triggers one bounded revision round before the draft is finalized. This is the workflow's HITL checkpoint: the point at which a human economist's judgment about narrative and emphasis is incorporated before the numbers are locked in for verification.

 \textbf{Quality, Journal Targeting \& Formatting}
 The number-consistency checker looks up the numbers in the draft among the numeric fields of the named upstream artifacts and the numbers the report template derives from them; numbers inside generated prose are not sources. Small integers used as counts, ordinals, or list labels, the confidence levels 90, 95, and 99 printed next to ``CI'' or ``confidence'', and digits inside URLs and DOIs are exempt. For each number, the checker records the artifact it was found in and the kind of match (exact, rounded, same significant figures, or a percentage rescaling), and it annotates the report with a verified/total count; it establishes that a number occurs upstream, not that it is used in the right place. \texttt{Proofreader} then narrates wording suggestions on the annotated draft, \texttt{JournalAdvisor} shortlists candidate journals and narrates an adaptation strategy, and \texttt{Formatter} writes a separately formatted, submission-styled copy. Economists receive the annotated report, the journal shortlist, and the formatted copy together as the pipeline's final output.

\subsection{Cross-Team Pipeline Integration}
\label{sec:pipeline-design}

While each team described above can run on its own, a full study requires them to execute in dependency order: a research question is formulated, the literature is reviewed, a model is specified and reconciled against the data that actually exists, and the reconciled model is then implemented, estimated, and reported. A cross-team orchestrator formalizes this dependency chain into an automated, resumable pipeline over the seven teams.

The orchestrator resolves the team dependencies into an execution order and runs in two phases. A linear pass carries the research topic through the \texttt{IdeationTeam} (prioritized research questions), the \texttt{LiteratureTeam} (a review, gap analysis, and knowledge graph), the \texttt{ModelTeam} (a model specification), and the \texttt{DataTeam} (a validated dataset). The three analytic teams are held back until the model and data have been reconciled: the \texttt{CodeTeam} turns the specification into an executable model, the \texttt{EstimationTeam} produces estimation results on the assembled data, and the \texttt{ReportingTeam} assembles the research report; the \texttt{CodeTeam} is optional and off by default. When it runs, it precedes the \texttt{EstimationTeam}, which reads the lead model's executable module and records which of its variables each estimated series measures. Each of these outputs is registered as a named artifact in a shared store, alongside the feasibility contracts introduced below. Each artifact is signed with a hash-based message authentication code (HMAC) when it is registered. Before a team starts, the orchestrator retrieves the upstream artifacts it depends on and checks each against the signature and a content fingerprint recorded when it was produced; an artifact that fails the signature check fails the team that reads it, a fingerprint mismatch is logged as a warning, and the run manifest records the number of verified reads and every signature failure.

The Model--Data boundary is a bounded \emph{feasibility cycle} rather than a one-way hand-off. The \texttt{ModelTeam} emits the DRS; the \texttt{DataTeam} answers with the DAR, grading each requirement and issuing a feasibility verdict; and where the data cannot support the specification, an MRR carries the binding constraints back to the \texttt{ModelTeam} for re-specification (Section~\ref{sec:data-collection}). These three contracts are themselves provenance-stamped artifacts, so the final study records why each variable is a proxy or why the model was narrowed. The cycle is bounded (at most three reconciliations by default), stops when a revision does not shrink the set of unmet essential requirements, and on non-convergence it terminates by proceeding with a documented data limitation rather than failing the run. The cycle's decisions are not checkpointed, so a resumed run repeats the cycle from its start. Escalation to premium or user-uploaded data is off by default, so the reconciliation turns on the substantive choice between accepting a proxy and revising the model. That choice is made at the Feasibility Review, a human-in-the-loop checkpoint owned by an economist.

Resource use and failures are handled as the pipeline runs. Each team's cost is metered separately and rolled up into a running total; budget limits, per team and overall, are checked before each team and halt the run if an allocation would be exceeded. After each team completes, its outputs and timing are saved, so the pipeline can be resumed from the last completed team rather than restarted. This matters for long research campaigns, where a failure late in the pipeline should not force re-execution of the earlier, more expensive teams.

LLM selection is configured per stage. The same configuration the standalone workflows use binds a chosen LLM to each team and stage, supporting cost-aware deployment: a stronger reasoning LLM for the \texttt{ModelTeam}'s theory stage, a lighter one for routine extraction in the \texttt{DataTeam}. The whole sequence runs from a single command, given a research-topic seed, an optional global LLM override, a choice of data workflow (open-source APIs, premium subscriptions, or user-uploaded data), and a resume option for continuing an interrupted run.

\section{An End-to-End Demonstration}
\label{sec:Examples}

We demonstrate the framework by tracing one chained run through all seven stages of the pipeline, laid out in Figure~\ref{fig:pipeline-flow}. It is one of 26 runs made with the same version of the code and the same seed topic, chosen by the rule stated below. Seeded only with the topic ``agent-based and heterogeneous-agent modeling in macroeconomics and monetary policy'', it carried one line of inquiry to a number-checked research report in 7.4 hours of wall-clock time, with every choice after the seed (the questions, the models, the data, the specification) made inside the pipeline. The stack throughout is entirely open (Table~\ref{tab:tech-stack}): the reasoning teams run on an open-weight LLM served locally by vLLM, with web crawling and human-in-the-loop review enabled; the checkpoint committee comes from three further model families and the evaluation judges from two others; data arrives through open-source-API connectors (FRED with a free key, the others without one);\footnote{This demonstration exercises the open-source-API tier end to end, so that it needs no paid subscription. The premium-subscribed and user-uploaded tiers are validated separately (the premium path against a real Refinitiv terminal, the upload path on a sample CSV) rather than folded into this particular run. With escalation to a premium source out of scope here, the workflow's Feasibility Review checkpoint chooses among accepting a proxy series, revising the model, narrowing the question, and accepting a documented limitation.} and web content through an open crawl chain. Table~\ref{tab:ex-summary} collects the per-stage results discussed below.

\paragraph{Choosing the run.} The 26 runs differ in their questions, models, data, and estimates. One run had been designated as the demonstration when the jobs were submitted; its estimation stage was refused for too few overlapping observations (12, against the 32 its four parameters required). The run shown here is instead chosen by a rule that a script in the replication package applies to the stored artifacts. A run qualifies if (G1)~all seven teams completed; (G2)~every final research question passes the seed-topic check; (G3)~the estimation harness returned estimated, not fragile or inestimable; (G4)~every series in the reported regression was retrieved live and its source title matches the requested variable; and (G5)~the report's number-consistency check, re-run on the stored report with the checker as corrected after the runs, returned consistent. Qualifying runs are ranked by (R1)~the mean of the two judges' Economic Rigor and Soundness scores on the Ideation and Estimation stages, then by (R2)~the share of hypothesis tests with a definite outcome, (R3)~the number of models with harness-estimated parameters, and (R4)~the estimation sample size. Of the 26 runs, all 26 pass G1 and G2, 15 pass G3, 19 pass G4, and 14 pass G5; four pass all five gates, with R1 values of $0.732$, $0.724$, $0.708$, and $0.708$, and the run shown here is the first of them. The four R1 values span $0.023$; on the same eight scores, the two judges differ from each other by $0.124$ on average. Appendix~\ref{app:typical-case} summarizes the run closest to the median of the 26 runs on R1--R4 and the estimation verdict, together with the originally designated run.

The Model stage proposes three candidate formal models, one for each final research question, and the calibration harness returns a verdict for each: the first is point-calibrated in closed form, the second partially calibrated, and the third uncalibratable because it contains learned components. The first model, built on the top-ranked question, is carried through Estimation and Reporting; the Code team derives a module for all three. Figure~\ref{fig:pipeline-flow} marks where the candidates part; Appendix~\ref{appendix:example-model} gives the lead model in full and the verdicts on the other two.

\begin{figure}[!htbp]
\centering
\scalebox{\aelFigScale}{\begin{tikzpicture}[ael]

\node[aelsource] (direction) at (1.5,0)
  {Research\\Direction};
\node[aelagentcol, text width=22mm, minimum height=11mm, right=8mm of direction] (ideation) {\aelTeam{Ideation}};
\node[aelagentcol, text width=22mm, minimum height=11mm, right=18.3mm of ideation] (literature) {\aelTeam{Literature}};
\node[aelagentcol, text width=28mm, minimum height=11mm, right=22.9mm of literature] (modelcal)
  {\aelTeam{Model Design\\\& Calibration}};

\draw[aelflow] (direction) -- (ideation);
\draw[aelflow] (ideation)  -- node[aellabel, above] {questions} (literature);
\draw[aelflow] (literature) -- node[aellabel, above] {review + gaps} (modelcal);

\node[aelverdict, above=2mm of ideation]   {finalized};
\node[aelverdict, above=2mm of literature] {verified citations};
\node[aelverdict, above=2mm of modelcal]   {calibration verdicts};

\node[aelagentcol, text width=28mm, minimum height=11mm] (data) at (modelcal |- 0,-4.7)
  {\aelTeam{Data\\(Open Connectors)}};
\node[aelagentcol, text width=22mm, minimum height=11mm, anchor=west] (reporting)
  at (direction.west |- data) {\aelTeam{Reporting}};
\node[aelagentcol, text width=22mm, minimum height=11mm] (estimation)
  at ($(reporting.east)!0.5!(data.west)$) {\aelTeam{Estimation}};

\draw[aelflow] (data)       -- node[aellabel, above] {dataset}   (estimation);
\draw[aelflow] (estimation) -- node[aellabel, above] {estimates} (reporting);

\node[aelverdict, below=2mm of data]       {QA certification};
\node[aelverdict, below=2mm of estimation] {estimation verdicts};
\node[aelverdict, below=2mm of reporting]  {consistency check};

\node[aelagentcol, text width=28mm, minimum height=11mm] (code)
  at ($(estimation)!0.5!(estimation |- modelcal)$) {\aelTeam{Code\\(Derived Modules)}};
\draw[aelflow] (code) -- node[aellabel, right=1pt] {modules} (estimation);
\node[aelverdict, left=2mm of code] {validation verdicts};

\node[aelagentopt, text width=26mm, anchor=west] (a2) at ($(code.east)+(6mm,0)$)
  {\aelTeam{Other candidates}\\[1.5pt]own calibration verdicts\\and code modules};
\draw[aelflowopt] (a2) -- (code);

\coordinate (specout) at ($(modelcal.south west)+(0.3,0)$);
\coordinate (forky) at ($(modelcal.south)!0.5!(code.north)$);
\draw[aelflow] (specout) -- (specout |- forky) -| (code.north);
\node[aellabel, above] at ($(code.north |- forky)!0.5!(a2.north |- forky)$)
  {specification};
\draw[aelflowopt] (a2.north |- forky) -- (a2.north);

\draw[aelflow] ($(modelcal.south)+(-0.55,0)$) -- node[aellabel, left=1pt, pos=0.8] {DRS}
               ($(data.north)+(-0.55,0)$);
\draw[aelflow] ($(data.north)+(0.55,0)$) -- node[aellabel, right=1pt, pos=0.2] {DAR/MRR}
               ($(modelcal.south)+(0.55,0)$);

\coordinate (drsx) at ($(modelcal.south)+(-0.55,0)$);
\node[aelverdict, anchor=north east, align=right] (feasnote) at ($(drsx |- a2.south)+(-11.7mm,-2mm)$)
  {feasibility cycle:\\at most three rounds;\\unmet requirements\\recorded};

\begin{scope}[on background layer]
  \node[aelcheckframe,
        fit=(direction)(modelcal)(data)(reporting)] (checkframe) {};
\end{scope}
\node[aelverdict, above=1.5mm of checkframe.north, anchor=south]
  {committee-resolved human-in-the-loop checkpoints
   (three model families, lensed, 2/3 vote, ballots logged)};
\node[aelverdict, below=1.5mm of checkframe.south, anchor=north] (lowcap)
  {every stage: a language model proposes, a deterministic harness
   disposes $\rightarrow$ a scoped verdict; typed artifacts with provenance};

\node[aelout, anchor=north] (out) at ($(estimation |- lowcap.south)+(0,-4mm)$) {Report \& Audit Artifacts};
\coordinate (lane) at ($(checkframe.west)+(-6mm,0)$);
\draw[aelflow] (reporting.west) -- (lane |- reporting) |- (out.west);

\end{tikzpicture}}
\caption{Integration of the seven teams into one pipeline. Each solid box is a team (internal workflow detailed in Section~\ref{sec:workflow}); arrows carry the artifact each stage hands the next, italic tags mark each stage's verdict, and the dashed frame marks the checkpoint layer. The dotted box marks the other candidate models: each receives its own calibration verdict and its own derived code module; Estimation fits one specification proposed for the lead question, and the report tabulates the calibration and code results of all three models.}
\label{fig:pipeline-flow}
\end{figure}

\begin{table}[!htbp]
\centering
\begin{threeparttable}
\caption{Technical Stack of the Demonstration Run}
\label{tab:tech-stack}
\small
\begin{tabularx}{\textwidth}{@{}l Y@{}}
\toprule
Layer & Open components \\
\midrule
Model serving & vLLM 0.22.1, on a node of four A100 (40GB) GPUs \\
Generating model & Qwen3.6-27B-FP8 \\
Checkpoint committee & gpt-oss-20b, Phi-4-mini-instruct, Olmo-3.1-32B-Instruct \\
Evaluation judges & Mistral-Small-3.2-24B, Gemma-3-27B \\
Orchestration & AEL orchestrator: artifact store with HMAC-signed hand-offs, checkpoint/resume, budgets \\
Web acquisition & SearXNG, DDGS; trafilatura, crawl4ai, pypdf \\
Scholarly indices & OpenAlex (keyed), arXiv, Semantic Scholar, Crossref \\
Symbolic \& numeric & sympy, numpy, scipy \\
Data connectors & FRED, Yahoo Finance, World Bank, DBnomics (OECD/IMF/BIS), ECB, Eurostat, BLS \\
Econometrics & statsmodels, linearmodels \\
Reporting & deterministic Markdown assembly; matplotlib; number-consistency verifier \\
Evaluation & two-tier framework: structural metrics + dual-judge review, team-specific rubrics \\
Compute & SLURM batch on A100 nodes; full execution and observability logs \\
\bottomrule
\end{tabularx}
\begin{tablenotes}[flushleft]\footnotesize
\item \textit{Notes:} Every layer is open-source or open-weight; commercial services, including an institutional abstract API, serve only as fallbacks.
\end{tablenotes}
\end{threeparttable}
\end{table}

\paragraph{The LLM-Economist Committee.} The methodology keeps a real economist at each human-in-the-loop checkpoint the workflow defines. A real economist cannot sit an autonomous batch run, so for this demonstration those decisions were made by an \emph{LLM-Economist committee} standing in for the reviewer: a testing convenience that changes only who answers a checkpoint, never whether it exists. Following the same ``no model grades its own work'' principle as the dual-judge evaluation, the committee is three open-weight models from distinct families (OpenAI's gpt-oss, Microsoft's Phi-4-mini, and AllenAI's Olmo), none shared with the generating model or the two evaluation judges, each given the decision packet a human reviewer would see together with a shared economist rubric; in the chained runs the Model team's theory and calibration checkpoints are logged without a vote. Each member additionally reviews through a distinct economist lens (theoretical soundness, empirical realism, policy relevance), and across the 26 runs of Section~\ref{sec:robustness-checks}, 40.1\% of the 1{,}341 voted ballots were split rather than unanimous, with the minority vote coming from the data lens in 268 ballots, the theory lens in 250, and the policy lens in 20. Every yes/no proposition is carried on a two-of-three majority, and in a list selection (which questions to merge or discard) a merge or discard takes effect only if two of the three members chose it; a multi-option gate such as the Feasibility Review is resolved as a priority-ordered cascade of such votes with a documented-limitation default that guarantees resolution. Checkpoints that ask for free-text feedback are not voted on: the first member (gpt-oss) consolidates the three members' comments as chair. Every ballot is logged with each member's vote and rationale, with its prompt recorded in its first 500 characters, and tagged as committee-resolved, so these runs stay separable from a real reviewer's calls.

\paragraph{Ideation.} The Ideation team returned three prioritized research questions, all passing the seed-topic check: how AI-driven adaptive regulatory controllers mitigate systemic risk in a hybrid heterogeneous-agent New Keynesian (HANK) and agent-based model of financial networks under monetary-policy uncertainty; how the adoption of stablecoins by high-MPC households in high-inflation emerging markets alters monetary transmission in a HANK model; and how machine-learning surrogates can be calibrated to reproduce the distributional outcomes of HANK models under extreme fiscal shocks. Each of two refinement rounds turned 50 retrieved source papers into eight concept clusters and six candidate questions; at the integration checkpoint the committee cut the slate from five questions to three, and the \texttt{Finalizer}'s ranking placed the regulatory-controller question first (priority score $0.88$, tied with the stablecoin question and ahead of it in the ranked list). The pipeline carries that question forward.

\paragraph{Literature.} The Literature team sent 27 queries each to Crossref, arXiv, and OpenAlex. OpenAlex answered every query with HTTP status 429 (too many requests), so the candidate pool came from Crossref and arXiv alone: 257 papers, of which 182 carried a usable abstract and 70 were kept in the working base. The final report's 70 references are these 70 papers. Gap detection recorded 16 gaps (five methodological, four theoretical, four empirical, three on data) and a knowledge graph of 336 nodes (70 papers, 244 methods, 6 trends, 16 gaps) and 659 links.

\paragraph{Model.} The lead model couples a heterogeneous-household New Keynesian block (an Euler equation with a borrowing-constraint multiplier, a budget constraint, labor supply, Cobb--Douglas production, a Phillips curve, and a Taylor rule) with a bank-distress process on an interbank network and a reinforcement-learning regulator whose loss penalizes the output gap, aggregate bank distress, and abrupt changes in regulation: 16 equations and 13 listed parameters. The calibration harness parsed 5 of the 16 equations to closed form and refused the other 11, each with a recorded reason (Appendix~\ref{app:lead-calibration}). Three cited targets bear on the parameters through steady-state identities. The real interest rate ($2\%$) pins the discount factor at $\beta = 1/1.02 = 0.980$, against the language model's proposed $0.99$; the labor share ($0.60$) and the capital share ($0.36$) both restrict the capital elasticity, which the harness estimates at $\alpha = 0.38$ (proposed: $0.37$), missing each share by $0.02$ because the two targets do not sum to one. The harness classified all three moments as pinning moments and ran no over-identification test; the verdict is point-calibrated. The second model was partially calibrated: $\beta$ again from the real rate, while the share of high-MPC households stopped at a bound of its declared range without reproducing the MPC target. The third, which approximates the HANK distribution with a learned embedding, was declared uncalibratable, since its learned components would have to be trained before any moment could be computed. The Model--Data feasibility cycle ran three rounds, issued two revision requests (asking the models to drop their dependence on, among others, bank-distress frequency, systemic-risk contagion speed, and a foreign interest rate), and ended unresolved: of the 34 data requirements of the three models, 2 were graded feasible, 15 available only through a proxy, and 17 unavailable, and the committee answered the unmet requirements by voting to narrow the question (12 decisions, recorded without returning to the Ideation team) or to accept a proxy (2).

\paragraph{Data.} The Data team assembled 17 series. Thirteen were retrieved live, twelve from FRED and one from the World Bank. One of them, the 10-year real interest rate, was fetched on demand after the feasibility scout matched the requirement for a real interest rate to a FRED series outside the retrieval plan. Another, government consumption expenditures and gross investment, is disclosed as a proxy for the requested federal current expenditures, and two were retrieved after a disclosed repair of the requested series identifier. The other four (a CPI research series, a stablecoin market capitalization, bank deposit credit, and federal debt held by the public) had no successful live retrieval and are carried as disclosed simulations, which the estimation harness excludes. Quality assurance graded the dataset at 60 of 100 (Bronze); the certification logic caps the grade whenever a simulated series is present.

\paragraph{Code.} From all three formal models, the Code team derived executable steady-state modules directly from the parsed equations, with no free-hand model code from a language model anywhere in the artifact. All three carry a partial verdict: the lead model's module resolves 5 of 16 equations against 17 unknowns, the second model's 14 of 20 against 27, and the third's 2 of 20 against 4; each module compiles and passes the parameter-completeness and initial-residual checks, and its steady-state solve is declined because the parsed system is non-square. A chained run fits one regression, for the lead question, and assembles one report, which tabulates the calibration and code results of all three models (Figure~\ref{fig:pipeline-flow}).

\paragraph{Estimation.} \texttt{Estimator} proposed a quarterly regression of the St.~Louis Fed Financial Stress Index on a real policy rate (the federal funds rate minus year-over-year CPI inflation), the quarterly change in the federal funds rate (labeled monetary-policy uncertainty), and real GDP growth (labeled an output-gap proxy). The committee rejected the proposal and its revision; its objections named the absence of any variable measuring AI-driven regulation, the use of a rate change for uncertainty, and the use of a growth rate for the output gap. Under the review rule (Section~\ref{sec:workflow}), the original proposal was estimated with the objections recorded: OLS with heteroskedasticity-and-autocorrelation-consistent (HAC) standard errors, 1994Q4--2026Q2, $n=127$, $R^{2}=0.29$. None of the regression's variables is linked to a variable of the lead model. The real-rate coefficient is $-0.035$ ($p=0.36$) and the hypothesis that it is negative is inconclusive (one-sided $p=0.18$); the coefficient on the change in the funds rate is $-0.94$ ($p=0.025$), contradicting the hypothesis that it is positive. The diagnostics battery returned caveats with no severe failure (Durbin--Watson $0.66$, residual first-order autocorrelation $0.67$, RESET $p=0.008$), and the real-rate coefficient kept its sign in all five robustness variants (each other regressor dropped, each half of the sample, an alternative covariance estimator).

\paragraph{Reporting.} The terminal report runs to about 3{,}200 words before its 70 references, with two figures. Its number-consistency check traced 315 of the 332 numbers in the body to upstream artifacts; the other 17 (16 small integers and one confidence level) are exempt by the checker's rules; the verdict is consistent. The bibliography is the retrieved literature base, and the limitations section is collected automatically from every stage's recorded caveats: the four simulated series and the proxy, the two repaired identifiers, the committee's objections at the data, estimation, and report-draft checkpoints, the three calibration verdicts, the two hypothesis outcomes, and the unresolved feasibility cycle.

\begin{table}[!htbp]
\centering
\begin{threeparttable}
\caption{Per-Stage Outputs of the Demonstration Run}
\label{tab:ex-summary}
\small
\begin{tabularx}{\textwidth}{@{}l Y c@{}}
\toprule
Stage & Principal outputs & Tier-2 (M / G) \\
\midrule
Ideation & 3 prioritized questions, all on the seed topic; 2 refinement rounds; committee cut the slate from 5 to 3 & $0.802$ / $0.868$ \\
\addlinespace
Literature & 257 candidates (Crossref, arXiv; OpenAlex rate-limited) $\rightarrow$ 182 with abstracts $\rightarrow$ 70 in the working base; 16 gaps & $0.712$ / $0.812$ \\
\addlinespace
Model & 3 formal models: lead model point-calibrated ($\beta{=}0.980$, $\alpha{=}0.38$; 5 of 16 equations parsed), second partially calibrated, third uncalibratable; feasibility cycle: 3 rounds, 2 revision requests, unresolved & $0.740$ / $0.750$ \\
\addlinespace
Data & 17 series: 13 live (FRED, World Bank), 1 of them a disclosed proxy; 4 disclosed simulations, excluded from estimation; quality assurance 60/100 & $0.618$ / $0.562$ \\
\addlinespace
Code & 3 executable steady-state modules derived from parsed equations (5/16, 14/20, 2/20 equations resolved); all three partial & $0.528$ / $0.441$ \\
\addlinespace
Estimation & OLS--HAC, $n=127$ (1994Q4--2026Q2), $R^{2}=0.29$; 2 hypotheses: 1 inconclusive, 1 contradicted; specification rejected twice by the committee & $0.660$ / $0.608$ \\
\addlinespace
Reporting & $\sim$3{,}200-word report, 70 references; 315 of 332 numbers traced, consistent; auto-collected limitations & $0.663$ / $0.681$ \\
\bottomrule
\end{tabularx}
\begin{tablenotes}[flushleft]\footnotesize
\item \textit{Notes:} M = Mistral-Small-3.2-24B, G = Gemma-3-27B; each score is the mean over the six content dimensions of the dual-judge protocol (Section~\ref{sec:evaluations}), scored at the stage level rather than separately per candidate model. Scores apply to this run; the Correctness, Soundness, and Economic Rigor rubrics are team-specific, and the Innovation Potential, Transparency, and Decision Quality rubrics are shared across stages; the per-dimension breakdown is in Appendix~\ref{appendix:example-model}, and the distribution over the 26 runs in Section~\ref{sec:robustness-checks}.
\end{tablenotes}
\end{threeparttable}
\end{table}

Across the seven stages, a language model proposed and deterministic machinery decided: the calibration harness returned three different verdicts on the three candidate models; the estimation harness estimated the specification the committee had rejected twice, with the objections attached, and reported one hypothesis inconclusive and the other contradicted; the feasibility cycle ended by recording what the open data could not supply; and the four series that could not be retrieved were carried as disclosed simulations and kept out of the regression. In the terminal report, the numbers trace to artifacts, the references to retrieved papers, and the limitations to the stages that raised them.

\section{Evaluation Framework}
\label{sec:evaluations}

Evaluating agentic workflows for economic research presents methodological challenges distinct from standard AI benchmarking. Unlike classification or generation tasks amenable to controlled experiments, research workflows involve extended, context-dependent processes where output quality depends on interactions between system design, domain knowledge, and researcher judgment.
Following established frameworks for evaluating complex socio-technical systems \citep{Hollnagel2006, Woods2017}, we develop an evaluation framework that assesses agentic workflows across twelve dimensions applicable to economic research generally, together with subfield-specific dimensions illustrated here by \textit{Simulation Fidelity} and \textit{Causal Validity}, which cover the optional agent-based simulation and causal-foundation-model stages and are reported as not evaluated when those stages do not run; further such dimensions can be developed as a study requires.
The framework relates architectural choices to research outcomes, and Section~\ref{sec:empirical-validation} applies it to workflow executions.

\subsection{Theoretical Foundations}
\label{sec:theoretical-foundations}

Our evaluation framework integrates three theoretical perspectives that inform how workflow design relates to research quality.
\textbf{Distributed cognition theory} conceptualizes research workflows as systems where cognitive labor is allocated across human and artificial agents \citep{Hutchins1995, Hollan2000}.
This perspective generates specific predictions for workflow evaluation: task decomposition across specialized agents should improve \textit{Reliability} through role-constrained behavior and \textit{Correctness} through domain-focused verification, while coordination overhead creates \textit{Efficiency} trade-offs that depend on the degree of task interdependence and communication protocol design.
The same allocation of labor across modular agents governs \textit{Scalability}, since performance under a growing workload depends on how far components execute independently, and \textit{Innovation Potential}, since exploration distributed over agents with different sources and vantage points is the mechanism through which the architecture pursues the cross-domain recombination that accounts of scientific innovation emphasize \citep{Kuhn1962, Thagard1992}.
\textbf{Epistemic reliability theory} analyzes workflows as inference chains where each step transforms inputs into outputs for subsequent steps \citep{Goldman1999, Kitcher1993}.
This framework predicts that sequential pipeline stages can either amplify or attenuate errors depending on inter-stage verification mechanisms, directly informing \textit{Soundness} (whether reasoning chains preserve logical validity across stages) and \textit{Robustness} (whether error detection and recovery mechanisms prevent cascading failures).
The same lens covers the domain-specific dimensions that turn on the validity of a single inference step: \textit{Decision Quality} (whether the method chosen at a stage is a reliable process for the question asked), \textit{Economic Rigor} (whether identification and calibration steps are valid in the discipline's own terms), and \textit{Causal Validity} (whether a causal claim rests on a sound identification strategy).
\textbf{Scientific norm theory} emphasizes transparency, replicability, and scrutability as essential properties for knowledge production \citep{Merton1973, Popper1959}.
Merton's norms of communalism and organized skepticism map directly to \textit{Transparency} (making operations inspectable for scrutiny) and \textit{Reproducibility} (enabling independent replication), while the norm of universalism, that claims should be evaluated on merit regardless of source, motivates \textit{Traceability} (preserving provenance so that AI-generated outputs can be verified against their inputs).
Popper's falsificationism grounds \textit{Simulation Fidelity}: a simulated economy is held to the stylized facts of its empirical counterpart and actively searched for counterexamples.

These theoretical foundations map to twelve dimensions for general economic research, organized into five clusters: \textit{Quality} dimensions (Reliability, Correctness, Soundness) address output dependability; \textit{Operational} dimensions (Efficiency, Scalability, Robustness) address system performance; \textit{Epistemic} dimensions (Transparency, Traceability, Reproducibility) address scientific norms; the \textit{Innovation} dimension addresses generative capability; and the general \textit{Domain-Specific} dimensions (Decision Quality, Economic Rigor) address the appropriateness of methodological choices and the domain validity of reasoning. Within the Domain-Specific cluster, two further, subfield-specific dimensions (\textit{Simulation Fidelity} and \textit{Causal Validity}) illustrate how the framework extends to a particular subfield (here agent-based macroeconomics), capturing whether simulation outputs' macro-moments match stylized facts and whether causal claims rest on sound identification; additional such dimensions can be developed as a study requires.
Figure~\ref{fig:theoretical-framework} illustrates this mapping.

\begin{figure}[!htbp]
\centering
\scalebox{\aelFigScale}{\begin{tikzpicture}[ael,
  dimrow/.style={
    execute at begin node=\aelNoHyphen,
    anchor=west, inner sep=0pt, text=aelInk, font=\sffamily\scriptsize,
  },
  dimrowopt/.style={dimrow, font=\sffamily\scriptsize\itshape},
  boxtitle/.style={
    anchor=north west, inner sep=0pt, text=aelBlue1,
    font=\sffamily\footnotesize\bfseries,
  },
  fbadge/.style={
    circle, draw=aelBlue3, fill=aelBlue4, text=aelBlue1, line width=\aelThin,
    inner sep=0pt, minimum size=3mm, font=\sffamily\tiny\bfseries,
  },
  tone/.style={rectangle, draw=aelBlue2, fill=aelBlue2, inner sep=0pt,
               minimum size=2mm, line width=0.5pt},
  ttwo/.style={rectangle, draw=aelBlue2, fill=white, inner sep=0pt,
               minimum size=2mm, line width=0.5pt},
  cbox/.style={inner xsep=2.5mm, inner ysep=2.4mm, outer sep=0pt},
  fbox/.style={aelart, minimum width=64.83mm, minimum height=11.5mm},
  tbox/.style={aelagent, minimum width=98.75mm, minimum height=11.5mm},
  layertitle/.style={inner sep=0pt, text=aelBlue1, font=\sffamily\footnotesize\bfseries,
                     text height=1.6ex, text depth=0.45ex,
                     execute at begin node=\aelNoHyphen},
  layernote/.style={inner sep=0pt, text=aelInk, font=\sffamily\scriptsize,
                    execute at begin node=\aelNoHyphen},
  bar/.style={draw=aelAccent, line width=\aelBold},
]

\foreach \i/\x in {1/0,2/40.7,3/81.4,4/122.1,5/162.8} {
  \coordinate (c\i x0) at (\x mm, 0);
  \coordinate (c\i bd) at (\x mm+24.2mm, 0);
  \coordinate (c\i t1) at (\x mm+28.7mm, 0);
  \coordinate (c\i t2) at (\x mm+32.7mm, 0);
  \coordinate (c\i tm) at (\x mm+30.7mm, 0);
}
\coordinate (row1) at (0,-5.6mm);
\coordinate (row2) at (0,-9.8mm);
\coordinate (row3) at (0,-14.0mm);
\coordinate (row4) at (0,-18.2mm);
\coordinate (rowend) at (0,-20.2mm);

\node[boxtitle] (t1) at (c1x0) {Quality};
\node[dimrow] (r1) at (c1x0 |- row1) {Reliability};
\node[dimrow] (r2) at (c1x0 |- row2) {Correctness};
\node[dimrow] (r3) at (c1x0 |- row3) {Soundness};
\node[fbadge] at (c1bd |- row1) {1};   \node[tone] at (c1t1 |- row1) {};
\node[fbadge] at (c1bd |- row2) {1};   \node[tone] at (c1t1 |- row2) {}; \node[ttwo] at (c1t2 |- row2) {};
\node[fbadge] at (c1bd |- row3) {2};   \node[tone] at (c1t1 |- row3) {}; \node[ttwo] at (c1t2 |- row3) {};
\begin{scope}[on background layer]
\node[aelstageA, cbox, fit=(t1)(c1t2 |- row1)(c1x0 |- rowend)] (b1) {};
\end{scope}

\node[boxtitle] (t2) at (c2x0) {Operational};
\node[dimrow] (r4) at (c2x0 |- row1) {Efficiency};
\node[dimrow] (r5) at (c2x0 |- row2) {Scalability};
\node[dimrow] (r6) at (c2x0 |- row3) {Robustness};
\node[fbadge] at (c2bd |- row1) {1};   \node[tone] at (c2t1 |- row1) {};
\node[fbadge] at (c2bd |- row2) {1};   \node[tone] at (c2t1 |- row2) {};
\node[fbadge] at (c2bd |- row3) {2};   \node[tone] at (c2t1 |- row3) {};
\begin{scope}[on background layer]
\node[aelstageA, cbox, fit=(t2)(c2t2 |- row1)(c2x0 |- rowend)] (b2) {};
\end{scope}

\node[boxtitle] (t3) at (c3x0) {Epistemic};
\node[dimrow] (r7) at (c3x0 |- row1) {Transparency};
\node[dimrow] (r8) at (c3x0 |- row2) {Traceability};
\node[dimrow] (r9) at (c3x0 |- row3) {Reproducibility};
\node[fbadge] at (c3bd |- row1) {3};   \node[tone] at (c3t1 |- row1) {}; \node[ttwo] at (c3t2 |- row1) {};
\node[fbadge] at (c3bd |- row2) {3};   \node[tone] at (c3t1 |- row2) {};
\node[fbadge] at (c3bd |- row3) {3};   \node[tone] at (c3t1 |- row3) {};
\begin{scope}[on background layer]
\node[aelstageA, cbox, fit=(t3)(c3t2 |- row1)(c3x0 |- rowend)] (b3) {};
\end{scope}

\node[boxtitle] (t4) at (c4x0) {Innovation};
\node[dimrow] (r10) at (c4x0 |- row1) {Innovation Potential};
\node[fbadge] at (c4bd |- row1) {1};   \node[tone] at (c4t1 |- row1) {}; \node[ttwo] at (c4t2 |- row1) {};
\begin{scope}[on background layer]
\node[aelstageA, cbox, fit=(t4)(c4t2 |- row1)(c4x0 |- rowend)] (b4) {};
\end{scope}

\node[boxtitle] (t5) at (c5x0) {Domain-Specific};
\node[dimrow]    (r11) at (c5x0 |- row1) {Decision Quality};
\node[dimrow]    (r12) at (c5x0 |- row2) {Economic Rigor};
\node[dimrowopt] (r13) at (c5x0 |- row3) {Simulation Fidelity};
\node[dimrowopt] (r14) at (c5x0 |- row4) {Causal Validity};
\node[fbadge] at (c5bd |- row1) {2};   \node[tone] at (c5t1 |- row1) {}; \node[ttwo] at (c5t2 |- row1) {};
\node[fbadge] at (c5bd |- row2) {2};   \node[tone] at (c5t1 |- row2) {}; \node[ttwo] at (c5t2 |- row2) {};
\node[fbadge] at (c5bd |- row3) {3};   \node[tone] at (c5t1 |- row3) {};
\node[fbadge] at (c5bd |- row4) {2};   \node[tone] at (c5t1 |- row4) {};
\begin{scope}[on background layer]
\node[aelstageC, cbox, fit=(t5)(c5t2 |- row1)(c5x0 |- rowend)] (b5) {};
\end{scope}

\coordinate (midx) at ($(b1.west)!0.5!(b5.east)$);
\coordinate (topbar) at ($(midx |- b1.north)+(0,4.5mm)$);     %
\coordinate (collect) at ($(topbar)+(0,6mm)$);                %
\coordinate (fbase) at ($(collect)+(0,3mm)$);
\node[fbox, anchor=south west] (f1) at (b1.west |- fbase) {};
\node[fbox, anchor=south]      (f2) at (midx |- fbase) {};
\node[fbox, anchor=south east] (f3) at (b5.east |- fbase) {};
\foreach \i/\name in {1/Distributed Cognition,2/Epistemic Reliability,3/Scientific Norms} {
  \node[layertitle] (fl\i) at ($(f\i.center)+(2mm,0)$) {\name};
  \node[fbadge, anchor=east] at ($(fl\i.west)+(-1.5mm,0)$) {\i};
}
\draw[bar] (f1.south |- collect) -- (f3.south |- collect);
\foreach \i in {1,2,3} \draw[bar] (f\i.south) -- (f\i.south |- collect);
\draw[bar] (midx |- collect) -- (topbar);
\draw[bar] (c1bd |- topbar) -- (c5bd |- topbar);
\foreach \i in {1,...,5} \draw[aelflow] (c\i bd |- topbar) -- (c\i bd |- b\i.north);
\node[aellabel] at ($(midx |- collect)!0.5!(topbar)$) {grounds};

\coordinate (lowbar) at ($(midx |- b1.south)+(0,-4.5mm)$);    %
\coordinate (dist)   at ($(lowbar)+(0,-6mm)$);                %
\coordinate (mbase) at ($(dist)+(0,-3mm)$);
\node[tbox, anchor=north west] (m1) at (b1.west |- mbase) {};
\node[tbox, anchor=north east] (m2) at (b5.east |- mbase) {};
\foreach \i/\name/\note in {1/{Tier 1: structural}/{metrics from the execution logs},
                              2/{Tier 2: content}/{scores from the two LLM judges}} {
  \node[layertitle] (ml\i) at ($(m\i.center)+(1.5mm,1.9mm)$) {\name};
  \node[layernote] at ($(m\i.center)+(0,-2.3mm)$) {\note};
}
\node[tone, anchor=east] at ($(ml1.west)+(-1.5mm,0)$) {};
\node[ttwo, anchor=east] at ($(ml2.west)+(-1.5mm,0)$) {};
\draw[bar] (c1tm |- lowbar) -- (c5tm |- lowbar);
\foreach \i in {1,...,5} \draw[bar] (c\i tm |- b\i.south) -- (c\i tm |- lowbar);
\draw[bar] (lowbar) -- (dist);
\draw[bar] (m1.north |- dist) -- (m2.north |- dist);
\foreach \i in {1,2} \draw[aelflow] (m\i.north |- dist) -- (m\i.north);
\node[aellabel] at ($(lowbar)!0.5!(dist)$) {is measured by};

\end{tikzpicture}}
\caption{Evaluation framework: theoretical foundations, dimensions, and measurement. Top: the three theoretical foundations of Section~\ref{sec:theoretical-foundations}, numbered. Middle: the fourteen dimensions in their five clusters, twelve that apply to economic research generally and, in italics, the two subfield extensions this example illustrates. The circled number beside each dimension is the foundation in which Section~\ref{sec:theoretical-foundations} grounds it; a filled square marks a dimension scored by Tier~1 (structural metrics from the execution logs) and a hollow square one scored by Tier~2 (the two-judge LLM-as-reviewer ensemble, Section~\ref{sec:two-tier-methodology}); for a dimension with both squares the reported score is the Tier-2 score and the Tier-1 baseline is not blended in (Appendix~\ref{appendix:implemented-metrics}). Bottom: the two measurement tiers. The italicized extensions, Simulation Fidelity and Causal Validity, illustrate one subfield (agent-based macroeconomics); they are computed structurally when their optional stage runs, from the simulation output or the causal estimate rather than from the execution logs, and neither was evaluated in the reported runs. Further domain-specific dimensions can be developed as a study requires.}
\label{fig:theoretical-framework}
\end{figure}

\subsection{Evaluation Dimensions}
\label{sec:evaluation-dimensions}

The evaluation dimensions, with their definitions, supporting design features, and target metrics, are summarized in Table~\ref{tab:dimension-summary} (Appendix~\ref{appendix:quantitative-methods}).
The following subsections provide integrated discussion of each cluster.

\paragraph{Quality Dimensions: Reliability, Correctness, and Soundness.}

The quality dimensions address the fundamental requirement that workflow outputs meet research standards.
\textbf{Reliability} concerns consistency of outputs under varying conditions; it is supported through explicit agent specifications that constrain behavior within defined limits, structured communication protocols that standardize information exchange, and version control that tracks system state.
\textbf{Correctness} addresses accuracy relative to domain standards. It is supported through specialized agents with domain-focused capabilities (e.g., \texttt{TopicCrawler} for literature discovery, \texttt{Estimator} for econometric analysis), verification agents that check outputs against specified criteria, and human-in-the-loop checkpoints at critical decision points.
\textbf{Soundness} concerns logical and methodological validity, and is supported through staged workflows that enforce appropriate sequencing (e.g., theoretical grounding before model specification) and explicit reasoning chains maintained across agent interactions.

These dimensions share common design features: multi-layer verification, explicit error escalation, and human oversight at critical junctures.
However, they face distinct limitations: Reliability (output consistency) is bounded by LLM stochasticity; Correctness is limited by the completeness of verification criteria and the possibility of plausible but incorrect LLM outputs; and Soundness is constrained by the inherent opacity of agent reasoning processes.
The architecture provides multiple layers of quality assurance without guaranteeing error elimination.

\paragraph{Operational Dimensions: Efficiency, Scalability, and Robustness.}

The operational dimensions address system performance characteristics essential for practical deployment.
\textbf{Efficiency} concerns the relationship between resources consumed and outputs produced; it is supported through parallelization capabilities enabling concurrent agent execution, task automation for routine operations, and caching mechanisms for previously computed results.
\textbf{Scalability} concerns performance maintenance as workload increases. It is supported through modular agent design enabling independent component scaling, distributed execution support, and incremental processing strategies for large datasets.
\textbf{Robustness} concerns the ability to handle errors and unexpected conditions without catastrophic degradation. It is supported through error escalation pathways that route anomalies to appropriate resolution mechanisms, validation checkpoints that detect invalid inputs early, and fallback mechanisms that enable alternative execution paths when primary components fail.

The operational dimensions involve inherent trade-offs.
Parallelization provides efficiency gains only when tasks are genuinely independent; coordination overhead increases with system complexity.
Scalability in research workflows must be evaluated not only in terms of computational performance but also in terms of maintaining research quality.
Robustness mechanisms may not detect all errors, and graceful degradation may produce outputs of uncertain quality.

\paragraph{Epistemic Dimensions: Transparency, Traceability, and Reproducibility.}

The epistemic dimensions directly address scientific norms essential for trustworthy research.
\textbf{Transparency} refers to the inspectability of system operations; it is supported through explicit agent definitions specifying roles and constraints, logged inter-agent communications, and documented decision points with associated rationales.
Transparency enables real-time or post-hoc inspection of system behavior at multiple levels: architectural (what agents exist), execution (what happened during a run), and decision (why particular choices were made).
\textbf{Traceability} concerns the ability to track output provenance. It is supported through logging, version control for all artifacts, explicit provenance metadata, and checkpoint preservation enabling workflow replay.
Traceability enables reconstruction of workflow history when outputs are questioned.
\textbf{Reproducibility} concerns obtaining consistent results when repeating identical procedures, and is supported through deterministic workflow orchestration, configuration documentation, version-controlled dependencies, and automated generation of replication packages.

The epistemic dimensions face bounded completeness.
Transparency is limited by the interpretability of underlying LLM operations; while we can log agent outputs, internal reasoning remains largely opaque.
Traceability depends on logging comprehensiveness; external operations may not be fully captured.
Complete reproducibility faces inherent challenges: LLM outputs exhibit stochasticity, external resources may change, and proprietary model access may vary.

\paragraph{Innovation Dimension.}

\textbf{Innovation Potential} (the generation of novel, valuable ideas) represents a particularly challenging dimension for automated systems.
Theoretical accounts of scientific innovation emphasize analogical reasoning, cross-domain transfer, and creative recombination \citep{Kuhn1962, Thagard1992}.
The architecture includes features potentially relevant to innovation: multi-agent exploration of diverse information sources, explicit gap identification agents (e.g., \texttt{GapFinder}), and cross-domain integration capabilities.

We measure Innovation Potential through a two-tier approach (see Appendix~\ref{appendix:implemented-metrics}).
The Tier-1 structural component captures output diversity (normalized Shannon entropy of item distribution across pipeline stages) and coverage breadth (total item count relative to a reference threshold). These are proxies for the range of exploration, not for novelty per se.
The Tier-2 content quality component uses LLM-as-reviewer assessment against four sub-criteria: \textit{Novelty} (whether outputs identify genuinely under-explored research directions), \textit{Cross-Domain Integration} (whether outputs connect concepts across economic subfields), \textit{Research Significance} (whether identified questions address substantive gaps), and \textit{Specificity and Actionability}.
These sub-criteria operationalize specific facets of innovation potential, but the resulting scores are preliminary indicators that have not been calibrated against human expert judgments or benchmarked against human-generated research outputs.
Without such calibration, the absolute magnitude of innovation scores (e.g., 0.70 vs.\ 0.45) should not be interpreted as reflecting a known level of innovativeness; only relative comparisons across configurations within the same evaluation framework are meaningful.

We explicitly refrain from claiming that agentic workflows enhance breakthrough innovation.
The design supports systematic exploration and comprehensive coverage, potentially valuable for identifying incremental research opportunities, while recognizing that transformative innovation likely depends on human insight, serendipity, and factors beyond architectural control.

\paragraph{Domain-Specific Cluster.}
The four domain-specific dimensions (Decision Quality, Economic Rigor, Simulation Fidelity, and Causal Validity) extend the framework beyond generic AI evaluation to capture economics-discipline requirements.
Decision Quality assesses whether agents select appropriate methodological approaches given the research question (e.g., choosing instrumental variables over OLS when endogeneity is present, or selecting DSGE over reduced-form models when general equilibrium effects matter). This is evaluated through LLM-as-reviewer scoring against sub-criteria for decision coherence, information utilization, and scope management.
Economic Rigor assesses domain-specific validity: whether identification strategies are sound, whether calibration follows established discipline (e.g., matching steady-state ratios, using Bayesian priors from the literature), and whether robustness checks cover standard sensitivity analyses.
Simulation Fidelity applies only when the Model team's optional Simulation Stage is active; it is the share of the scenario's benchmark moments that the simulation reproduces within $5\%$ of the target, or within $0.05$ when the target is smaller than one in absolute value (Appendix~\ref{appendix:implemented-metrics}); impulse-response coverage remains a target metric and the Falsifier cross-check a design feature that is not yet scored (Table~\ref{tab:dimension-summary}).
Causal Validity applies when \texttt{CausalFMClient} emits a \texttt{causal\_estimate} artifact; it compares the estimate with a known effect or, failing that, a prior mean (Appendix~\ref{appendix:implemented-metrics}); the soundness of the identification strategy (back-door, front-door, or instrumental-variable strategies), the divergence between the zero-shot and classical estimates, and the sensitivity to unobserved confounding remain target metrics (Table~\ref{tab:dimension-summary}).
Decision Quality and Economic Rigor are scored exclusively through Tier-2 LLM-as-reviewer assessment; Simulation Fidelity and Causal Validity are computed structurally from the simulation output and the causal estimate (Appendix~\ref{appendix:implemented-metrics}); no LLM judge scores them.
Absolute values on Simulation Fidelity and Causal Validity have not yet been calibrated against human-expert judgment and should be interpreted relatively across configurations rather than as standalone grades.

\paragraph{Agent-Level Metrics.}
In addition to the fourteen workflow-level dimensions, the code defines four Tier-1 diagnostics, computed from per-call observability records without further LLM calls:
\textit{Tool correctness} combines whether a team called the tools expected for its stages, avoided the tools declared inappropriate for it, and had its calls succeed;
\textit{Step efficiency} combines the number of LLM calls against an expected range per stage, the share of output tokens, and the balance of stage durations;
\textit{Plan adherence} combines stage completion, stage ordering, and the presence of each stage's expected output;
and \textit{Argument correctness} is the share of individual argument checks passed (required arguments present and non-empty, typed arguments of the declared type).
They locate execution problems within a stage; they do not assess the quality of reasoning (Appendix~\ref{appendix:agent-metrics}). The execution logs of the runs reported here store aggregated rather than per-call observability, so these diagnostics were not computed for them.

\subsection{Empirical Validation}
\label{sec:empirical-validation}

To evaluate the framework empirically, we apply the two-tier scheme of Section~\ref{sec:two-tier-methodology} to the seven-stage chained run of Section~\ref{sec:Examples}. This is the deployment behavior the framework is meant to exhibit, run in its single active configuration: \texttt{ModeWithWcWithHITL} for the reasoning teams paired with the \texttt{OpenSourceAPI} data workflow. The configuration was executed 26 times with the same version of the code and the same seed topic (one run designated in advance as the demonstration and 25 replicates, submitted together); all 26 completed the seven stages. Section~\ref{sec:Examples} presents one of them, chosen by the published rule stated there, and Section~\ref{sec:robustness-checks} reports the distribution over all 26.
Both Tier-2 judges scored all seven stages of all 26 runs (2{,}184 judge calls, none failed).
The entire stack is fully open-weight: the subject workflows are served locally by Qwen3.6-27B-FP8 and Tier-2 content review is a two-judge ensemble of Mistral-Small-3.2-24B and Gemma-3-27B, all on vLLM. An earlier project phase evaluated fourteen team-level configurations (four combinations of web-crawl integration and human-in-the-loop review for each of the Ideation, Literature, and Model teams, and two data-source workflows) and selected this one.

\subsubsection{Two-Tier Evaluation}
\label{sec:two-tier-methodology}

Evaluating workflow quality requires assessment at two distinct levels: (i)~\textit{structural compliance}: whether the system executes correctly, produces expected outputs, and maintains execution traces; and (ii)~\textit{content quality}: whether the outputs are factually accurate, logically sound, and scientifically valuable.
Structural properties can be measured automatically from execution logs, but content quality inherently requires domain judgment.
Since human expert evaluation at scale is infeasible prior to peer review, we employ an \textit{LLM-as-reviewer} approach \citep{Zheng2023} as a systematic proxy for expert assessment.

\paragraph{Tier~1: Structural Assessment.}
We analyze the execution log of the seven-stage chained run.
The workflows are instrumented using the shared \texttt{WorkflowLogger} module, which captures per-stage timing information, item counts, error occurrences, and output file tracking.
A companion \texttt{MetricsCollector} observability layer records every LLM API call (model, token counts, latency, cost) and external tool invocation (HTTP requests, database queries, search-API calls) with per-agent and per-stage attribution.
The orchestrator hands the same collector to every agent it creates, and the collector writes an \texttt{observability} section into each execution log that records resource use by agent and stage without changes to the agents' own code.
Each of the seven stages generates a structured \texttt{execution\_log.json} file containing execution metadata, and the pipeline-level log tracks the chain across stages.
Tier-1 metrics assess six dimensions automatically: Reliability (stage completion rate, output file coverage, and error penalty), Efficiency (normalized execution time), Scalability (throughput relative to a reference rate), Robustness (error recovery rate), Traceability (provenance metadata completeness), and Reproducibility (configuration documentation coverage).
Importantly, these structural metrics measure \textit{design-level compliance} (whether the pipeline executes as specified) rather than empirical properties that require repeated trials (e.g., output consistency across runs for Reliability) or external validation (e.g., expert-rated correctness).
Across the 26 runs, stage completion, stage ordering, output coverage, provenance completeness, and the presence of configuration metadata are all $1.0$, with no stage error and no failed artifact-signature check (996 verified reads); Table~\ref{tab:multirun-ci} reports these values with execution time and efficiency.

\paragraph{Tier~2: Content Quality via LLM-as-Reviewer Ensemble.}
For six dimensions that require domain judgment (Correctness, Soundness, Innovation Potential, Transparency, Decision Quality, and Economic Rigor), we employ a multi-model ensemble evaluation.
Each workflow's actual output files (research questions, literature reviews, theoretical frameworks, model specifications, data requirements) are evaluated against rubrics with 3--4 sub-criteria per dimension, scored on a 1--5 Likert scale; the Correctness, Soundness, and Economic Rigor rubrics are team-specific.
To mitigate evaluator stochasticity identified in preliminary single-pass analyses, we use a two-judge ensemble of open-weight reviewers drawn from different model families, Mistral-Small-3.2-24B and Gemma-3-27B, each at temperature~$=0.0$ with structured output in JavaScript Object Notation (JSON).
Both reviewers are distinct from the Qwen3.6-27B-FP8 model that produced the workflow outputs, and neither sits on the checkpoint committee (Section~\ref{sec:Examples}), so no model grades its own work or its own checkpoint decisions.
Each judge's scores are reported separately; where a single figure is given (Table~\ref{tab:human-validation}), it is the arithmetic mean of the two judges' normalized dimension scores.
This design keeps the entire evaluation on local open-weight inference and surfaces between-judge scoring bias: Gemma scores systematically higher than Mistral on the chained run's reasoning stages. This is why we report both judges rather than a single conflated figure.
Sub-criterion scores are averaged and normalized to a 0--1 scale.
The reported score for each of these six dimensions is its Tier-2 score. Tier~1 also computes a structural proxy for each of them; in each of the 26 runs five of the six proxies equal $1.0$ and the sixth (Innovation Potential) equals $0.584$, so they are not blended in by default.
Exact formulae for all metric computations are provided in Appendix~\ref{appendix:implemented-metrics}.

This two-tier approach addresses a fundamental limitation: structural metrics (stage completion rates, file existence) confirm that a workflow ran successfully but cannot assess whether its outputs are correct, logically sound, or novel.
Using open-weight reviewers from families distinct from the subject model (Mistral and Gemma vs.\ Qwen) as the Tier-2 evaluators provides a degree of evaluator independence.
Reporting both judges makes between-judge disagreement visible, which a single reviewer would conceal; the inter-evaluator analysis of Section~\ref{sec:robustness-checks} quantifies this agreement per stage.
However, LLM-as-reviewer evaluation serves as a systematic proxy for, not a replacement of, human expert review, and does not by itself constitute the independent external validation that would require domain experts, standardized benchmarks, or blind evaluation protocols; three co-author raters completed the preliminary expert check reported in Section~\ref{sec:robustness-checks}, and independent external validation remains outstanding (see Section~\ref{sec:limitations}).

\subsubsection{Evaluating the End-to-End Run}
\label{sec:cross-workflow-comparison}

We apply the two-tier framework to the seven-stage chained run, the deployment behavior the framework is meant to exhibit, rather than to teams in isolation. Tier-2 content scores are reported below for the demonstration run of Section~\ref{sec:Examples} and for the 25 replicates, which include it; Section~\ref{sec:robustness-checks} reports their spread across runs and the agreement between the two judges.

\paragraph{Scoring the seven stages.} The dual-judge Tier-2 protocol scores all seven stages of every run on their actual artifacts, including the Code, Estimation, and Reporting teams; the demonstration run's per-stage means are in Table~\ref{tab:ex-summary} and its per-dimension breakdown in Appendix~\ref{appendix:example-model}. The Correctness, Soundness, and Economic Rigor rubrics are team-specific (code fidelity and explicit scoping for Code; specification fit and identification discipline for Estimation; number fidelity and caveat proportionality for Reporting); the Innovation Potential, Transparency, and Decision Quality rubrics are shared across all stages. On the demonstration run, Gemma scores above Mistral on Ideation, Literature, Model, and Reporting and below it on Data, Code, and Estimation. Code is the lowest-scored stage under both judges ($0.528$ under Mistral, $0.441$ under Gemma): its Correctness scores $0.67$ under both, while its Decision Quality scores $0.42$ and $0.25$. The two judges differ most on the Estimation stage's soundness ($0.75$ against $0.50$).

\subsubsection{Key Findings}

The findings below use the 25 replicates, that is, all runs except the one designated in advance; the run shown in Section~\ref{sec:Examples} is one of them; Table~\ref{tab:inter-evaluator} gives each stage's mean under each judge with its 95\% confidence interval across runs.
\paragraph{Stage Profiles.}
Ideation posts the highest Tier-2 mean of the seven stages under both judges ($0.754$ under Mistral, $0.810$ under Gemma), followed by Literature ($0.707$/$0.776$) and Model ($0.698$/$0.778$), whose order differs between the judges. The four execution stages score lower: Estimation $0.627$/$0.613$, Reporting $0.606$/$0.631$, Data $0.558$/$0.565$, and Code $0.546$/$0.458$, the lowest under both judges. The 95\% confidence intervals across runs are at most $0.054$ wide, and Code's interval under Gemma ($[0.442, 0.475]$) lies below every other stage's interval under either judge.

\paragraph{Two-Tier Insights.}
Tier-1 structural compliance is uniform: stage completion, ordering, and provenance are $1.0$ with zero errors in all 26 runs (Table~\ref{tab:multirun-ci}). Tier-2 content quality is not uniform: the replicate stage means span $0.458$ (Code, Gemma) to $0.810$ (Ideation, Gemma), a $0.351$-point spread on the same $0$--$1$ scale on which Tier-1 records none.

\paragraph{Content Quality Patterns.}
Pooled across the seven stages, both judges, and the 25 replicates, Transparency is the highest-scoring dimension (mean $0.678$) and Economic Rigor the lowest ($0.612$); Soundness, Correctness, Decision Quality, and Innovation Potential fall between the two ($0.669$, $0.667$, $0.649$, $0.637$). The judges diverge most on Estimation's Soundness ($0.70$ under Mistral, $0.50$ under Gemma) and Estimation's Decision Quality ($0.60$ against $0.80$); averaged over the six dimensions, the absolute between-judge gap is smallest on Data ($0.035$) and largest on Estimation ($0.092$).

The evaluation's own resource usage (LLM calls, tokens, and latency per Tier-2 dimension) is reported in Table~\ref{tab:eval-cost} below.

\subsubsection{Robustness Checks}
\label{sec:robustness-checks}

To probe the stability of the primary results, we conducted three supplementary analyses: a multi-run study, a multi-evaluator agreement assessment, and a human-expert validation of the LLM-as-reviewer scores.
All three are preliminary and limited in scope; they are reported here to characterize the evaluation infrastructure and identify directions for future work, not to establish definitive conclusions.

\paragraph{Multi-Run Variability of the Chained Run ($n=26$).}
To measure run-to-run variability, the seven-stage chained pipeline was executed 26 times with the same version of the code, each run from its own copy of the code tree, with identical settings and the same seed topic. All 26 completed every stage. Table~\ref{tab:multirun-ci} reports the Tier-1 structural metrics. Completion, ordering, output coverage, and provenance metrics take the value $1.0$ in every run. Wall-clock time ranges from 4.0 to 7.4 hours (mean $5.38$, standard deviation $1.10$). The nine runs that the scheduler started last, overnight, took 5.9 to 7.4 hours; the Literature stage, which issues most external search requests, accounts for most of the spread (34 to 185 minutes, against 62 to 77 minutes for the Model stage). In those nine runs OpenAlex answered every literature query with HTTP status 429 (too many requests), and the Literature team drew its candidates from Crossref and arXiv alone.

\begin{table}[!htbp]
\centering
\begin{threeparttable}
\caption{Tier-1 Structural Metrics of the Chained Run Across 26 Executions}
\label{tab:multirun-ci}
\small
\begin{tabular*}{\textwidth}{@{\extracolsep{\fill}} l c c @{}}
\toprule
Metric & Mean $\pm$ s.d. & Range \\
\midrule
Execution time (h) & $5.38 \pm 1.10$ & $4.01$--$7.41$ \\
Stage completion rate & $1.00 \pm 0.00$ & $1.00$ \\
Stage ordering & $1.00 \pm 0.00$ & $1.00$ \\
Errors & $0.00 \pm 0.00$ & $0.00$ \\
Reliability & $1.00 \pm 0.00$ & $1.00$ \\
Output coverage (transparency) & $1.00 \pm 0.00$ & $1.00$ \\
Provenance completeness & $1.00 \pm 0.00$ & $1.00$ \\
Configuration metadata present & $1.00 \pm 0.00$ & $1.00$ \\
Efficiency & $0.57 \pm 0.05$ & $0.49$--$0.64$ \\
\bottomrule
\end{tabular*}
\begin{tablenotes}[flushleft]\footnotesize
\item \textit{Notes:} Pipeline-level Tier-1 metrics recomputed from the execution log of each of the 26 runs (one designated demonstration and 25 replicates, \texttt{ModeWithWcWithHITL}/\texttt{OpenSourceAPI}). Efficiency is $T_{\text{ref}}/(T_{\text{ref}}+T)$ with $T_{\text{ref}}$ one hour per team (Appendix~\ref{appendix:implemented-metrics}).
\end{tablenotes}
\end{threeparttable}
\end{table}

\paragraph{Pipeline Outcomes Across Runs.}
Table~\ref{tab:run-outcomes} reports what the deterministic components decided in each of the 26 runs, stage by stage. Calibration returned a harness verdict for each of the three candidate models per run: 11 of the 78 models were point-calibrated, 20 partially calibrated, 4 calibrated against a registered canonical model, and 43 were declared uncalibratable: 22 because the model's moments have no closed form and no runnable simulator of its own equations exists, 17 because the canonical model's estimate fell on a bound of the parameter's declared range without reproducing the target, and 4 because the model contains learned components that would have to be trained. The estimation harness estimated 15 specifications, downgraded 8 to fragile on its diagnostics, and refused 3 for too few overlapping observations. Of the 35 theory-derived hypotheses tested, 5 were supported, 6 contradicted, and 24 inconclusive. The Model--Data feasibility loop did not converge in any run: after two or three cycles a median of one requirement per run was graded feasible, 10 available only through a proxy, and 23 unavailable, most of them calibration targets that no open data source publishes as a series. Across runs the Data team assembled 316 series; 39 of them, in 16 runs, could not be retrieved and were marked simulated, and the estimation harness excluded every simulated series. In none of the 26 runs did the estimated specification link a series to a variable of the executable model. The number-consistency check, re-run offline on the stored reports after three extraction errors in the checker were corrected (Unicode minus signs, rounding at the printed precision, declared-range strings), matched every non-exempt body number to the upstream artifacts in 14 of the 26 reports; the other 12 reports contain 24 numbers without a source, 16 of them the ``500'' in ``S\&P 500'', for which the stored artifacts carry no source title, and the rest publication years, unit conversions written by the language model, and one restatement of the 32-observation minimum.

\begin{table}[!htbp]
\centering
\begin{threeparttable}
\caption{Deterministic Outcomes Across the 26 Chained Runs}
\label{tab:run-outcomes}
\small
\begin{tabularx}{\textwidth}{@{}l Y@{}}
\toprule
Stage & Outcome (count over 26 runs) \\
\midrule
Calibration (78 models) & 11 point-calibrated; 20 partially calibrated; 4 archetype-calibrated; 43 uncalibratable \\
Feasibility loop & 0 converged; median per run: 1 feasible, 10 proxy-only, 23 unavailable requirements \\
Data retrieval & 316 series; 39 simulated (16 runs), excluded from estimation; 19 flagged as proxies \\
Code & 77 modules partial, 1 fragment solves; none complete \\
Estimation & 15 estimated; 8 fragile; 3 inestimable \\
Hypothesis tests (35) & 5 supported; 6 contradicted; 24 inconclusive \\
Report consistency & 14 consistent; 12 with unverified numbers (24 numbers in total) \\
\bottomrule
\end{tabularx}
\begin{tablenotes}[flushleft]\footnotesize
\item \textit{Notes:} Counts from the run artifacts (extraction script in the replication package). Consistency verdicts are the offline recheck with the corrected checker (\texttt{pipeline/recheck\_consistency.py}); run with the uncorrected checker, the same script reproduces the verdicts stored by the runs exactly (10 consistent, 16 with unverified numbers).
\end{tablenotes}
\end{threeparttable}
\end{table}

\paragraph{Checkpoint Committee Across Runs.}
The LLM-Economist committee cast 4{,}023 member votes on 1{,}341 voted checkpoint ballots across the 26 runs, with no abstention and no failed call. 538 ballots (40.1\%) were split rather than unanimous. Minority votes came from the data lens (Phi-4-mini) in 268 ballots, the theory lens (gpt-oss-20b) in 250, and the policy lens (Olmo-3.1-32B) in 20. At the estimation checkpoint the committee rejected the proposed specification in both review rounds in every run, so each run passed the original proposal to the harness with the committee's objections recorded; the harness estimated it in 23 runs and refused it in 3.

\paragraph{Inter-Evaluator Agreement.}
To characterize sensitivity to evaluator choice, and to justify reporting both judges rather than a single conflated score, we compare the two open-weight reviewers (Mistral-Small-3.2-24B and Gemma-3-27B) across the stages of the chained run under identical rubrics, pooling the six content dimensions across the 25 replicates (150 dimension-run pairs per stage, six from each of the 25 runs).
Table~\ref{tab:inter-evaluator} reports each judge's per-stage mean with its 95\% confidence interval across runs, and the between-judge Pearson correlation. Agreement ranges from $r=0.05$ on Literature to $r=0.76$ on Code across the seven stages; on Literature and Estimation ($r=0.19$), the two judges' scores for the same run and dimension are close to uncorrelated even though their stage means differ by less than $0.07$.

\begin{table}[!htbp]
\centering
\begin{threeparttable}
\caption{Inter-Evaluator Agreement Between the Two Judges (25 Replicates)}
\label{tab:inter-evaluator}
\small
\begin{tabular*}{\textwidth}{@{\extracolsep{\fill}} l S[table-format=1.3] c S[table-format=1.3] c S[table-format=1.2] @{}}
\toprule
 & \multicolumn{2}{c}{Mistral-Small-3.2-24B} & \multicolumn{2}{c}{Gemma-3-27B} & {Agreement} \\
\cmidrule(lr){2-3}\cmidrule(lr){4-5}
Stage & {Mean} & {95\% CI} & {Mean} & {95\% CI} & {$r$} \\
\midrule
Ideation   & 0.754 & [0.743, 0.766] & 0.810 & [0.790, 0.829] & 0.51 \\
Literature & 0.707 & [0.702, 0.712] & 0.776 & [0.768, 0.783] & 0.05 \\
Model      & 0.698 & [0.692, 0.704] & 0.778 & [0.767, 0.789] & 0.61 \\
Data       & 0.558 & [0.545, 0.571] & 0.565 & [0.546, 0.585] & 0.58 \\
Code       & 0.546 & [0.541, 0.552] & 0.458 & [0.442, 0.475] & 0.76 \\
Estimation & 0.627 & [0.600, 0.654] & 0.613 & [0.595, 0.631] & 0.19 \\
Reporting  & 0.606 & [0.591, 0.622] & 0.631 & [0.609, 0.652] & 0.51 \\
\bottomrule
\end{tabular*}
\begin{tablenotes}[flushleft]\footnotesize
\item \textit{Notes:} Per-stage mean of the six content-quality dimensions from each open-weight judge, with the 95\% confidence interval of the mean across the 25 replicate runs of the seven-stage chained pipeline (the run designated in advance excluded; the run shown in Section~\ref{sec:Examples} included), and the between-judge Pearson correlation over the 150 dimension-run pairs per stage (six dimensions in each of the 25 runs).
\end{tablenotes}
\end{threeparttable}
\end{table}

\paragraph{Human-Expert Validation of LLM-as-Reviewer Scores.}
Because the Tier-2 scores rely on LLM reviewers standing in for human experts, their credibility depends on whether they track expert judgment rather than merely tracking one another.
To anchor them, we conduct a small expert-validation study: a stratified sample of workflow outputs spanning the seven stages (research questions, literature reviews, model specifications, datasets, code modules, estimation results, and the final report) is rated independently by economists with one 1--5 score per dimension, given holistically against the sub-criteria of the LLM judges' rubrics, blind to the model-generated scores.
We then compare the expert ratings against the two-judge ensemble on the same items, reporting the mean expert--LLM gap and their rank correlation, per stage and pooled across items (Table~\ref{tab:human-validation}).
The 28 items were drawn from the 26 runs reported above (they come from 16 of them), four per stage, chosen to span each stage's range of ensemble scores. Three raters completed the full sample; all three are co-authors of this paper rather than independent external economists, so the result is a preliminary domain-expert check rather than blind external validation (Section~\ref{sec:limitations}), and recruitment of additional, independent external raters is ongoing. Pooled across all 168 (item, dimension) pairs, the expert mean was $0.440$ against an LLM-ensemble mean of $0.645$ (gap $+0.205$), Spearman $\rho=0.46$. The ensemble scored above the experts in every stage, by $0.135$ (Data) to $0.301$ (Model), and on every dimension, by $0.04$ on Transparency and by $0.13$ to $0.28$ on the other five. Averaged over the six dimensions, the expert and ensemble scores of the 28 items correlate at $\rho=0.70$.
Mean pairwise inter-rater reliability across the three raters (Spearman) was $\rho=0.38$. Two raters correlate with each other at $\rho=0.82$ and average $2.3$ and $2.4$ on the 1--5 scale; the third averages $3.6$, correlates with them at $0.13$ and $0.19$, and gave all four items of a stage identical scores in three of the seven stages. Dropping any one rater leaves the pooled correlation between $0.43$ and $0.46$; dropping the third raises the gap to $0.31$. The two LLM judges (Mistral vs.\ Gemma) correlate at $\rho=0.72$ on the same 168 pairs. Their agreement does not by itself establish accuracy: the two open-weight models may share training-derived bias, a possibility the self-referential-evaluation limitation discussed in Section~\ref{sec:limitations} already raises. Per-stage correlations range from $0.04$ (Reporting) to $0.64$ (Data) (Table~\ref{tab:human-validation}); with only 4 sampled items per stage, and the six dimensions of a given item highly correlated with each other, each stage's correlation rests on effectively 4 independent observations, not the 24 (item, dimension) pairs it is computed over. We treat the pooled correlation as the more stable summary, though it may partly reflect between-stage agreement on overall score level (e.g.\ all three raters and both judges rank Ideation above Code) rather than purely within-stage rank agreement.
The pooled correlation is positive, but $28$ items are too few to fit the calibration step below (the \texttt{JudgeCalibratorGLM}) on human-rated data.
By design this is a minimal anchor: a small-sample check, not the blind independent-panel evaluation that full external validation would require (Section~\ref{sec:limitations}).

\begin{table}[!htbp]
\centering
\begin{threeparttable}
\caption{Human-Expert Validation: Expert Ratings vs.\ LLM-as-Reviewer Ensemble Scores}
\label{tab:human-validation}
\small
\begin{tabular*}{\textwidth}{@{\extracolsep{\fill}} l S[table-format=1.3] S[table-format=1.3] S[table-format=1.3] S[table-format=-1.2] @{}}
\toprule
Stage & {Expert mean} & {LLM mean} & {Gap (LLM $-$ expert)} & {Correlation} \\
\midrule
Ideation & 0.576 & 0.775 & 0.198 & 0.27 \\
Literature & 0.503 & 0.741 & 0.238 & 0.33 \\
Model & 0.434 & 0.735 & 0.301 & 0.11 \\
Data & 0.424 & 0.559 & 0.135 & 0.64 \\
Code & 0.337 & 0.500 & 0.164 & 0.10 \\
Estimation & 0.427 & 0.605 & 0.178 & 0.22 \\
Reporting & 0.378 & 0.601 & 0.222 & 0.04 \\
\midrule
Pooled & 0.440 & 0.645 & 0.205 & 0.46 \\
\bottomrule
\end{tabular*}
\begin{tablenotes}[flushleft]\footnotesize
\item \textit{Notes:} Expert ratings are the mean over three raters (all co-authors; a preliminary domain-expert check, not blind external validation) scoring a 28-item spread-stratified sample (4 items per stage, drawn from the 26 runs of Table~\ref{tab:multirun-ci}) with one 1--5 score per dimension against the LLM judges' rubrics (normalized to 0--1), blind to the model-generated scores; mean pairwise inter-rater Spearman $\rho=0.38$. LLM mean is the two-judge ensemble (Mistral-Small-3.2-24B and Gemma-3-27B) on the same items. Gap is the mean signed difference (LLM $-$ expert), computed from unrounded means; Correlation is the Spearman rank correlation between expert and ensemble scores, computed on 24 (item, dimension) pairs per stage (168 pooled) but resting on only 4 independent items per stage; per-stage correlations should be read with that caveat (see prose above).
\end{tablenotes}
\end{threeparttable}
\end{table}

\paragraph{Evaluation Compute Transparency.}
Because the evaluation framework itself consumes LLM resources, we report the compute of the Tier-2 scoring process to support replication budgeting and meta-evaluation transparency.
Table~\ref{tab:eval-cost} breaks down the evaluation compute by dimension for the demonstration run, scored with the two-judge open-weight ensemble (Mistral-Small-3.2-24B and Gemma-3-27B) across the seven stages.
Scoring one chained run costs 84 judge calls (7 stages $\times$ 6 dimensions $\times$ 2 judges) and about 713K tokens (676K to 723K across the 26 runs); since both judges are local open-weight inference, this carries no API cost, only compute time: the judge calls of one run take 6 to 9 minutes for Mistral and 22 to 29 minutes for Gemma, excluding model start-up, and Gemma's mean per-call latency is 3.5 times Mistral's on average over the 26 runs (2.3 to 4.6).

\begin{table}[!htbp]
\centering
\begin{threeparttable}
\caption{Tier-2 Evaluation Resource Breakdown by Dimension (Demonstration Run)}
\label{tab:eval-cost}
\small
\begin{tabular*}{\textwidth}{@{\extracolsep{\fill}} l S[table-format=2.0] S[table-format=3.1] S[table-format=2.1] @{}}
\toprule
Dimension & {Calls} & {Tokens (K)} & {Avg.\ latency (s)} \\
\midrule
Correctness          & 14 & 119.1 & 22.1 \\
Soundness            & 14 & 118.2 & 21.2 \\
Innovation Potential & 14 & 123.5 & 23.3 \\
Transparency         & 14 & 117.2 & 19.2 \\
Decision Quality     & 14 & 117.7 & 19.2 \\
Economic Rigor       & 14 & 117.2 & 19.1 \\
\midrule
Total & 84 & 712.9 & 20.7 \\
\bottomrule
\end{tabular*}
\begin{tablenotes}[flushleft]\footnotesize
\item \textit{Notes:} The demonstration run of Section~\ref{sec:Examples}, all seven stages, both open-weight judges (Mistral-Small-3.2-24B, Gemma-3-27B), all six content dimensions (the same six pooled in Table~\ref{tab:inter-evaluator}). Calls and tokens are summed across both judges; latency is the mean per-call latency across both judges' calls for that dimension. The stack runs on open-weight local inference, so figures report compute (calls, tokens, latency) only, not API cost.
\end{tablenotes}
\end{threeparttable}
\end{table}

\subsection{Extended Evaluation Capabilities}

Beyond the fourteen-dimension framework and the two-judge evaluator ensemble, AEL provides additional evaluation infrastructure.

\paragraph{Enhanced Observability.}
The runtime observability layer has been extended beyond the \texttt{WorkflowLogger} and \texttt{MetricsCollector} described above.
All LLM calls, tool invocations, and API requests are recorded with per-agent attribution and can be exported in the OpenTelemetry tracing format, with 12 model providers recognized.
The \texttt{evaluation\_observability} key in result JSONs captures evaluation-specific metrics including evaluator model, temperature, token counts, and per-dimension latency, enabling meta-evaluation of the evaluation process itself.

\paragraph{Princeton Reliability Dimensions.}
Complementing the fourteen primary dimensions, AEL implements a four-dimension reliability framework \citep{rabanser2026reliability}: \textit{consistency} (variance of quality scores across $K$ runs of the same input), \textit{robustness} (quality maintenance under prompt paraphrasing), \textit{predictability} (correlation between agent confidence and actual quality), and \textit{safety} (absence of hallucinations, harmful content, and factual errors).
These dimensions provide complementary reliability diagnostics for deployment decisions.

\paragraph{Judge Calibration.}
A \texttt{JudgeCalibratorGLM} module can calibrate LLM-as-reviewer scores against a reference dataset of human-rated outputs using generalized linear models, producing calibrated probability estimates in place of raw Likert scores, as a response to the systematic biases documented in LLM evaluation~\citep{Zheng2023}.
It requires a human-rated reference set larger than the one available here, so the scores in this paper are uncalibrated and its effect on bias was not evaluated.

\subsection{Limitations and Future Directions}
\label{sec:limitations}

This evaluation framework has important limitations.
First, the Tier-2 LLM-as-reviewer approach is a proxy for human expert evaluation rather than a substitute.
LLM evaluators exhibit known biases including verbosity preference, position effects, and limited ability to verify factual claims against ground truth \citep{Zheng2023}.
The evaluation uses a two-judge ensemble from different model families (Mistral-Small-3.2-24B and Gemma-3-27B), deterministic settings (temperature~$=0.0$), structured rubrics with explicit scoring criteria, and multi-criterion averaging.
The ensemble partially addresses between-model bias but does not eliminate systematic biases.
Future work should validate LLM-based scores against human expert panels to establish calibration.
Second, the analysis assumes design features operate as intended; implementation gaps and emergent behaviors may cause actual performance to diverge from specifications.
Third, the value of different dimensions varies across research contexts: reproducibility may be paramount for empirical research but less critical for theoretical exploration.
Fourth, LLM capabilities evolve rapidly; design features appropriate for current capabilities may become unnecessary or insufficient as underlying technologies develop.
Fifth, the Tier-1 structural metrics for Reliability, Robustness, Traceability, and Reproducibility confirm design-level compliance (that stages complete, outputs are produced, errors are recovered, provenance is recorded, and configurations are documented) but do not empirically demonstrate the full scope of these dimensions.
In particular, Reliability (as defined in Table~\ref{tab:dimension-summary}) encompasses output consistency under varying conditions, which would require repeated executions with identical inputs and measurement of output similarity; our structural metric captures only the necessary condition of successful execution, not the sufficient condition of consistent outputs.
The multi-run study ($n=26$, Section~\ref{sec:robustness-checks}) measures structural consistency at the pipeline level and reports the distribution of the deterministic outcomes; it does not measure whether repeated runs reach the same research question, model, or estimate. With the same seed topic, the final research questions differed across runs (24 distinct leading questions in 26 runs), because the agents sample at temperatures between 0.2 and 0.7 without a fixed sampling seed, the literature and trend searches are run live, and committee votes vary; the downstream models, data, and estimates differ accordingly (Table~\ref{tab:run-outcomes}).
Content quality scores use a two-judge ensemble to reduce scoring variance; Table~\ref{tab:inter-evaluator} reports per-stage means with confidence intervals across the 25 replicates, which describe the precision of each mean across runs, not whether the runs reach the same content.
Similarly, Reproducibility is assessed through the presence of configuration metadata rather than actual replication studies.
Sixth, 26 runs of one configuration on one seed topic remain a small sample for confirmatory analysis.
The Tier-2 content quality scores use a two-judge ensemble, which reduces but does not eliminate scoring stochasticity.
Small between-judge differences may fall within the natural variation of LLM evaluation rather than reflecting a genuine dimension effect.
Consequently, the empirical analysis should be interpreted as an exploratory characterization rather than a confirmatory study.
Seventh, the evaluation is \textit{self-referential}: the same research team designed both the workflows and the evaluation framework, and the evaluation was conducted on the team's own system without large-scale independent external validation.
While the Tier-2 LLM-as-reviewer introduces partial evaluator independence (the open-weight judge models were involved neither in generating workflow outputs nor in the checkpoint committee, and belong to different model families than both), this does not substitute for the rigorous external validation protocols recommended in the evaluation methodology literature, including blind evaluation, independent expert panels, and comparison against standardized benchmarks with known ground-truth answers.
The inter-evaluator analysis (Section~\ref{sec:robustness-checks}, Table~\ref{tab:inter-evaluator}) characterizes agreement across stages; absolute score magnitudes are evaluator-dependent and should be interpreted cautiously.
The small-sample human-expert check described there has been run (Table~\ref{tab:human-validation}); its raters are co-authors, so it is a preliminary domain-expert check, not the blind, full-scale external validation described above, which remains outstanding.
The absence of large-scale human expert validation means that systematic biases in framework design (e.g., dimensions defined to favor the system's strengths) or in rubric construction (e.g., criteria that the system is designed to satisfy) cannot be ruled out.

Future work should pursue several directions:
(i) \textit{external validation}: independent expert panels conducting blind evaluation of workflow outputs to validate both the LLM-based scores and the evaluation framework itself (the preliminary multi-evaluator study and the small-sample human-expert validation in Section~\ref{sec:robustness-checks} are first steps toward evaluator triangulation, but scaling the expert panel to a blind, full-sample study remains essential);
(ii) \textit{comparative benchmarking}: evaluating the workflows against prompt-based human--LLM interaction and traditional manual research on identical tasks, with multiple executions per condition, to characterize where multi-agent orchestration adds the most value;
(iii) \textit{larger-scale multi-run studies}: the 26-run study in Section~\ref{sec:robustness-checks} covers one configuration and one seed topic; runs across several seed topics and configurations, with significance tests between them, remain outstanding;
(iv) \textit{standardized benchmarks}: developing task suites with known ground-truth answers (e.g., literature review tasks with pre-established corpora) to measure absolute rather than relative performance;
(v) longitudinal studies tracking workflow performance in production research settings;
(vi) development of automated evaluation tools for real-time quality monitoring;
(vii) investigation of how evaluation dimension trade-offs vary across economic research subfields;
and (viii) calibration of the two subfield-specific dimensions (Simulation Fidelity and Causal Validity) against human-expert judgment, since absolute magnitudes on these dimensions have not been validated against domain-expert ratings and should be interpreted only as within-framework relative indicators.

\section{Conclusion}
\label{sec:conclusion}

This paper has presented a methodology for agentic workflows in economic research and an implementation of it, the Agentic Econ Lab (AEL). Seven teams, from ideation to reporting, run in the order of the methodological dependencies of economic research under a cross-team orchestrator. A bounded feasibility cycle connects the Model and Data teams, and the design places economist review at checkpoints in every stage. Every LLM used in the reported evaluation, including the workflow, committee, and judge models, was served locally on open weights.

All 26 executions of the chained pipeline completed the seven stages. Completion, ordering, and provenance metrics were $1.0$ in every run, and mean wall-clock time was $5.38$ hours (standard deviation $1.10$). An LLM-Economist committee provided checkpoint review in place of economists; the Model team's theory and calibration checkpoints were logged without a vote. The harnesses declared 43 of the 78 candidate models uncalibratable and refused 3 of the 26 estimation specifications. Of the 35 theory-derived hypotheses tested, 24 were inconclusive. An offline recheck with the corrected number-consistency checker found at least one unmatched, non-exempt number in 12 of the 26 reports.

The two-tier, fourteen-dimension evaluation combined execution-log metrics with Tier-2 content scores from two LLM judges. In a preliminary human check, three raters scored a 28-item sample drawn from the 26 runs at a pooled mean of $0.440$, compared with $0.645$ for the LLM ensemble. The ensemble scored higher in all seven stages; the pooled Spearman correlation was $\rho=0.46$ over 168 item--dimension pairs. Three limits apply to these results (Section~\ref{sec:limitations}): the runs share one configuration and one seed topic, the judges are themselves LLMs, and all three human raters are co-authors rather than independent external economists.

The reported workflow automates the production of specifications, datasets, estimates, and reports, while recording refusals and unresolved limitations for economist review. These runs do not measure how human economists would revise the outputs or how much of their work the workflow would replace.

Future research should pursue six directions:
(i) blind validation of the Tier-2 scores by an independent panel of external economists, which would also allow the judge-calibration step to be fit on human-rated data;
(ii) calibration of the Simulation Fidelity and Causal Validity dimensions against expert judgment, so that their absolute magnitudes can be interpreted rather than only compared within the framework;
(iii) comparative benchmarks against prompt-based use of LLMs and manual research on identical tasks, together with task suites that have known ground-truth answers;
(iv) adaptive orchestration that selects between sequential, parallel, and hierarchical team execution according to task dependencies and resource constraints;
(v) application of the workflow in other fields of economics, such as public policy evaluation, financial markets, and development economics;
and (vi) bringing the inputs that economics already measures through purpose-built pipelines, such as satellite imagery, audio, video, and large document collections, into the agentic workflow.

The contribution claimed here is workflow-structural: the framework organizes and records the production of a first research prototype. Domain experts remain responsible for assessing the substantive validity of its models and interpretations.
The use we designed it for is entry into an unfamiliar field: a researcher crossing a boundary, whether between subfields of economics, between economics and the other social sciences, or into interdisciplinary work, begins from a prototype whose sources, calibration verdict, and provenance are recorded.

\paragraph{Code Availability and Reproducibility}
All code, configuration files, and documentation are available at \url{https://github.com/AgenticEconLab/agentic-econ-lab}.
The repository includes implementations of all seven research teams, a cross-team pipeline orchestrator, 97 shared infrastructure modules, 81 of them in sixteen packages, a fourteen-dimension evaluation framework, and 2{,}972 automated tests.
The repository README documents installation, configuration, and use.
We encourage the economic research community to build upon, extend, and contribute to these open-source tools.
All code is released under the MIT License.

\paragraph{Data Availability}
The demonstration uses only publicly available data: the macroeconomic series the workflow retrieves come from FRED and the World Bank, and the literature metadata from the open scholarly indices it queries (OpenAlex, Crossref, arXiv, and Semantic Scholar). The retrieval code is part of the repository above. The outputs of the 26 runs reported in Section~\ref{sec:evaluations}, the scripts that compute the reported numbers from them, and the second-round expert ratings, with rater identities replaced by neutral identifiers, are archived on Zenodo (\href{https://doi.org/10.5281/zenodo.23258640}{doi:10.5281/zenodo.23258640}).

\paragraph{Acknowledgments}
Part of this work was carried out while Hankui Wang was at Bielefeld University. The computing time was provided through a project of the Chair for Economic Theory and Computational Economics (ETACE) at Bielefeld University. The authors gratefully acknowledge the funding of this project by computing time provided by the Paderborn Center for Parallel Computing (PC2).

\paragraph{Funding}
This work was supported by the French government through the France 2030 investment plan managed by the National Research Agency (ANR), as part of the Initiative of Excellence Universit\'e C\^ote d'Azur under reference number ANR-15-IDEX-01.

\paragraph{Competing Interests}
The authors declare no competing interests.

\paragraph{Use of AI Tools}
Large language models are the subject of this research: the workflows and the Tier-2 judges run on the open-weight models named in the text. In preparing the manuscript and the software, the authors also used Claude Code (Anthropic) to assist with code development, LaTeX and figure preparation, and editing. The authors reviewed all such output and take full responsibility for the content of the paper.

\section*{Appendices}

\appendix

\section{The Demonstration Run in Detail}
\label{appendix:example-model}

This appendix reproduces the economic content behind the end-to-end demonstration of Section~\ref{sec:Examples}: the model-development, calibration, data, estimation, and reporting detail of one chained pass through the seven-stage pipeline (configuration \texttt{ModeWithWcWithHITL}, committee-resolved checkpoints). The run's lead research question asked how AI-driven adaptive regulatory controllers mitigate systemic risk in a hybrid heterogeneous-agent New Keynesian (HANK) and agent-based model of financial networks during periods of monetary-policy uncertainty. Of the Model team's three candidate models, we present the lead model in full (\S\ref{app:lead-model}) and the verdicts on the other two (\S\ref{app:sibling-models}); \S\ref{app:typical-case} summarizes, for comparison, the run closest to the median of the 26 runs and the run originally designated as the demonstration. The material below is the workflow's own output; it demonstrates what the system generates and how it assesses its own work, not a validated economic contribution.

\subsection{The Lead Model: An AI-Regulated HANK--Agent-Based Framework}
\label{app:lead-model}

\subsubsection{Theoretical Framework}
The framework combines a HANK household sector with an agent-based model of an interbank network. Monetary-policy uncertainty enters as an exogenous stochastic process. A regulator modeled as a reinforcement-learning agent observes household liquidity constraints and network stress indicators and adjusts regulatory parameters to contain contagion, without foresight of future policy regimes. The equilibrium concept is a Markov perfect equilibrium in which households and banks optimize given prices, constraints, and regulation, and the regulator's policy is stationary and optimal given its value function. The framework's stated testable predictions include that the regulator tightens capital or liquidity requirements preemptively in high-uncertainty regimes, and that liquidity-constrained households experience smaller consumption volatility under the learned regime than under static rules.

\subsubsection{Formal Model}
The model has 20 variables, 13 listed parameters, 16 equations, and 10 constraints. Its core consists of the household Euler equation with a borrowing-constraint multiplier, the Phillips curve, the policy rule, the bank-distress process, and the regulator's problem:
\begin{align}
u'(c_{i,t}) &= \beta\,\mathbb{E}_t\!\left[\frac{u'(c_{i,t+1})}{1+\pi_{t+1}}\,(1+r_t+\lambda_{i,t})\right], \\[2pt]
\pi_t &= \beta\,\mathbb{E}_t[\pi_{t+1}] + \kappa\,y_t, \qquad
r_t = \phi_\pi\,\pi_t + \phi_y\,y_t + \epsilon_t, \qquad
\epsilon_t = \rho_\epsilon\,\epsilon_{t-1} + \sigma_\epsilon\,\eta_t, \\[2pt]
D_{j,t} &= \max\!\big(0,\; \rho_D\,D_{j,t-1} + \xi_t - \theta\,a_t\big), \qquad
\xi_t = \rho_\xi\,\xi_{t-1} + \sigma_\xi\,\zeta_t, \\[2pt]
V(s_t) &= \max_{a_t}\,\mathbb{E}_t\big[R(s_t,a_t) + \gamma\,V(s_{t+1})\big], \qquad
R(s_t,a_t) = -\Big(\omega_1\,y_t^{2} + \omega_2 \textstyle\sum_j D_{j,t}^{2} + \omega_3\,\lVert a_t - a_{t-1}\rVert^{2}\Big),
\end{align}
where $D_{j,t}$ is the distress of bank $j$, $\xi_t$ a network shock, $a_t$ the regulator's action, and $R$ its reward (written $r(s_t,a_t)$ in the model output). Around this core the model carries the household budget constraint and labor supply, aggregate labor, Cobb--Douglas production with the implied wage and capital rental rate, capital accumulation, and goods-market clearing. Six symbols in the equations ($\delta$, $\theta$, $\rho_\epsilon$, $\rho_D$, $\rho_\xi$, $\sigma_\xi$) have no entry in the parameter list, and two listed parameters (the interbank exposure weight $\alpha_{ij}$ and the exploration decay $\sigma_t$) appear in no equation. The parameter labeled ``financial friction transmission coefficient'' ($\kappa$) enters the model only as the slope of the Phillips curve. Table~\ref{tab:app-params} reports the proposed parameterization and the harness's estimates.

\begin{table}[!htbp]
\centering
\begin{threeparttable}
\caption{Parameters of the Lead Model: Proposed Values and Harness Estimates}
\label{tab:app-params}
\small
\begin{tabularx}{\textwidth}{@{}c Y c c@{}}
\toprule
Symbol & Parameter & Proposed & Harness \\
\midrule
$\beta$ & Household discount factor & $0.99$ & $0.980$ \\
$\alpha$ & Capital share & $0.37$ & $0.38$ \\
$\sigma$ & Relative risk aversion & $2.5$ & -- \\
$\phi_\pi,\ \phi_y$ & Policy-rule responses to inflation, output gap & $1.5,\ 0.125$ & -- \\
$\sigma_\epsilon$ & Policy-shock volatility & $0.002$ & -- \\
$\kappa$ & Phillips-curve slope (labeled financial-friction transmission) & $0.05$ & -- \\
$\alpha_{ij}$ & Interbank exposure weight & $0.15$ & -- \\
$\gamma$ & Regulator's discount factor & $0.95$ & -- \\
$\omega_1,\ \omega_2,\ \omega_3$ & Reward weights: output gap, bank distress, regulatory change & $0.4,\ 0.4,\ 0.2$ & -- \\
$\sigma_t$ & Exploration decay & $0.99$ & -- \\
\bottomrule
\end{tabularx}
\begin{tablenotes}[flushleft]\footnotesize
\item \textit{Notes:} Proposed values are the language model's; the harness estimated $\beta$ and $\alpha$ and did not estimate the others (--).
\end{tablenotes}
\end{threeparttable}
\end{table}

\subsubsection{Calibration}
\label{app:lead-calibration}
The deterministic calibration harness parsed 5 of the 16 equations to closed algebraic form: the budget constraint, the wage equation, the policy rule, and the two shock processes. It refused the other 11, each with a recorded reason: the Euler equation and the Phillips curve contain expectation operators; labor supply is written with utility derivatives that are not defined as functions; aggregate labor and the reward contain sums over agents; the distress process and the value function contain max operators; capital accumulation and goods-market clearing use an investment symbol $I_t$ that the model never declares and that collides with a reserved symbol; the capital rental rate carries a superscript label ($r^{k}_t$); and the production function $Y_t = K_t^{\alpha} H_t^{1-\alpha}$ was refused because the parser read the exponent $\alpha$ as a superscript label, which is an error of the parser rather than a property of the model.

Three cited targets bear on the parameters through steady-state identities: the real interest rate ($2\%$) determines $\beta = 1/1.02 = 0.980$, and the labor share ($0.60$) and capital share ($0.36$) both restrict $\alpha$. The harness's least-squares estimate is $\alpha = 0.38$, which misses each share target by $0.02$ because the two targets do not sum to one; the real-rate target is matched exactly. The moment-match score is $0.97$. The harness classified all three moments as pinning moments, none as over-identifying, and so ran no over-identification test; each share target lies one standard error ($0.02$) from the model value, inside the harness's two-standard-error rule, and the verdict is point-calibrated: consistency with the targeted moments, not a certified structural fit.

\subsubsection{Derived Code}
\label{app:lead-code}
The Code team derived an executable module from this model's parsed equation system: 5 of the 16 equations parse to a closed algebraic residual against 17 unknowns, so the system is non-square and the module carries a partial verdict. Its validation battery passed module compilation, parameter completeness, and finite residuals at the starting point (largest absolute residual $0.81$), and declined the steady-state solve on the non-square system. The module takes the parameter values proposed by the language model, not the harness estimates. The interbank exposure weight $\alpha_{ij}$ and the capital share $\alpha$ map to the same symbol, so the module sets $\alpha=0.15$, the proposed value of $\alpha_{ij}$, and the report's parameter table for this model has no row for $\alpha$.

\subsection{How the Harness Chooses a Calibration Path}
\label{app:calibration-paths}
The harness organizes its verification by generality of mechanism rather than by economic discipline: a general research framework cannot enumerate the thousands of modeling traditions it may encounter, and the tiers determine only how a model is verified, never what the modeling stage may build. Tier~0, the default path, works from the model's own equations: it parses them with symbolic algebra, recognizes the signatures of a few textbook structures in them (an RBC or New Keynesian core, Cobb--Douglas or CES production, a hand-to-mouth household share), takes the closed-form steady-state identities of those structures as moments, and estimates the free parameters by bounded least squares against cited external targets from a curated target book, reporting calibrated (an over-identified, testable fit), point-calibrated (pinning moments matched within the harness's two-standard-error tolerance, with no over-identification test performed), or partially calibrated (some targets left unmatched, for example because a parameter stops at a bound of its declared range).

Tier~1 provides three general calibration routes for models the parser refuses: algebraic steady-state identities, reduced-form time-series dynamics, and agent-based or otherwise stochastic-algorithmic simulation estimated by the simulated method of moments against the same cited targets, reporting simulation-calibrated when a simulator of the model's own equations is supplied (the pipeline does not yet supply one). Tier~2 is a small registry of canonical worked examples of those routes (for example, a New Keynesian economy under a Taylor rule whose inflation response governs inflation volatility, or a kinetic wealth-exchange economy whose savings propensity governs the income Gini); a model enters the registry only if it belongs to a textbook-canonical family and comes with cited targets, explicit rules for accepting or declining a match, and unit tests. A fit on a registry example is reported as archetype-calibrated, and only the parameters that example estimates are harness values.

It is not expert-grade domain modeling, which remains the work of specialists in each field; the tiers give a rapid prototype, or a researcher's first model in an unfamiliar field, a calibration verdict scoped to the evidence. A model no tier can estimate is reported uncalibratable with a precise machine-readable reason (for instance, that the specification contains learned components with no numerical realization), and the placeholder fit is marked non-applicable.

\subsection{The Other Two Candidate Models}
\label{app:sibling-models}
The second model, on stablecoin adoption by high-MPC households in high-inflation emerging markets, has 20 equations and 15 parameters; the harness parsed 14 equations and refused 6. It estimated $\beta = 0.980$ from the real-rate target; the share of high-MPC households, which the MPC target ($0.25$) restricts, stopped at a bound of its declared range, leaving a model MPC of $0.43$. The verdict is partially calibrated, and the fit score is set to zero because the verdict is not a validated or point calibration. Its derived module resolves 14 of 20 equations against 27 unknowns (partial).

The third model, on machine-learning surrogates for HANK stress testing, represents the cross-sectional distribution by a learned embedding. The harness declared it uncalibratable with the recorded reason that the learned components have no numerical realization and would need training and simulation-based estimation rather than closed-form moment matching. Its derived module resolves 2 of 20 equations against 4 unknowns (partial).

\subsection{Feasibility Cycle and Data}
The Model--Data feasibility cycle ran three reconciliation rounds and issued two Model Revision Requests before ending unresolved after the final design. Across the three models it graded 34 data requirements: 2 feasible (government spending, matched to FRED's government consumption expenditures and gross investment, and the unemployment rate), 15 available only through a proxy, and 17 unavailable, with 11 essential requirements unmet. The unavailable requirements are mostly calibration targets that no open source publishes as a series (bank-distress frequency, systemic-risk contagion speed, the share of high-MPC households, the stablecoin share of liquid assets, and the surrogate's own error and latency targets). The committee resolved the unmet requirements by voting to narrow the question in 12 decisions and to accept a proxy in 2; a decision to narrow the question is recorded with the run's limitations, and the run does not return to the Ideation team.

The Data team assembled 17 series. Thirteen came live from open connectors: from FRED, the 10-year Treasury yield, government consumption expenditures and gross investment, personal consumption expenditures, the St.~Louis Fed and Chicago Fed financial-stress and financial-conditions indices, the federal funds rate, the unemployment rate, real GDP, total consumer credit, the 3-month Treasury bill rate, the CPI, and the 10-year real interest rate; from the World Bank, annual CPI inflation. Two of these (consumer credit and the bill rate) were retrieved after a disclosed repair of the requested identifier. The other four had no successful live retrieval and are carried as disclosed simulations; the estimation harness excludes them. Quality assurance graded the dataset at 60 of 100 (Bronze), the cap applied when a simulated series is present.

\subsection{Estimation}
\label{app:estimation}
A chained run produces one terminal estimation and report, built on the lead model. \texttt{Estimator}'s specification regresses the St.~Louis Fed Financial Stress Index (level) on a real policy rate (the federal funds rate minus year-over-year CPI inflation), the quarterly change in the federal funds rate, and the quarterly growth rate of real GDP, over 1994Q4--2026Q2 ($n=127$), by ordinary least squares with heteroskedasticity-and-autocorrelation-consistent (HAC) standard errors. The committee rejected this proposal and its revision; under the review rule the proposal was estimated and both sets of objections were recorded with the result. Table~\ref{tab:app-estimation} reports the estimates.

\begin{table}[!htbp]
\centering
\begin{threeparttable}
\caption{Estimation-Stage Output for the Demonstration Run}
\label{tab:app-estimation}
\small
\begin{tabular*}{\textwidth}{@{\extracolsep{\fill}} l S[table-format=-1.3] S[table-format=1.3] S[table-format=1.3] @{}}
\toprule
Regressor & {Estimate} & {Std.\ error} & {$p$-value} \\
\midrule
Constant & 0.117 & 0.195 & 0.548 \\
Real policy rate & -0.035 & 0.038 & 0.362 \\
Change in the federal funds rate (``uncertainty'') & -0.937 & 0.417 & 0.025 \\
Real GDP growth (``output-gap proxy'') & -0.179 & 0.149 & 0.231 \\
\bottomrule
\end{tabular*}
\begin{tablenotes}[flushleft]\footnotesize
\item \textit{Notes:} OLS with HAC standard errors. Dependent variable: St.~Louis Fed Financial Stress Index, quarterly, 1994Q4--2026Q2; $n=127$, $R^{2}=0.29$, adjusted $R^{2}=0.27$.
\end{tablenotes}
\end{threeparttable}
\end{table}

The output records three features. First, the two theory-derived hypotheses came out inconclusive (real-rate coefficient negative; one-sided $p=0.18$) and contradicted (coefficient on the change in the funds rate positive; the estimate is negative, $p=0.025$). The regression contains no variable measuring AI-driven regulation, and none of its variables is linked to a variable of the lead model; the committee's first-round objections named the missing regulation variable. Second, the diagnostics battery returned caveats with no severe failure: heteroskedasticity (Breusch--Pagan $p<0.001$, addressed by the robust covariance), residual autocorrelation (Durbin--Watson $0.66$, first-order coefficient $0.67$, Ljung--Box $p<0.001$), functional-form misspecification (RESET $p=0.008$), and non-normal residuals; the residuals are stationary and the largest variance-inflation factor is $1.1$. Third, the real-rate coefficient kept its sign in all five robustness variants the harness estimated (dropping each of the other two regressors, estimating on each half of the sample, and using the heteroskedasticity-consistent HC1 covariance estimator), for a sign-stability score of $1.0$.

\subsection{Report}
The Reporting team assembled the terminal report by deterministic assembly, with every table generated from artifacts and language-model prose confined to narrative slots; the committee objected to the first draft and to its revision; under the one-revision rule the revised draft was published with both sets of objections in its limitations section. The report runs to about 3{,}200 words before its 70 references, with two figures (the analysis series and the coefficient confidence intervals). Its number-consistency check traced 315 of the 332 numbers in the body to the artifact pool, exempted the remaining 17 (16 small integers and one confidence level), and returned consistent. Its limitations section lists the four simulated series, the proxy, the two repaired identifiers, the committee's objections at the data, estimation, and report-draft checkpoints, the three calibration verdicts, the two hypothesis outcomes, and the unresolved feasibility cycle.

\subsection{Dual-Judge Content Scores for This Run}
Table~\ref{tab:chained-tier2} reports the per-dimension Tier-2 content scores for this run, under each judge separately. Each judge reads the stage's actual artifacts (research questions, literature batch and synthesis, model design and calibration output, data provenance and quality assurance, generated modules and validation results, estimation and inference output, and the final research report) against the same team-specific rubrics; scores are normalized to $[0,1]$.

\begin{table}[!htbp]
\centering
\begin{threeparttable}
\caption{Tier-2 Content Scores for the Demonstration Run, by Judge and Dimension}
\label{tab:chained-tier2}
\small
\begin{tabular*}{\textwidth}{@{\extracolsep{\fill}} l c *{6}{S[table-format=1.2]} S[table-format=1.3] @{}}
\toprule
Stage & Judge & {Corr.} & {Snd.} & {Inn.} & {Trn.} & {Dec.} & {Rig.} & {Mean} \\
\midrule
Ideation & M & 0.75 & 0.81 & 0.75 & 0.83 & 0.83 & 0.83 & 0.802 \\
 & G & 0.94 & 0.88 & 0.81 & 0.83 & 0.83 & 0.92 & 0.868 \\
\addlinespace
Literature & M & 0.75 & 0.67 & 0.69 & 0.83 & 0.67 & 0.67 & 0.712 \\
 & G & 0.75 & 0.75 & 0.88 & 0.83 & 0.83 & 0.83 & 0.812 \\
\addlinespace
Model & M & 0.69 & 0.67 & 1.00 & 0.67 & 0.75 & 0.67 & 0.740 \\
 & G & 0.62 & 0.67 & 0.88 & 0.75 & 0.92 & 0.67 & 0.750 \\
\addlinespace
Data & M & 0.50 & 0.67 & 0.62 & 0.67 & 0.67 & 0.58 & 0.618 \\
 & G & 0.50 & 0.58 & 0.62 & 0.58 & 0.50 & 0.58 & 0.562 \\
\addlinespace
Code & M & 0.67 & 0.67 & 0.50 & 0.50 & 0.42 & 0.42 & 0.528 \\
 & G & 0.67 & 0.50 & 0.31 & 0.50 & 0.25 & 0.42 & 0.441 \\
\addlinespace
Estimation & M & 0.58 & 0.75 & 0.62 & 0.67 & 0.67 & 0.67 & 0.660 \\
 & G & 0.58 & 0.50 & 0.56 & 0.67 & 0.83 & 0.50 & 0.608 \\
\addlinespace
Reporting & M & 0.83 & 0.67 & 0.56 & 0.67 & 0.67 & 0.58 & 0.663 \\
 & G & 0.83 & 0.67 & 0.50 & 0.67 & 0.83 & 0.58 & 0.681 \\
\bottomrule
\end{tabular*}
\begin{tablenotes}[flushleft]\footnotesize
\item \textit{Notes:} M~=~Mistral-Small-3.2-24B, G~=~Gemma-3-27B. Corr.~=~Correctness, Snd.~=~Soundness, Inn.~=~Innovation Potential, Trn.~=~Transparency, Dec.~=~Decision Quality, Rig.~=~Economic Rigor. Dual-judge protocol of Section~\ref{sec:two-tier-methodology} applied to this run (subject model Qwen3.6-27B-FP8; both judges from non-subject families; temperature $=0.0$, structured output). The Correctness, Soundness, and Economic Rigor rubrics are team-specific, and the other three are shared across stages; for the execution stages the team-specific criteria target code fidelity and explicit scoping (Code), specification fit and identification discipline (Estimation), and number fidelity and caveat proportionality (Reporting).
\end{tablenotes}
\end{threeparttable}
\end{table}

\subsection{Comparison Runs: The Typical Run and the Originally Designated Run}
\label{app:typical-case}
\paragraph{The typical run.} Over the 26 runs, the run whose values of R1--R4 and estimation verdict lie closest to the medians (Section~\ref{sec:Examples}) took 6.1 hours. Its three research questions concern heterogeneous subjective expectations among high-MPC households in a behavioral HANK model, the conditions under which monetary expansion raises wealth inequality through asset prices, and the role of production and financial networks in transmitting monetary shocks to banked and underbanked households. The harness returned partially calibrated, uncalibratable (the canonical New Keynesian model's inflation response stopped at a bound of its declared range without reproducing the target), and point-calibrated ($\beta = 0.980$ from the real-rate target). The feasibility cycle ran two rounds with one revision request and ended unresolved (1 requirement feasible, 11 proxy-only, 23 unavailable). Of 11 series, one is a disclosed simulation and one a disclosed proxy. The estimated regression relates annual real GDP growth (labeled output volatility) to the World Bank Gini index and a real policy rate over 1964--2024 ($n=61$, $R^{2}=0.065$); both hypotheses are inconclusive, and the sign-stability score is $0.75$ over four variants. The committee rejected the specification in both review rounds. The report is consistent (235 of 252 numbers traced, 17 exempt). Its Tier-2 stage means range from $0.549$ (Data) to $0.747$ (Ideation) under Mistral and from $0.510$ (Data) to $0.795$ (Ideation) under Gemma.

\paragraph{The originally designated run.} The run designated as the demonstration when the jobs were submitted produced three questions on heterogeneous attention to climate risk under green monetary policy, informal-sector dynamics in an agent-based HANK hybrid for emerging economies, and inequality and cognitive load under contractionary policy. Its calibration verdicts were one partially calibrated and two uncalibratable, and 3 of its 14 series are disclosed simulations. Its specification regressed the growth of CO$_2$ emissions on a real policy rate, the returns of a clean-energy equity fund, and their interaction; after transformations the series overlapped in 12 observations, against the 32 the harness requires for four parameters, and the harness returned inestimable. A repair that would have changed the dependent variable was rejected. The report passes the consistency check after the offline recheck of Section~\ref{sec:robustness-checks}.

\section{Quantitative Evaluation Methodology}
\label{appendix:quantitative-methods}

This appendix provides the methodology for quantitative evaluation of agentic workflows, complementing the theory-grounded framework presented in Section~\ref{sec:evaluations}.
Section~\ref{appendix:implemented-metrics} documents the exact formulae used to compute the scores reported in the paper.

\subsection{Dimension Summary}

Table~\ref{tab:dimension-summary} summarizes the fourteen evaluation dimensions introduced in Section~\ref{sec:evaluations}: their definitions, the design features they assess, and the metrics used to operationalize them.

\begin{sidewaystable}
\centering
\begin{threeparttable}
\caption{Evaluation Dimensions: Definitions, Design Features, and Metrics}
\label{tab:dimension-summary}
\small
\begin{tabularx}{\textwidth}{@{}p{3.2cm} Y Y Y@{}}
\toprule
Dimension & Definition & Design features & Target metrics \\
\midrule
\multicolumn{4}{@{}l}{\textit{Quality Cluster}} \\
Reliability & Consistency under varying conditions & Agent specifications, structured protocols, version control & Output CV, path stability \\
Correctness & Accuracy against domain standards & Specialized agents, verification mechanisms, HITL checkpoints & Factual accuracy, error rate \\
Soundness & Logical and methodological validity & Staged workflows, explicit reasoning chains & Reasoning coherence, expert ratings \\
\addlinespace
\multicolumn{4}{@{}l}{\textit{Operational Cluster}} \\
Efficiency & Resource--output optimization & Parallelization, task automation, caching & Execution time, token usage, compute latency \\
Scalability & Performance under increasing load & Modular design, distributed execution & Scaling linearity, throughput \\
Robustness & Error handling and recovery & Escalation pathways, validation checkpoints, fallbacks & Error recovery rate, failure tolerance \\
\addlinespace
\multicolumn{4}{@{}l}{\textit{Epistemic Cluster}} \\
Transparency & Inspectability of operations & Explicit agent definitions, logged communications & Inspectability score, documentation coverage \\
Traceability & Provenance of outputs & Logging, version control, metadata & Provenance completeness, reconstruction accuracy \\
Reproducibility & Repeatable results & Deterministic orchestration, configuration documentation & Replication success rate \\
\addlinespace
\multicolumn{4}{@{}l}{\textit{Innovation Cluster}} \\
Innovation Potential & Novel idea generation & Multi-agent exploration, gap identification, cross-domain integration & Novelty score, diversity index \\
\addlinespace
\multicolumn{4}{@{}l}{\textit{Domain-Specific Cluster}} \\
Decision Quality & Appropriateness of methodological choices & Stage-gated methodology selection, HITL validation & Choice justification score, expert ratings \\
Economic Rigor & Domain-specific validity of reasoning and outputs & Identification-strategy checks, calibration discipline, robustness analysis & Identification validity, calibration accuracy, sensitivity coverage \\
Simulation Fidelity\tnote{a} & Match between simulated trajectories and stylized economic facts & Optional agent-based simulation stage; Falsifier cross-check & Macro-moment match, scenario residual, impulse-response coverage \\
Causal Validity\tnote{a} & Soundness of identification strategies and causal claims & Causal-foundation-model estimate (back-door, front-door, IV); classical econometric cross-check & Identification soundness, model-vs.-classical divergence, sensitivity to confounding \\
\bottomrule
\end{tabularx}
\begin{tablenotes}[flushleft]\footnotesize
\item \textit{Notes:} Fourteen dimensions grouped into five clusters. The last column lists the metrics each dimension targets; the scores reported in this paper are computed as defined in Section~\ref{appendix:implemented-metrics}, which implements a subset of them. CV denotes coefficient of variation; HITL, human-in-the-loop; IV, instrumental variable.
\item[a] Economics-domain dimensions, scored only when the corresponding optional stage (agent-based simulation; causal-foundation-model estimation) is active.
\end{tablenotes}
\end{threeparttable}
\end{sidewaystable}

\subsection{Implemented Metric Computations}
\label{appendix:implemented-metrics}

This subsection documents the exact formulae used to compute the scores reported in Section~\ref{sec:empirical-validation}.
All metrics are computed from structured \texttt{execution\_log.json} files (Tier~1) and from open-weight ensemble (Mistral-Small-3.2-24B and Gemma-3-27B) evaluations of workflow output files (Tier~2).

\paragraph{Tier-1 structural metrics.}
The Tier-1 metrics are derived from a small set of execution counts.
Let $n_s$ denote the number of pipeline stages, $n_{\text{succ}}$ the number of stages with status \texttt{success}, $n_{\text{out}}$ the number of stages producing output files, $n_e$ the number of errors recorded, $n_r$ the number of recovered errors, and $T$ the total execution time in seconds; let $\mathbf{c} = (c_1, \ldots, c_{n_s})$ be the item counts produced by each stage and $N = \sum_i c_i$ the total item count.
Reliability is the product of stage completion, output coverage, and an error penalty,
\begin{equation}
\text{Reliability} = C_s \cdot C_o \cdot P_e
\end{equation}
where the stage completion rate $C_s = n_{\text{succ}} / n_s$, the output coverage $C_o = n_{\text{out}} / n_s$, and the error penalty is
\begin{equation}
P_e = \begin{cases}
1.0 & \text{if } n_e = 0 \\
0.7 & \text{if } n_e > 0 \text{ and } C_s = 1.0 \\
\max\!\big(0,\; 1 - n_e / n_s\big) & \text{otherwise}
\end{cases}
\end{equation}
This measures structural execution compliance: whether all stages completed, produced outputs, and avoided or recovered from errors. It does not measure output consistency across repeated runs (see Section~\ref{sec:limitations}).
Efficiency maps execution time to a 0--1 score relative to a reference duration,
\begin{equation}
\text{Efficiency} = \frac{T_{\text{ref}}}{T_{\text{ref}} + T}, \qquad T_{\text{ref}} = 3600 \cdot n_{\text{team}}
\end{equation}
where $n_{\text{team}}$ is the number of teams in the logged run (one for a single team, seven for the chained pipeline); the score is $1.0$ for instantaneous completion, $0.5$ at $T = T_{\text{ref}}$, and decreases toward zero without reaching it.
Scalability normalizes throughput against a reference rate,
\begin{equation}
\text{Scalability} = \min\!\big(1.0,\; \theta / 10\big), \quad \text{where } \theta = \frac{N}{T / 60} \text{ (items per minute)}
\end{equation}
In a standalone team log, items are the team's output units (research questions for Ideation, literature entries for Literature, model components for Model, data series for Data); these units differ in granularity, so Scalability supports comparisons of one team across modes but not rankings across teams. In the chained-pipeline log, each team stage records the number of artifacts it hands on (eleven in total), so Scalability in the chained run is $0.002$--$0.005$ and is not comparable with standalone values.
Robustness measures the proportion of errors that were detected and recovered from without workflow failure,
\begin{equation}
\text{Robustness} = \begin{cases}
1.0 & \text{if } n_e = 0 \\
n_r / n_e & \text{otherwise}
\end{cases}
\end{equation}
Traceability is the fraction of required provenance metadata fields present in the log,
\begin{equation}
\text{Traceability} = \frac{|\{f \in F_p : f \text{ present in log}\}|}{|F_p|}
\end{equation}
where $F_p = \{\texttt{execution\_id}, \texttt{team}, \texttt{mode}, \texttt{started\_at}, \texttt{completed\_at}\}$ is the set of required provenance metadata fields.
Reproducibility records whether the execution log carries a non-empty metadata block (research topic, configuration name, mode); its contents are not checked,
\begin{equation}
\text{Reproducibility} = \begin{cases}
1.0 & \text{if the log's metadata block is non-empty} \\
0.5 & \text{otherwise}
\end{cases}
\end{equation}

\paragraph{Structural baselines for the Tier-2 dimensions.}
The six Tier-2 dimensions also have Tier-1 structural baselines ($S_{\text{T1}}$).
For Correctness the structural baseline is $C_s \times (1 \text{ if } n_e = 0; \; 0.5 \text{ otherwise})$, taking error-free stage completion as a necessary condition for correctness.
For Soundness it is $1.0$ if pipeline stages execute in valid sequential order and $0.0$ otherwise.
For Transparency it is the output file coverage $C_o = n_{\text{out}} / n_s$, reflecting whether each stage produces inspectable output files.
For Innovation Potential it is $(D + B) / 2$, where $D = H(\mathbf{p}) / \log K$ is the Shannon entropy of the item distribution across stages ($p_i = c_i / N$), normalized by its maximum over the $K$ stages with at least one item and $B = \min(1.0, N / 50)$ measures coverage breadth. For Decision Quality it is the mean of a stage-progression indicator ($1.0$ if every stage succeeded in valid order, $0.5$ otherwise) and the Robustness score; for Economic Rigor it is $C_s \cdot C_o$.

\paragraph{Tier-2 content-quality scores.}
For each Tier-2 dimension $d$ (Correctness, Soundness, Innovation Potential, Transparency, Decision Quality, and Economic Rigor), an open-weight two-judge ensemble (Mistral-Small-3.2-24B and Gemma-3-27B) evaluates the workflow's actual output files against a rubric with $k \in \{3, 4\}$ sub-criteria, each sub-criterion $i$ receiving a score $s_i \in \{1, 2, 3, 4, 5\}$. The Correctness, Soundness, and Economic Rigor rubrics are team-specific; the Innovation Potential, Transparency, and Decision Quality rubrics are shared across teams. Scores are clamped to 1--5, and a sub-criterion the judge omits is set to 3.
The raw average and normalized overall score are
\begin{equation}
\bar{s}_d = \frac{1}{k} \sum_{i=1}^{k} s_i, \qquad
S_{\text{T2},d} = \frac{\bar{s}_d - 1}{4}
\end{equation}
so that the normalization maps the 1--5 Likert scale onto a 0--1 range ($1 \mapsto 0.0$, $5 \mapsto 1.0$).

\paragraph{Combined two-tier scores.}
For dimensions carrying both Tier-1 and Tier-2 assessments, the implementation supports a convex combination
\begin{equation}
S_d = \alpha \cdot S_{\text{T1},d} + (1 - \alpha) \cdot S_{\text{T2},d}
\end{equation}
with $\alpha = 0$ by default, so the reported score is the Tier-2 score. In each of the 26 chained runs, five of the six structural baselines equal $1.0$ and the sixth (Innovation Potential) equals $0.584$, so a positive $\alpha$ would move every score toward a fixed baseline, shrinking differences between runs by the factor $1-\alpha$ without changing their order within a dimension. Dimensions assessed by Tier~1 only (Reliability, Efficiency, Scalability, Robustness, Traceability, Reproducibility) take
\begin{equation}
S_d = S_{\text{T1},d}
\end{equation}
The two subfield-specific dimensions are computed structurally; no LLM judge scores them. Simulation Fidelity is the share of a scenario's benchmark targets $m_j^{*}$ that the simulation reproduces within $5\%$ of the target, or within $0.05$ when the target is smaller than one in absolute value,
\begin{equation}
\text{SimulationFidelity} = \frac{1}{J}\sum_{j=1}^{J} \mathbf{1}\!\left[\frac{|m_j - m_j^{*}|}{\max(|m_j^{*}|, 1)} \le 0.05\right],
\end{equation}
not evaluated when no simulation ran and $0$ when the scenario has no benchmark. Causal Validity compares the estimated average treatment effect $\hat\tau$ with a known effect $\tau$ when one is supplied, $\max\!\big(0, 1 - |\hat\tau - \tau| / (0.15\,|\tau|)\big)$, with $|\tau|$ replaced by $1$ when $\tau = 0$; otherwise it is $1$ when a supplied prior mean lies inside the estimate's confidence interval and falls linearly with the prior's distance from the interval relative to its width; with neither a known effect nor a prior it is not evaluated. Neither stage ran in the reported runs.

\paragraph{Multi-run statistical aggregation.}
\label{appendix:multirun-methodology}
Table~\ref{tab:multirun-ci} reports Tier-1 means and sample standard deviations over the 26 runs. For Tier~2, let $S_{j,s,r}$ denote judge $j$'s mean over the six content dimensions for stage $s$ in run $r$; Table~\ref{tab:inter-evaluator} reports its mean over the $R = 25$ replicates,
\begin{equation}
\bar{S}_{j,s} = \frac{1}{R} \sum_{r=1}^{R} S_{j,s,r}, \qquad
\sigma_{j,s} = \sqrt{\frac{1}{R-1} \sum_{r=1}^{R} (S_{j,s,r} - \bar{S}_{j,s})^2},
\end{equation}
with the 95\% confidence interval from the $t$-distribution,
\begin{equation}
\text{CI}_{95} = \bar{S}_{j,s} \pm t_{R-1, 0.025} \cdot \frac{\sigma_{j,s}}{\sqrt{R}}.
\end{equation}
The code also provides a cross-run output-consistency score,
\begin{equation}
\text{Reliability}_{\text{output}} = 0.4 \cdot J_{\text{text}} + 0.3 \cdot O_{\text{struct}} + 0.3 \cdot \max(0,\, 1 - \text{CV}_{\text{count}}),
\end{equation}
with the weights renormalized over the components available, where $J_{\text{text}}$ is the mean pairwise Jaccard similarity of extracted key terms across runs, $O_{\text{struct}}$ averages the key-set Jaccard similarity of JSON outputs and the share of shared keys with similar values, and $\text{CV}_{\text{count}}$ is the coefficient of variation of total output item counts. The scores reported in this paper do not use it.

\paragraph{Inter-evaluator agreement.}
\label{appendix:inter-evaluator-methodology}
To assess evaluator sensitivity, the same workflow outputs are scored by $M \geq 2$ LLM reviewers using identical rubrics, each assigning a score $s \in \{1, \ldots, 5\}$ for every sub-criterion.
Agreement between two reviewers $a$ and $b$ is measured by the Pearson correlation of their dimension scores,
\begin{equation}
r_{ab} = \frac{\sum_{q} (x_{a,q} - \bar{x}_a)(x_{b,q} - \bar{x}_b)}{\sqrt{\sum_{q} (x_{a,q} - \bar{x}_a)^2} \sqrt{\sum_{q} (x_{b,q} - \bar{x}_b)^2}}
\end{equation}
where $q$ indexes the (dimension, run) pairs of a stage (Table~\ref{tab:inter-evaluator}).

\paragraph{Agent-level metrics.}
\label{appendix:agent-metrics}
In addition to the fourteen workflow-level dimensions, the code defines four Tier-1 diagnostics, computed per team from per-call observability records without further LLM calls. Expected tools, call ranges, and stage plans are declared for the Ideation, Literature, Model, and Data teams only. The execution logs of the reported runs store aggregated observability rather than per-call records, so these diagnostics were not computed for them.
Tool correctness weights tool selection, avoidance of tools declared inappropriate for the team, and call success,
\begin{equation}
\text{ToolCorrectness} = 0.50 \cdot \frac{n_{\text{expected}}}{n_{\text{calls}}} + 0.25 \cdot \Big(1 - \frac{n_{\text{forbidden}}}{n_{\text{calls}}}\Big) + 0.25 \cdot \frac{n_{\text{success}}}{n_{\text{calls}}}
\end{equation}
where the counts are over the team's tool calls, and the expected and inappropriate tool sets are declared per team.
Step efficiency weights the LLM-call count, the output-token share, and the balance of stage durations,
\begin{equation}
\text{StepEfficiency} = 0.40 \cdot E_{\text{calls}} + 0.30 \cdot E_{\text{tokens}} + 0.30 \cdot \max\!\big(0,\; 1 - \text{CV}_{\text{dur}}/2\big)
\end{equation}
where $E_{\text{calls}}$ is $1.0$ when the number of LLM calls lies in the range declared for the team's stages and falls off in proportion to the shortfall or excess (floored at $0.3$), $E_{\text{tokens}}$ is $1.0$ when output tokens make up 10--50\% of all tokens and $\max(0, 1 - 2\,|r - 0.3|)$ otherwise for output-token share $r$, and $\text{CV}_{\text{dur}}$ is the coefficient of variation of stage durations.
Plan adherence weights stage completion, ordering, and expected outputs,
\begin{equation}
\text{PlanAdherence} = 0.50 \cdot C_p + 0.25 \cdot O + 0.25 \cdot F
\end{equation}
where $C_p$ is the fraction of the team's planned stages that completed successfully, $O$ is $1.0$ if the executed stages appear in the expected order and $0.5$ otherwise, and $F$ is the fraction of the team's expected output files that were produced.
Argument correctness is the fraction of passed argument checks over all tool calls, where a check passes when a required argument is present and non-empty or a typed argument has the declared type.
These metrics locate execution problems within a stage; none of them assesses the quality of an agent's reasoning.

\bibliographystyle{apalike}
\bibliography{references_agentic_econ}  %

\begin{thebibliography}{}

\bibitem[{A2A Project}, 2025]{a2aprotocol2025}
{A2A Project} (2025).
\newblock {Agent2Agent (A2A) Protocol}.
\newblock \url{https://github.com/a2aproject/A2A}.
\newblock Accessed 2026-09-23.

\bibitem[Afonso et~al., 2026]{afonso2026deepresearch}
Afonso, S., Galiani, S., G{\'a}lvez, R.~H., and Sosa, R.~A. (2026).
\newblock Deep research on a loop: Using {AI} agents to construct economic
  datasets.
\newblock NBER Working Paper 35188, National Bureau of Economic Research.

\bibitem[Agarwal et~al., 2025a]{agarwal2025autodiscovery}
Agarwal, D., Majumder, B.~P., Adamson, R., Chakravorty, M., Gavireddy, S.~R.,
  Parashar, A., Surana, H., Mishra, B.~D., McCallum, A., Sabharwal, A., and
  Clark, P. (2025a).
\newblock {AutoDiscovery}: Open-ended scientific discovery via {Bayesian}
  surprise.
\newblock {\em arXiv preprint arXiv:2507.00310}.
\newblock Code: \url{https://github.com/allenai/autodiscovery-neurips}.

\bibitem[Agarwal et~al., 2025b]{agarwal2025designing}
Agarwal, N., Moehring, A., and Wolitzky, A. (2025b).
\newblock Designing human-{AI} collaboration: A sufficient-statistic approach.
\newblock NBER Working Paper 33949, National Bureau of Economic Research.

\bibitem[Akata et~al., 2025]{Akata2023}
Akata, E., Schulz, L., Coda-Forno, J., Oh, S.~J., Bethge, M., and Schulz, E.
  (2025).
\newblock Playing repeated games with large language models.
\newblock {\em Nature Human Behaviour}, 9(7):1380--1390.
\newblock Code: \url{https://github.com/eliaka/repeatedgames}.

\bibitem[Alizadeh et~al., 2026]{alizadeh2026socsci}
Alizadeh, M., Mosleh, M., Gilardi, F., Kasirzadeh, A., and Tucker, J. (2026).
\newblock {AI} coding agents can reproduce social science findings.
\newblock {\em arXiv preprint arXiv:2606.11447}.
\newblock Code: \url{https://github.com/malizad/SocSci-Repro-Bench}.

\bibitem[Allard et~al., 2024]{allard2024inflationnowcasting}
Allard, M.-A., Teiletche, P., and Zinebi, A. (2024).
\newblock Enhancing inflation nowcasting with {LLM}: Sentiment analysis on
  news.
\newblock {\em arXiv preprint arXiv:2410.20198}.
\newblock Code: \url{https://github.com/paultltc/InflaBERT}.

\bibitem[{Amazon Web Services}, 2026]{awsAgentCoreObservability2026}
{Amazon Web Services} (2026).
\newblock {Amazon Bedrock AgentCore}: Observability.
\newblock
  \url{https://docs.aws.amazon.com/bedrock-agentcore/latest/devguide/observability.html}.
\newblock Accessed 2026-09-17.

\bibitem[Anthropic, 2025]{Anthropic2025}
Anthropic (2025).
\newblock Privacy policy.

\bibitem[{Arunkumar V} et~al., 2026]{arunkumar2026agenticarchitecture}
{Arunkumar V}, {Gangadharan G. R.}, and Buyya, R. (2026).
\newblock Agentic artificial intelligence ({AI}): Architectures, taxonomies,
  and evaluation of large language model agents.
\newblock {\em arXiv preprint arXiv:2601.12560}.

\bibitem[Asai et~al., 2026]{asai2026openscholar}
Asai, A., He, J., Shao, R., Shi, W., Singh, A., Chang, J.~C., Lo, K., Soldaini,
  L., Feldman, S., D'Arcy, M., Wadden, D., Latzke, M., Sparks, J., Hwang,
  J.~D., Kishore, V., Tian, M., Ji, P., Liu, S., Tong, H., Wu, B., Xiong, Y.,
  Zettlemoyer, L., Neubig, G., Weld, D.~S., Downey, D., Yih, W.-t., Koh, P.~W.,
  and Hajishirzi, H. (2026).
\newblock Synthesizing scientific literature with retrieval-augmented language
  models.
\newblock {\em Nature}, 650:857--863.
\newblock Code: \url{https://github.com/AkariAsai/OpenScholar}.

\bibitem[Asatryan et~al., 2025]{Asatryan2024}
Asatryan, Z., Birkholz, C., and Heinemann, F. (2025).
\newblock Evidence-based policy or beauty contest? an {LLM}-based meta-analysis
  of {EU} cohesion policy evaluations.
\newblock {\em International Tax and Public Finance}, 32(2):625--655.

\bibitem[Bail, 2024]{Bail2024}
Bail, C.~A. (2024).
\newblock Can generative {AI} improve social science?
\newblock {\em Proceedings of the National Academy of Sciences},
  121(21):e2314021121.

\bibitem[Bastani and Cachon, 2025]{bastani2025human}
Bastani, H. and Cachon, G.~P. (2025).
\newblock The human-{AI} contracting paradox.
\newblock {\em Available at SSRN 5962739}.

\bibitem[Bauer et~al., 2024]{Bauer2024}
Bauer, M.~D., Huber, D., Offner, E., Renkel, M., and Wilms, O. (2024).
\newblock Corporate green pledges.
\newblock Working Paper 2024-36, Federal Reserve Bank of San Francisco.
\newblock Also CEPR Discussion Paper 19788.

\bibitem[Bertsch et~al., 2025]{bertsch2025central}
Bertsch, C., Hull, I., Lumsdaine, R.~L., and Zhang, X. (2025).
\newblock Central bank mandates and monetary policy stances: Through the lens
  of {Federal Reserve} speeches.
\newblock {\em Journal of Econometrics}, 249:105948.

\bibitem[Biancotti and Camassa, 2023]{Biancotti2023Loquacity}
Biancotti, C. and Camassa, C. (2023).
\newblock Loquacity and visible emotion: {ChatGPT} as a policy advisor.
\newblock Occasional Paper (Questioni di Economia e Finanza) 814, Bank of
  Italy.
\newblock arXiv:2309.03595.

\bibitem[Bonawitz et~al., 2022]{bonawitz2022federated}
Bonawitz, K., Kairouz, P., Mcmahan, B., and Ramage, D. (2022).
\newblock Federated learning and privacy.
\newblock {\em Communications of the ACM}, 65(4):90--97.

\bibitem[Bran et~al., 2024]{bran2024chemcrow}
Bran, A.~M., Cox, S., Schilter, O., Baldassari, C., White, A.~D., and
  Schwaller, P. (2024).
\newblock Augmenting large language models with chemistry tools.
\newblock {\em Nature Machine Intelligence}, 6(5):525--535.
\newblock Code: \url{https://github.com/ur-whitelab/chemcrow-public}.

\bibitem[Brand et~al., 2023]{Brand2024}
Brand, J., Israeli, A., and Ngwe, D. (2023).
\newblock Using {LLMs} for market research.
\newblock Working Paper 23-062, Harvard Business School.

\bibitem[Bricken et~al., 2025]{bricken2025building}
Bricken, T., Wang, R., and Bowman, S. (2025).
\newblock Building and evaluating alignment auditing agents.
\newblock \url{https://alignment.anthropic.com/2025/automated-auditing/}.
\newblock Accessed February 1, 2026.

\bibitem[Brookins and DeBacker, 2024]{Brookins2024}
Brookins, P. and DeBacker, J. (2024).
\newblock Playing games with {GPT}: What can we learn about a large language
  model from canonical strategic games?
\newblock {\em Economics Bulletin}, 44(1):25--37.

\bibitem[Capraro et~al., 2024]{Capraro2024}
Capraro, V., Lentsch, A., Acemoglu, D., Akgun, S., Akhmedova, A., Bilancini,
  E., Bonnefon, J.~F., Brañas-Garza, P., Butera, L., Douglas, K.~M., Everett,
  J.~A., Gigerenzer, G., Greenhow, C., Hashimoto, D.~A., Holt-Lunstad, J.,
  Jetten, J., Johnson, S., Kunz, W.~H., Longoni, C., Lunn, P., Natale, S.,
  Paluch, S., Rahwan, I., Selwyn, N., Singh, V., Suri, S., Sutcliffe, J.,
  Tomlinson, J., Linden, S. V.~D., Lange, P. A.~V., Wall, F., Bavel, J. J.~V.,
  and Viale, R. (2024).
\newblock The impact of generative artificial intelligence on socioeconomic
  inequalities and policy making.
\newblock {\em PNAS Nexus}, 3.

\bibitem[Chai et~al., 2025]{chai2025scimaster}
Chai, J., Tang, S., Ye, R., Du, Y., Zhu, X., Zhou, M., Wang, Y., Zhang, Y.,
  Zhang, L., Chen, S., et~al. (2025).
\newblock {SciMaster}: Towards general-purpose scientific {AI} agents, part
  {I}. {X-Master} as foundation: Can we lead on {Humanity's Last Exam}?
\newblock {\em arXiv preprint arXiv:2507.05241}.

\bibitem[Chang et~al., 2024]{chang2024survey}
Chang, Y., Wang, X., Wang, J., Wu, Y., Yang, L., Zhu, K., Chen, H., Yi, X.,
  Wang, C., Wang, Y., et~al. (2024).
\newblock A survey on evaluation of large language models.
\newblock {\em ACM transactions on intelligent systems and technology},
  15(3):1--45.

\bibitem[Chen et~al., 2026]{chen2026agenteconomist}
Chen, J., Piao, J., Xu, X., Li, S., Xia, T., He, X., and Li, Y. (2026).
\newblock {AgentEconomist}: An end-to-end agentic system translating economic
  intuitions into executable computational experiments.
\newblock {\em arXiv preprint arXiv:2604.27725}.
\newblock Code: \url{https://github.com/Jiaju-Chen/AgentEconomist}.

\bibitem[Chen et~al., 2025]{chen2025econometricsagent}
Chen, Q., Han, T., Li, J., Luo, Y., Wang, Z., Wu, Y., Zhang, X., and Zhou, T.
  (2025).
\newblock Can {AI} master econometrics? evidence from econometrics {AI} agent
  on expert-level tasks.
\newblock {\em arXiv preprint arXiv:2506.00856}.
\newblock Code: \url{https://github.com/FromCSUZhou/Econometrics-Agent}.

\bibitem[Chen et~al., 2024]{Chen2024a}
Chen, Y., Fang, H., Zhao, Y., and Zhao, Z. (2024).
\newblock Recovering overlooked information in categorical variables with
  {LLMs}: An application to labor market mismatch.
\newblock NBER Working Paper 32327, National Bureau of Economic Research.

\bibitem[Chen et~al., 2023]{Chen2023a}
Chen, Y., Liu, T.~X., Shan, Y., and Zhong, S. (2023).
\newblock The emergence of economic rationality of {GPT}.
\newblock {\em Proceedings of the National Academy of Sciences},
  120(51):e2316205120.

\bibitem[Chen et~al., 2021]{chen2021finqa}
Chen, Z., Chen, W., Smiley, C., Shah, S., Borova, I., Langdon, D., Moussa, R.,
  Beane, M., Huang, T.-H., Routledge, B.~R., et~al. (2021).
\newblock {FinQA}: A dataset of numerical reasoning over financial data.
\newblock In {\em Proceedings of the 2021 Conference on Empirical Methods in
  Natural Language Processing}, pages 3697--3711.

\bibitem[Cohen et~al., 2024]{cohen2024unleashing}
Cohen, S., Bitton, R., and Nassi, B. (2024).
\newblock Unleashing worms and extracting data: Escalating the outcome of
  attacks against {RAG}-based inference in scale and severity using
  jailbreaking.
\newblock {\em arXiv preprint arXiv:2409.08045}.
\newblock Code: \url{https://github.com/StavC/UnleashingWorms-ExtractingData}.

\bibitem[Corazzini et~al., 2025]{corazzini2025decision}
Corazzini, L., Deriu, E., and Guerzoni, M. (2025).
\newblock Decision and gender biases in large language models: A behavioral
  economic perspective.
\newblock {\em arXiv preprint arXiv:2511.12319}.

\bibitem[Curti and Kazinnik, 2023]{curti2023faceit}
Curti, F. and Kazinnik, S. (2023).
\newblock Let's face it: Quantifying the impact of nonverbal communication in
  {FOMC} press conferences.
\newblock {\em Journal of Monetary Economics}, 139:110--126.

\bibitem[Dawid et~al., 2025]{dawid2025agenticv1}
Dawid, H., Harting, P., Wang, H., Wang, Z., and Yi, J. (2025).
\newblock Agentic workflows for economic research: Design and implementation.
\newblock {\em arXiv preprint arXiv:2504.09736v1}.

\bibitem[Dell'Acqua et~al., 2026]{DellAcqua2023}
Dell'Acqua, F., McFowland, III, E., Mollick, E., Lifshitz, H., Kellogg, K.~C.,
  Rajendran, S., Krayer, L., Candelon, F., and Lakhani, K.~R. (2026).
\newblock Navigating the jagged technological frontier: Field experimental
  evidence of the effects of artificial intelligence on knowledge worker
  productivity and quality.
\newblock {\em Organization Science}, 37(2):403--423.

\bibitem[Demirer et~al., 2026]{demirer2026chaining}
Demirer, M., Horton, J.~J., Immorlica, N., Lucier, B., and Shahidi, P. (2026).
\newblock Chaining tasks, redefining work: A theory of {AI} automation.
\newblock NBER Working Paper 34859, National Bureau of Economic Research.
\newblock arXiv:2606.15960.

\bibitem[Doshi and Hauser, 2024]{Doshi2024}
Doshi, A.~R. and Hauser, O.~P. (2024).
\newblock Generative {AI} enhances individual creativity but reduces the
  collective diversity of novel content.
\newblock {\em Science Advances}, 10(28):eadn5290.

\bibitem[Dowling and Lucey, 2023]{Dowling2023}
Dowling, M. and Lucey, B. (2023).
\newblock {ChatGPT} for (finance) research: The {Bananarama} conjecture.
\newblock {\em Finance Research Letters}, 53:103662.

\bibitem[Du et~al., 2024]{du2023improving}
Du, Y., Li, S., Torralba, A., Tenenbaum, J.~B., and Mordatch, I. (2024).
\newblock Improving factuality and reasoning in language models through
  multiagent debate.
\newblock In {\em Forty-first International Conference on Machine Learning}.

\bibitem[Durante et~al., 2024]{Durante2024}
Durante, Z., Huang, Q., Wake, N., Gong, R., Park, J.~S., Sarkar, B., Taori, R.,
  Noda, Y., Terzopoulos, D., Choi, Y., Ikeuchi, K., Vo, H., Fei-Fei, L., and
  Gao, J. (2024).
\newblock Agent {AI}: Surveying the horizons of multimodal interaction.
\newblock arXiv:2401.03568.

\bibitem[Es et~al., 2024]{es2024ragas}
Es, S., James, J., Anke, L.~E., and Schockaert, S. (2024).
\newblock {RAGAS}: Automated evaluation of retrieval augmented generation.
\newblock In {\em Proceedings of the 18th Conference of the European Chapter of
  the Association for Computational Linguistics: System Demonstrations}, pages
  150--158.

\bibitem[Fang et~al., 2025]{fang2025decoding}
Fang, H., Li, M., and Lu, G. (2025).
\newblock Decoding {China}'s industrial policies.
\newblock NBER Working Paper 33814, National Bureau of Economic Research.

\bibitem[Ferrara, 2024]{ferrara2024fairness}
Ferrara, E. (2024).
\newblock Fairness and bias in artificial intelligence: A brief survey of
  sources, impacts, and mitigation strategies.
\newblock {\em Sci}, 6(1):3.

\bibitem[Floridi and Cowls, 2019]{floridi2019unified}
Floridi, L. and Cowls, J. (2019).
\newblock A unified framework of five principles for {AI} in society.
\newblock {\em Harvard Data Science Review}, 1(1).

\bibitem[Gaddipati et~al., 2026]{gaddipati2026mlreplicate}
Gaddipati, S.~K., Muhammed, D., Keya, F., Rabby, G., and Auer, S. (2026).
\newblock {MLReplicate}: Benchmarking autonomous research systems for machine
  learning reproducibility.
\newblock {\em arXiv preprint arXiv:2605.16616}.
\newblock Code: \url{https://github.com/gsasikiran/MLReplicate-benchmarking}.

\bibitem[Galiani et~al., 2026]{galiani2026measuring}
Galiani, S., G{\'a}lvez, R.~H., {Mettola La Giglia}, F., and Sosa, R.~A.
  (2026).
\newblock Measuring efficiency and equity framing in economics research:
  {LLM}-based evidence from 1950 to 2021.
\newblock NBER Working Paper 34714, National Bureau of Economic Research.

\bibitem[Gambacorta et~al., 2024]{Gambacorta2024}
Gambacorta, L., Qiu, H., Shan, S., and Rees, D.~M. (2024).
\newblock Generative {AI} and labour productivity: A field experiment on
  coding.
\newblock BIS Working Paper 1208, Bank for International Settlements.

\bibitem[Goldman, 1999]{Goldman1999}
Goldman, A.~I. (1999).
\newblock {\em Knowledge in a Social World}.
\newblock Oxford University Press, Oxford.

\bibitem[Gorodnichenko et~al., 2023]{gorodnichenko2023voice}
Gorodnichenko, Y., Pham, T., and Talavera, O. (2023).
\newblock The voice of monetary policy.
\newblock {\em American Economic Review}, 113(2):548--584.

\bibitem[Grace et~al., 2026]{grace2026demystifying}
Grace, M., Hadfield, J., Olivares, R., and De~Jonghe, J. (2026).
\newblock Demystifying evals for {AI} agents.
\newblock
  \url{https://www.anthropic.com/engineering/demystifying-evals-for-ai-agents}.
\newblock Accessed February 1, 2026.

\bibitem[Gu et~al., 2025]{gu2025giffluence}
Gu, M., Hirshleifer, D., Teoh, S.~H., and Wu, S. (2025).
\newblock {GIFfluence}: A visual approach to investor sentiment and the stock
  market.
\newblock {\em arXiv preprint arXiv:2512.20027}.

\bibitem[Guo et~al., 2024]{guo2024large}
Guo, T., Chen, X., Wang, Y., Chang, R., Pei, S., Chawla, N.~V., Wiest, O., and
  Zhang, X. (2024).
\newblock Large language model based multi-agents: A survey of progress and
  challenges.
\newblock {\em arXiv preprint arXiv:2402.01680}.

\bibitem[Hartley, 2025]{hartley2025narratives}
Hartley, E. (2025).
\newblock Narratives to numbers: Large language models and economic policy
  uncertainty.
\newblock {\em arXiv preprint arXiv:2511.17866}.

\bibitem[Heikkilä, 2025]{Heikkil2025}
Heikkilä, J.~T. (2025).
\newblock Human intelligence versus artificial intelligence in classifying
  economics research articles: exploratory evidence.
\newblock {\em Journal of Documentation}, 81.

\bibitem[Henderson et~al., 2012]{henderson2012outerspace}
Henderson, J.~V., Storeygard, A., and Weil, D.~N. (2012).
\newblock Measuring economic growth from outer space.
\newblock {\em American Economic Review}, 102(2):994--1028.

\bibitem[Hendrycks et~al., 2020]{hendrycks2020measuring}
Hendrycks, D., Burns, C., Basart, S., Zou, A., Mazeika, M., Song, D., and
  Steinhardt, J. (2020).
\newblock Measuring massive multitask language understanding.
\newblock {\em arXiv preprint arXiv:2009.03300}.

\bibitem[Hollan et~al., 2000]{Hollan2000}
Hollan, J., Hutchins, E., and Kirsh, D. (2000).
\newblock Distributed cognition: Toward a new foundation for human-computer
  interaction research.
\newblock {\em ACM Transactions on Computer-Human Interaction}, 7(2):174--196.

\bibitem[Hollnagel et~al., 2006]{Hollnagel2006}
Hollnagel, E., Woods, D.~D., and Leveson, N., editors (2006).
\newblock {\em Resilience Engineering: Concepts and Precepts}.
\newblock Ashgate Publishing, Aldershot, UK.

\bibitem[Hopkins et~al., 2025]{hopkins2025ai}
Hopkins, A., Cen, S.~H., Struckman, I., Ilyas, A., Videgaray, L., and
  M{\k{a}}dry, A. (2025).
\newblock {AI} supply chains: An emerging ecosystem of {AI} actors, products,
  and services.
\newblock In {\em Proceedings of the AAAI/ACM Conference on AI, Ethics, and
  Society}, volume~8, pages 1266--1277.

\bibitem[Horton et~al., 2023]{Horton2023}
Horton, J.~J., Filippas, A., and Manning, B.~S. (2023).
\newblock Large language models as simulated economic agents: What can we learn
  from homo silicus?
\newblock NBER Working Paper 31122, National Bureau of Economic Research.
\newblock arXiv:2301.07543. Replication code:
  \url{https://github.com/johnjosephhorton/homo_silicus}.

\bibitem[Hutchins, 1995]{Hutchins1995}
Hutchins, E. (1995).
\newblock {\em Cognition in the Wild}.
\newblock MIT Press, Cambridge, MA.

\bibitem[Jean et~al., 2016]{jean2016satellite}
Jean, N., Burke, M., Xie, M., Davis, W.~M., Lobell, D.~B., and Ermon, S.
  (2016).
\newblock Combining satellite imagery and machine learning to predict poverty.
\newblock {\em Science}, 353(6301):790--794.

\bibitem[Jiang et~al., 2025]{jiang2025scp}
Jiang, Y., Lou, W., Wang, L., Tang, Z., Feng, S., Lu, J., Sun, H., Pan, Y., Gu,
  S., Su, H., et~al. (2025).
\newblock {SCP}: Accelerating discovery with a global web of autonomous
  scientific agents.
\newblock {\em arXiv preprint arXiv:2512.24189}.

\bibitem[Jobin et~al., 2019]{jobin2019global}
Jobin, A., Ienca, M., and Vayena, E. (2019).
\newblock The global landscape of {AI} ethics guidelines.
\newblock {\em Nature machine intelligence}, 1(9):389--399.

\bibitem[Karpas et~al., 2022]{karpas2022mrkl}
Karpas, E., Abend, O., Belinkov, Y., Lenz, B., Lieber, O., Ratner, N., Shoham,
  Y., Bata, H., Levine, Y., Leyton-Brown, K., et~al. (2022).
\newblock {MRKL} systems: A modular, neuro-symbolic architecture that combines
  large language models, external knowledge sources and discrete reasoning.
\newblock {\em arXiv preprint arXiv:2205.00445}.

\bibitem[Kim et~al., 2024]{Kim2024}
Kim, J., Kovach, M., Lee, K.-M., Shin, E., and Tzavellas, H. (2024).
\newblock Can an {LLM} learn preferences from choice data?
\newblock {\em arXiv preprint arXiv:2401.07345}.
\newblock Earlier title: ``Learning to be Homo Economicus''.

\bibitem[Kitcher, 1993]{Kitcher1993}
Kitcher, P. (1993).
\newblock {\em The Advancement of Science: Science without Legend, Objectivity
  without Illusions}.
\newblock Oxford University Press, New York.

\bibitem[Korinek, 2023]{korinek2023generative}
Korinek, A. (2023).
\newblock Generative {AI} for economic research: Use cases and implications for
  economists.
\newblock {\em Journal of Economic Literature}, 61(4):1281--1317.

\bibitem[Korinek, 2025]{korinek2025aiagents}
Korinek, A. (2025).
\newblock {AI Agents for Economic Research}.
\newblock NBER Working Paper 34202, National Bureau of Economic Research.

\bibitem[Kuhn, 1962]{Kuhn1962}
Kuhn, T.~S. (1962).
\newblock {\em The Structure of Scientific Revolutions}.
\newblock University of Chicago Press, Chicago.

\bibitem[Kumar et~al., 2024]{kumar2024ethics}
Kumar, A., Murthy, S.~V., Singh, S., and Ragupathy, S. (2024).
\newblock The ethics of interaction: Mitigating security threats in {LLMs}.
\newblock {\em arXiv preprint arXiv:2401.12273}.

\bibitem[Leigh, 2024]{Leigh2024}
Leigh, A. (2024).
\newblock Using artificial intelligence for economic research: An agricultural
  odyssey.
\newblock {\em Australian Journal of Agricultural and Resource Economics},
  68(3):521--529.

\bibitem[Leslie et~al., 2021]{leslie2021artificial}
Leslie, D., Burr, C., Aitken, M., Cowls, J., Katell, M., and Briggs, M. (2021).
\newblock Artificial intelligence, human rights, democracy, and the rule of
  law: a primer.
\newblock {\em arXiv preprint arXiv:2104.04147}.

\bibitem[Li et~al., 2026]{aidev2026codingagents}
Li, H., Zhang, H., and Hassan, A.~E. (2026).
\newblock {AIDev}: Studying {AI} coding agents on {GitHub}.
\newblock {\em arXiv preprint arXiv:2602.09185}.
\newblock Dataset: \url{https://huggingface.co/datasets/hao-li/AIDev}.

\bibitem[Li et~al., 2023]{Li2023b}
Li, L., Chang, T.-Y., and Wang, H. (2023).
\newblock Multimodal {Gen-AI} for fundamental investment research.
\newblock {\em arXiv preprint arXiv:2401.06164}.
\newblock Code: \url{https://github.com/Firenze11/finance_lm}.

\bibitem[Li et~al., 2025]{li2025rdagentquant}
Li, Y., Yang, X., Chen, X., Wang, M., Li, W., Liu, X., and Bian, J. (2025).
\newblock {R\&D-Agent-Quant}: A multi-agent framework for data-centric factors
  and model joint optimization.
\newblock {\em arXiv preprint arXiv:2505.15155}.
\newblock Code: \url{https://github.com/microsoft/RD-Agent}.

\bibitem[Liang et~al., 2022]{liang2022holistic}
Liang, P., Bommasani, R., Lee, T., Tsipras, D., Soylu, D., Yasunaga, M., Zhang,
  Y., Narayanan, D., Wu, Y., Kumar, A., et~al. (2022).
\newblock Holistic evaluation of language models.
\newblock {\em arXiv preprint arXiv:2211.09110}.

\bibitem[Lin et~al., 2022]{lin2022truthfulqa}
Lin, S., Hilton, J., and Evans, O. (2022).
\newblock {TruthfulQA}: Measuring how models mimic human falsehoods.
\newblock In {\em Proceedings of the 60th annual meeting of the association for
  computational linguistics (volume 1: long papers)}, pages 3214--3252.

\bibitem[Liu et~al., 2023]{liu2023agentbench}
Liu, X., Yu, H., Zhang, H., Xu, Y., Lei, X., Lai, H., Gu, Y., Ding, H., Men,
  K., Yang, K., et~al. (2023).
\newblock {AgentBench}: Evaluating {LLMs} as agents.
\newblock {\em arXiv preprint arXiv:2308.03688}.

\bibitem[Ludwig et~al., 2026]{ludwig2025large}
Ludwig, J., Mullainathan, S., and Rambachan, A. (2026).
\newblock Large language models: An applied econometric framework.
\newblock {\em Annual Review of Economics}, 18(1):283--316.
\newblock NBER Working Paper 33344.

\bibitem[Ma et~al., 2026]{ma2026maestro}
Ma, T., Chen, Y., Anand, V., Cornacchia, A., Faustino, A.~R., Liu, G., Zhang,
  S., Luo, H., Fahmy, S.~A., Qazi, Z.~A., and Canini, M. (2026).
\newblock {MAESTRO}: Multi-agent evaluation suite for testing, reliability, and
  observability.
\newblock {\em arXiv preprint arXiv:2601.00481}.
\newblock Code: \url{https://github.com/sands-lab/maestro}.

\bibitem[Manakul et~al., 2023]{manakul2023selfcheckgpt}
Manakul, P., Liusie, A., and Gales, M. (2023).
\newblock {SelfCheckGPT}: Zero-resource black-box hallucination detection for
  generative large language models.
\newblock In {\em Proceedings of the 2023 conference on empirical methods in
  natural language processing}, pages 9004--9017.

\bibitem[Manning and Horton, 2025]{manning2026generalsocialagents}
Manning, B.~S. and Horton, J.~J. (2025).
\newblock General social agents.
\newblock {\em arXiv preprint arXiv:2508.17407}.

\bibitem[Merton, 1973]{Merton1973}
Merton, R.~K. (1973).
\newblock {\em The Sociology of Science: Theoretical and Empirical
  Investigations}.
\newblock University of Chicago Press, Chicago.

\bibitem[{Microsoft}, 2026]{microsoftAgentFramework2026}
{Microsoft} (2026).
\newblock {Microsoft Agent Framework}: Overview.
\newblock
  \url{https://learn.microsoft.com/en-us/agent-framework/overview/agent-framework-overview}.
\newblock Accessed 2026-09-17.

\bibitem[Min et~al., 2023]{min2023factscore}
Min, S., Krishna, K., Lyu, X., Lewis, M., Yih, W.-t., Koh, P., Iyyer, M.,
  Zettlemoyer, L., and Hajishirzi, H. (2023).
\newblock {FActScore}: Fine-grained atomic evaluation of factual precision in
  long form text generation.
\newblock In {\em Proceedings of the 2023 Conference on Empirical Methods in
  Natural Language Processing}, pages 12076--12100.

\bibitem[Motoki et~al., 2025]{Motoki2025}
Motoki, F. Y.~S., Pinho~Neto, V., and Rangel, V. (2025).
\newblock Assessing political bias and value misalignment in generative
  artificial intelligence.
\newblock {\em Journal of Economic Behavior \& Organization}, 234:106904.
\newblock Replication data and code: \url{https://doi.org/10.7910/DVN/VZRKWP}.

\bibitem[Nakano et~al., 2021]{nakano2021webgpt}
Nakano, R., Hilton, J., Balaji, S., Wu, J., Ouyang, L., Kim, C., Hesse, C.,
  Jain, S., Kosaraju, V., Saunders, W., et~al. (2021).
\newblock {WebGPT}: Browser-assisted question-answering with human feedback.
\newblock {\em arXiv preprint arXiv:2112.09332}.

\bibitem[{National Institute of Standards and Technology},
  2026]{nistAgentStandards2026}
{National Institute of Standards and Technology} (2026).
\newblock {AI} agent standards initiative.
\newblock \url{https://www.nist.gov/caisi/ai-agent-standards-initiative}.
\newblock Accessed 2026-09-18.

\bibitem[Noy and Zhang, 2023]{Noy2023}
Noy, S. and Zhang, W. (2023).
\newblock Experimental evidence on the productivity effects of generative
  artificial intelligence.
\newblock {\em Science}, 381(6654):187--192.

\bibitem[{NVIDIA}, 2026]{nvidiaAIQBlueprint2026}
{NVIDIA} (2026).
\newblock {NVIDIA AI-Q Blueprint}: An open reference example for enterprise
  research agents.
\newblock \url{https://github.com/NVIDIA-AI-Blueprints/aiq}.
\newblock Accessed 2026-09-17.

\bibitem[OpenAI, 2025]{OpenAI2025}
OpenAI (2025).
\newblock Privacy policy.

\bibitem[{OpenAI}, 2026]{openaiAgentsSDK2026}
{OpenAI} (2026).
\newblock {OpenAI Agents SDK}: Documentation.
\newblock \url{https://openai.github.io/openai-agents-python/}.
\newblock Accessed 2026-09-17.

\bibitem[{OWASP GenAI Security Project}, 2025]{owasptop10agentic2026}
{OWASP GenAI Security Project} (2025).
\newblock {OWASP} top 10 for agentic applications for 2026.
\newblock
  \url{https://genai.owasp.org/resource/owasp-top-10-for-agentic-applications/}.
\newblock Accessed 2026-09-18.

\bibitem[Peng et~al., 2023]{Peng2023}
Peng, S., Kalliamvakou, E., Cihon, P., and Demirer, M. (2023).
\newblock {The Impact of AI on Developer Productivity: Evidence from GitHub
  Copilot}.
\newblock {\em arXiv preprint arXiv:2302.06590}.

\bibitem[Plaat et~al., 2025]{plaat2025agentic}
Plaat, A., van Duijn, M., van Stein, N., Preuss, M., van~der Putten, P., and
  Batenburg, K.~J. (2025).
\newblock Agentic large language models, a survey.
\newblock {\em arXiv preprint arXiv:2503.23037}.

\bibitem[Popper, 1959]{Popper1959}
Popper, K.~R. (1959).
\newblock {\em The Logic of Scientific Discovery}.
\newblock Hutchinson, London.
\newblock Originally published in German as Logik der Forschung, 1934.

\bibitem[Rabanser et~al., 2026]{rabanser2026reliability}
Rabanser, S., Kapoor, S., Kirgis, P., Liu, K., Utpala, S., and Narayanan, A.
  (2026).
\newblock Towards a science of {AI} agent reliability.
\newblock {\em arXiv preprint arXiv:2602.16666}.
\newblock Code: \url{https://github.com/princeton-pli/hal-harness}.

\bibitem[Ross et~al., 2024]{Ross2024}
Ross, J., Kim, Y., and Lo, A.~W. (2024).
\newblock {LLM} economicus? {Mapping} the behavioral biases of {LLMs} via
  utility theory.
\newblock In {\em First Conference on Language Modeling (COLM)}.
\newblock arXiv:2408.02784.

\bibitem[Ruan et~al., 2024]{ruan2024llmrdf}
Ruan, Y., Lu, C., Xu, N., He, Y., Chen, Y., Zhang, J., Xuan, J., Pan, J., Fang,
  Q., Gao, H., Shen, X., Ye, N., Zhang, Q., and Mo, Y. (2024).
\newblock An automatic end-to-end chemical synthesis development platform
  powered by large language models.
\newblock {\em Nature Communications}, 15:10160.
\newblock Code: \url{https://github.com/Ruan-Yixiang/LLM-RDF}.

\bibitem[Schmidgall et~al., 2025]{schmidgall2025agentlaboratory}
Schmidgall, S., Su, Y., Wang, Z., Sun, X., Wu, J., Yu, X., Liu, J., Liu, Z.,
  and Barsoum, E. (2025).
\newblock Agent laboratory: Using {LLM} agents as research assistants.
\newblock {\em arXiv preprint arXiv:2501.04227}.
\newblock Code: \url{https://github.com/SamuelSchmidgall/AgentLaboratory}.

\bibitem[Schneider, 2025]{schneider2025generative}
Schneider, J. (2025).
\newblock Generative to agentic {AI}: Survey, conceptualization, and
  challenges.
\newblock {\em arXiv preprint arXiv:2504.18875}.

\bibitem[Schreyer et~al., 2024]{Schreyer2025}
Schreyer, M., Gu, H., Moffitt, K., and Vasarhelyi, M.~A. (2024).
\newblock Artificial intelligence agentic auditing.
\newblock SSRN Working Paper 4909147.

\bibitem[Sellam et~al., 2020]{sellam2020bleurt}
Sellam, T., Das, D., and Parikh, A. (2020).
\newblock {BLEURT}: Learning robust metrics for text generation.
\newblock In {\em Proceedings of the 58th annual meeting of the association for
  computational linguistics}, pages 7881--7892.

\bibitem[Shao et~al., 2026]{shao2025sciscigpt}
Shao, E., Wang, Y., Qian, Y., Pan, Z., Liu, H., and Wang, D. (2026).
\newblock {SciSciGPT}: advancing human--{AI} collaboration in the science of
  science.
\newblock {\em Nature Computational Science}, 6(3):301--315.
\newblock Code: \url{https://github.com/Northwestern-CSSI/SciSciGPT}.

\bibitem[Shao et~al., 2024]{shao2024storm}
Shao, Y., Jiang, Y., Kanell, T.~A., Xu, P., Khattab, O., and Lam, M.~S. (2024).
\newblock Assisting in writing {Wikipedia}-like articles from scratch with
  large language models.
\newblock In {\em Proceedings of the 2024 Conference of the North American
  Chapter of the Association for Computational Linguistics (NAACL)}.
\newblock arXiv:2402.14207. Code: \url{https://github.com/stanford-oval/storm}.

\bibitem[Skarlinski et~al., 2024]{skarlinski2024paperqa2}
Skarlinski, M.~D., Cox, S., Laurent, J.~M., Braza, J.~D., Hinks, M.,
  Hammerling, M.~J., Ponnapati, M., Rodriques, S.~G., and White, A.~D. (2024).
\newblock Language agents achieve superhuman synthesis of scientific knowledge.
\newblock {\em arXiv preprint arXiv:2409.13740}.
\newblock Code: \url{https://github.com/Future-House/paper-qa}.

\bibitem[Swanson et~al., 2025]{swanson2025virtuallab}
Swanson, K., Wu, W., Bulaong, N.~L., Pak, J.~E., and Zou, J. (2025).
\newblock The {Virtual Lab} of {AI} agents designs new {SARS-CoV-2} nanobodies.
\newblock {\em Nature}, 646:716--723.
\newblock Code: \url{https://github.com/zou-group/virtual-lab}.

\bibitem[Tang et~al., 2025]{tang2025autoagent}
Tang, J., Fan, T., and Huang, C. (2025).
\newblock {AutoAgent}: A fully-automated and zero-code framework for {LLM}
  agents.
\newblock {\em arXiv preprint arXiv:2502.05957}.

\bibitem[Thagard, 1992]{Thagard1992}
Thagard, P. (1992).
\newblock {\em Conceptual Revolutions}.
\newblock Princeton University Press, Princeton, NJ.

\bibitem[Tomasev et~al., 2025]{tomasev2025virtual}
Tomasev, N., Franklin, M., Leibo, J.~Z., Jacobs, J., Cunningham, W.~A.,
  Gabriel, I., and Osindero, S. (2025).
\newblock Virtual agent economies.
\newblock {\em arXiv preprint arXiv:2509.10147}.

\bibitem[Tranchero et~al., 2024]{Tranchero2024}
Tranchero, M., Brenninkmeijer, C.-F., Murugan, A., and Nagaraj, A. (2024).
\newblock Theorizing with large language models.
\newblock NBER Working Paper 33033.

\bibitem[Verma and Unnikrishnan, 2025]{verma2025ai}
Verma, P. and Unnikrishnan, B. (2025).
\newblock When {AI} democratizes exploitation: {LLM}-assisted strategic
  manipulation of fair division algorithms.
\newblock {\em arXiv preprint arXiv:2511.14722}.

\bibitem[Vijayvargiya et~al., 2026]{vijayvargiya2025openagentsafety}
Vijayvargiya, S., Soni, A.~B., Zhou, X., Wang, Z.~Z., Dziri, N., Neubig, G.,
  and Sap, M. (2026).
\newblock {OpenAgentSafety}: A comprehensive framework for evaluating
  real-world {AI} agent safety.
\newblock In {\em International Conference on Learning Representations (ICLR)}.
\newblock arXiv:2507.06134. Code:
  \url{https://github.com/Open-Agent-Safety/OpenAgentSafety}.

\bibitem[Wang et~al., 2024a]{wang2024survey}
Wang, L., Ma, C., Feng, X., Zhang, Z., Yang, H., Zhang, J., Chen, Z., Tang, J.,
  Chen, X., Lin, Y., et~al. (2024a).
\newblock A survey on large language model based autonomous agents.
\newblock {\em Frontiers of Computer Science}, 18(6):186345.

\bibitem[Wang et~al., 2024b]{wang2024executable}
Wang, X., Chen, Y., Yuan, L., Zhang, Y., Li, Y., Peng, H., and Ji, H. (2024b).
\newblock Executable code actions elicit better {LLM} agents.
\newblock In {\em Forty-first International Conference on Machine Learning}.

\bibitem[Wang et~al., 2025]{wang2025empowering}
Wang, X., Tang, Z., Guo, J., Meng, T., Wang, C., Wang, T., and Jia, W. (2025).
\newblock Empowering edge intelligence: A comprehensive survey on on-device
  {AI} models.
\newblock {\em ACM Computing Surveys}, 57(9):1--39.

\bibitem[Waugh, 2026]{waugh2026tradeai}
Waugh, M.~E. (2026).
\newblock Trade in {AI}-related products.
\newblock NBER Working Paper 35053, National Bureau of Economic Research.
\newblock Code: \url{https://github.com/tradewartracker/ai-trade-index}.

\bibitem[Wei et~al., 2022]{wei2022chain}
Wei, J., Wang, X., Schuurmans, D., Bosma, M., Xia, F., Chi, E., Le, Q.~V.,
  Zhou, D., et~al. (2022).
\newblock Chain-of-thought prompting elicits reasoning in large language
  models.
\newblock {\em Advances in neural information processing systems},
  35:24824--24837.

\bibitem[Woods, 2017]{Woods2017}
Woods, D.~D. (2017).
\newblock {STELLA}: Report from the {SNAFUcatchers} workshop on coping with
  complexity.
\newblock The Ohio State University.

\bibitem[Wu et~al., 2024]{wu2024autogen}
Wu, Q., Bansal, G., Zhang, J., Wu, Y., Li, B., Zhu, E., Jiang, L., Zhang, X.,
  Zhang, S., Liu, J., et~al. (2024).
\newblock {AutoGen}: Enabling next-gen {LLM} applications via multi-agent
  conversations.
\newblock In {\em First Conference on Language Modeling}.

\bibitem[Xi et~al., 2025]{xi2025rise}
Xi, Z., Chen, W., Guo, X., He, W., Ding, Y., Hong, B., Zhang, M., Wang, J.,
  Jin, S., Zhou, E., et~al. (2025).
\newblock The rise and potential of large language model based agents: A
  survey.
\newblock {\em Science China Information Sciences}, 68(2):121101.

\bibitem[Xue et~al., 2024]{xue2024demonstration}
Xue, S., Qi, D., Jiang, C., Shi, W., Cheng, F., Chen, K., Yang, H., Zhang, Z.,
  He, J., Zhang, H., et~al. (2024).
\newblock Demonstration of {DB-GPT}: Next generation data interaction system
  empowered by large language models.
\newblock {\em arXiv preprint arXiv:2404.10209}.
\newblock Code: \url{https://github.com/eosphoros-ai/DB-GPT}.

\bibitem[Yamada et~al., 2025]{yamada2025aiscientistv2}
Yamada, Y., Lange, R.~T., Lu, C., Hu, S., Lu, C., Foerster, J., Clune, J., and
  Ha, D. (2025).
\newblock The {AI} scientist-v2: Workshop-level automated scientific discovery
  via agentic tree search.
\newblock {\em arXiv preprint arXiv:2504.08066}.
\newblock Code: \url{https://github.com/SakanaAI/AI-Scientist-v2}.

\bibitem[Yao et~al., 2023]{yao2022react}
Yao, S., Zhao, J., Yu, D., Du, N., Shafran, I., Narasimhan, K.~R., and Cao, Y.
  (2023).
\newblock {ReAct}: Synergizing reasoning and acting in language models.
\newblock In {\em The eleventh international conference on learning
  representations}.

\bibitem[Ye et~al., 2024]{ye2024mirai}
Ye, C., Hu, Z., Deng, Y., Huang, Z., Ma, M.~D., Zhu, Y., and Wang, W. (2024).
\newblock {MIRAI}: Evaluating {LLM} agents for event forecasting.
\newblock {\em arXiv preprint arXiv:2407.01231}.

\bibitem[Zhang et~al., 2019]{zhang2019bertscore}
Zhang, T., Kishore, V., Wu, F., Weinberger, K.~Q., and Artzi, Y. (2019).
\newblock {BERTScore}: Evaluating text generation with {BERT}.
\newblock {\em arXiv preprint arXiv:1904.09675}.

\bibitem[Zhao et~al., 2024]{Zhao2024}
Zhao, Q., Wang, J., Zhang, Y., Jin, Y., Zhu, K., Chen, H., and Xie, X. (2024).
\newblock {CompeteAI}: Understanding the competition dynamics of large language
  model-based agents.
\newblock In {\em Proceedings of the 41st International Conference on Machine
  Learning (ICML)}, volume 235 of {\em PMLR}.
\newblock arXiv:2310.17512. Code: \url{https://github.com/microsoft/competeai}.

\bibitem[Zheng et~al., 2023]{Zheng2023}
Zheng, L., Chiang, W.-L., Sheng, Y., Zhuang, S., Wu, Z., Zhuang, Y., Lin, Z.,
  Li, Z., Li, D., Xing, E.~P., Zhang, H., Gonzalez, J.~E., and Stoica, I.
  (2023).
\newblock {Judging LLM-as-a-Judge with MT-Bench and Chatbot Arena}.
\newblock In {\em Advances in Neural Information Processing Systems},
  volume~36, pages 46595--46623.

\bibitem[Zheng et~al., 2024]{Zheng2024}
Zheng, X., Li, J., Lu, M., and Wang, F.-Y. (2024).
\newblock New paradigm for economic and financial research with generative
  {AI}: Impact and perspective.
\newblock {\em IEEE Transactions on Computational Social Systems},
  11(3):3457--3467.

\bibitem[Zheng et~al., 2025]{agentsight2025}
Zheng, Y., Hu, Y., Yu, T., and Quinn, A. (2025).
\newblock {AgentSight}: System-level observability for {AI} agents using
  {eBPF}.
\newblock In {\em Proceedings of the 4th Workshop on Practical Adoption
  Challenges of ML for Systems (PACMI '25)}.
\newblock arXiv:2508.02736. Code:
  \url{https://github.com/agent-sight/agentsight}.

\end{thebibliography}

\end{document}